\documentclass{amsart}
\usepackage{yfonts}
\usepackage{amsmath}%
\usepackage{amsfonts}%
\usepackage{amssymb}%
\usepackage{graphicx}
\usepackage{enumitem}
\usepackage{setspace}
\usepackage[dvipsnames]{xcolor}
\usepackage{comment}
\usepackage[utf8]{inputenc}
\usepackage{physics}

\theoremstyle{plain}
\newtheorem{theorem}{Theorem}[section]
\newtheorem{definition}[theorem]{Definition}

\newtheorem{corollary}[theorem]{Corollary}

\newtheorem{example}[theorem]{Example}

\newtheorem{lemma}[theorem]{Lemma}

\newtheorem{proposition}[theorem]{Proposition}
\newtheorem{remark}[theorem]{Remark}

\numberwithin{equation}{section}

\begin{document}
\title[Spin  correspondences: localization \& quantization]{Asymptotic localization of\\symbol correspondences for spin systems\\and sequential quantizations of $S^2$}
\author[P.A.S. Alc\^antara]{P.A.S. Alc\^antara}
\author[P. de M. Rios]{P. de M. Rios}
\thanks{The authors thank CAPES (finance code 001) for support.}
\address{Instituto de Ci\^encias Matem\'aticas e de Computa\c{c}\~ao, Universidade de S\~ao Paulo. \newline
S\~ao Carlos, SP, Brazil.}
\email{pedro.antonio.alcantara@usp.br}
\email{prios@icmc.usp.br}



\begin{abstract}
Quantum or classical mechanical systems symmetric under $SU(2)$ are called spin systems. A $SU(2)$-equivariant map from $(n+1)$-square matrices to functions on the $2$-sphere $S^2$, satisfying some basic properties, is called a spin-$j$ symbol correspondence ($n=2j\in\mathbb N$). Given a spin-$j$ symbol correspondence, the matrix algebra induces a twisted $j$-algebra of symbols. 
In the first part of this paper, we establish a more intuitive criterion for when 
the Poisson algebra of smooth functions on $S^2$ emerges asymptotically ($n\to\infty$) from the sequence of twisted $j$-algebras. This more geometric criterion, which in many cases is equivalent to the numerical criterion obtained in \cite{prios} for describing symbol correspondence sequences of (anti-)Poisson type, is now given in terms of a classical (asymptotic) localization of symbols of all projectors (quantum pure states) in a certain family. For some important kinds of symbol correspondence sequences, such a classical localization condition is equivalent to  asymptotic emergence of the Poisson algebra. But in general, the classical localization condition 
is stronger than Poisson emergence. We thus also consider some weaker notions of asymptotic localization of projector-symbols. In the second part of this paper, for each sequence of symbol correspondences of (anti-)Poisson type, we define the sequential quantization of a smooth function on $S^2$ and its asymptotic operator acting on a ground Hilbert space. Then, after presenting some concrete examples of these constructions, we obtain some relations between asymptotic localization of a symbol correspondence sequence and the asymptotics of its sequential quantization of smooth functions on $S^2$.

\end{abstract}
\maketitle

\tableofcontents

\section{Introduction}

Quantum or classical mechanical systems which are symmetric under $SU(2)$ are called spin systems. While the Poisson algebra of the classical spin systems is infinite dimensional (smooth functions on the $2$-sphere $S^2$), the operator algebra of a quantum spin-$j$ system is finite dimensional (linear operators on $\mathbb C^{n+1}$, where $n=2j\in\mathbb N$). Thus, one is naturally led to ask if, or under which conditions, the Poisson algebra of the classical spin system emerges as the asymptotic limit of the operator algebra of quantum spin-$j$ systems, as $j\to\infty$. But in full generality, this question can only be well posed if we first specify a sequence of injective $SU(2)$-equivariant linear maps from linear operators on $\mathbb C^{n+1}$ to smooth functions on $S^2$ (satisfying a few other properties), a so-called sequence of {\it symbol correspondences}.    

This question has been addressed and answered in the research monograph \cite{prios}.
An important conclusion of \cite{prios} is that, given a generic sequence of symbol correspondences, 
the sequence of 
spin-$j$ quantum systems do not approach the classical spin system as $j$ grows indefinitely. 
Actually, for spin systems,
a necessary and sufficient condition was found 
in \cite{prios} 
for the asymptotic emergence of the classical Poisson algebra from a sequence of  {\it twisted algebras} of the corresponding symbols. Symbol correspondence sequences with this property are said to be of Poisson or anti-Poisson type and the necessary and sufficient condition for such  
was stated in \cite{prios} in numerical terms, as a non-generic asymptotic condition on the sequence of {\it characteristic numbers} of the respective symbol correspondences, as follows.

For $n=2j\in\mathbb N$,\footnote{In this paper, we adopt the usual convention $\mathbb N=\{1,2,3,\cdots\}$ and denote by $\mathbb{N}_0=\mathbb N\cup\{0\}$.} a \emph{quantum spin-$j$ system} is a complex Hilbert space $\mathcal{H}_j\simeq \mathbb{C}^{n+1}$ together with an irreducible unitary representation $\varphi_j : SU(2)\to U(n+1)$. Then, a  {\it spin-$j$ symbol correspondence} $W^j$ is a linear injective and $SO(3)$-equivariant map\footnote{The action of $SU(2)$ on both spaces, $M_{\mathbb C}(n+1)$ and $C^{\infty}_{\mathbb C}(S^2)$, is effectively an $SO(3)$ action.} 
\begin{equation}\label{Wj1} 
W^j:M_{\mathbb C}(n+1)\to C^{\infty}_{\mathbb C}(S^2) \ , \ P\mapsto W^j_P \ ,
\end{equation}
such that $W^j_{P^*}=\overline{W^j_P}$ and $W^j_I=1$,  {\it cf.} Definition \ref{corresp}. By decomposing the operator space  into $SO(3)$-invariant subspaces (indexed by $l$, \ $0\leq l\leq n$), we establish a standard basis $\{\boldsymbol{e}^j(l,m)\}_{-l\leq m\leq l\leq n}$ of $M_{\mathbb C}(n+1)$, {\it cf.} Theorem \ref{ejlm}, so that 
\begin{equation}\label{Wjc}
W^j: \sqrt{n+1} \boldsymbol{e}^j(l,m) \mapsto c_l^nY_l^m \ , \end{equation}
where $Y_l^m$ are the spherical harmonics, and the $n$ nonzero real numbers\footnote{$W^j_I\equiv 1 \iff c_0^n\equiv 1$.} 
\begin{equation}\label{char}
c_l^n\in\mathbb R^* \ , \ \ 1\leq l\leq n=2j \ ,
\end{equation} 
are the {\it characteristic numbers} of $W^j$, {\it cf.} Theorem \ref{symb_c}. 

Given a  symbol correspondence $W^j$, the operator product is ``imported'' as a product of functions on the sphere, the {\it twisted product of symbols}, $\star^n_{\vec{c}}$ , defined by 
\begin{equation}\label{prodchar}
W^j_P\star^n_{\vec{c}} W^j_Q = W^j_{PQ} \ ,
\end{equation} 
where $\vec{c}=(c_1^n,\cdots,c^n_n)$ denotes the $n$-tuple of characteristic numbers of $W^j$. 

Now, the standard area form on homogeneous  $S^2$ is a $SO(3)$-invariant symplectic form, 
then, a {\it symbol correspondence sequence} $\boldsymbol W=(W^j)_{n\in\mathbb N}$ is of {\it Poisson type} if the asymptotic $n\to\infty$ limit of the twisted product of symbols always coincides with their pointwise product and if, to first order in $1/n$, the  asymptotic limit of the {\it twisted commutator} of symbols equals $\sqrt{-1}$ times their Poisson bracket (being of {\it anti-Poisson type} if 
it equals $-\sqrt{-1}$ times their Poisson bracket).

In fact, care is needed with this asymptotic limit because, although the image of any operator by (\ref{Wj1})-(\ref{Wjc}) is a smooth function, the limit of images of operators in a sequence may not belong to $C^{\infty}_{\mathbb C}(S^2)$, and this is intrinsically related to the subject of this paper. But this subtlety can be circumvented in order to well define symbol correspondence sequences of (anti-)Poisson type, {\it cf.} Definition \ref{p_ap}.

Then it was shown in \cite{prios} that, for any symbol correspondence sequence $\boldsymbol W$, 
\begin{equation}\label{RS}
\begin{aligned} 
\mbox{Poisson type}  &\iff
\lim_{n\to\infty}c_l^n=1 \ , \ \forall l\in\mathbb N \ , \\
\mbox{anti-Poisson type}  &\iff
\lim_{n\to\infty}c_l^n=(-1)^l \ , \ \forall l\in\mathbb N \ , \end{aligned}
\end{equation} 
 {\it cf.} Theorem \ref{conviso} below \cite[Theorem 8.2.21]{prios}. Thus, being of (anti-)Poisson type is a nongeneric condition for a symbol correspondence sequence because the sequence of  symbol products, defined by (\ref{Wjc})-(\ref{prodchar}), clearly depends (and only depends) on the bi-sequence $(c_l^n)_{l\leq n\in\mathbb N}$ of characteristic numbers, but condition (\ref{RS}) is obviously too stringent for a generic bi-sequence $(c_l^n)_{l\leq n\in\mathbb N}$ satisfying just (\ref{char}).

 Furthermore, as also shown in \cite{prios}, this asymptotic condition fails to be generic even in each of two very important subsets of symbol correspondence sequences: the subset consisting of sequences of 
 {\it isometric} (Stratonovich-Weyl) correspondences, for which the map (\ref{Wjc}) is an isometry\footnote{With respect to the normalized inner products on both spaces, {\it cf.} Definition \ref{SW}.}, and the subset of sequences of {\it mapping-positive} (coherent-state) correspondences, for which the map (\ref{Wjc}) is positive.
 
For isometric correspondences there is the additional requirement: $|c_l^n|\equiv 1$, but because the signs can generically be anything, condition (\ref{RS}) is still too stringent. The requirements on the $c_l^n$'s for a symbol correspondence to be mapping-positive are less strict, hence condition (\ref{RS}) is more stringent on this subset. In fact, these two subsets, isometric and mapping-positive, 
 were conjectured in \cite{prios} to be disjoint. In this paper, this important conjecture is proved in Theorem \ref{propgood}.

However, given the relevance of singling out the symbol correspondence sequences of Poisson or anti-Poisson type, we were still missing a  geometrical or ``physically intuitive'' interpretation for when a sequence of symbol correspondences takes the sequence of spin-$j$ quantum systems to the classical spin system in the asymptotic $j\to\infty$ limit. In other words, we wanted a better answer to the question: 
\begin{equation}\label{mainQ}
\mbox{How can we ``see'' condition (\ref{RS})?}
\end{equation}
This question is answered in this paper by looking at the asymptotic localization of the symbols of pure $J_3$-invariant states\footnote{For affine quantum systems, {\it i.e.} ordinary quantum mechanics with Hilbert space $L^2(\mathbb R^k)$, it's well known  that Ehrenfest's theorem fails to hold in general for pure states which are not sufficiently localized in $\mathbb R^k$ ({\it cf. e.g.} \cite{cohen}), thus the notion that emergence of Poisson mechanics from quantum mechanics should be related to (asymptotic) localization of pure states is somewhat ``physically intuitive''. Because we look at $J_3$-invariant pure states in the case of spin systems, the analogy with affine systems is perhaps clearer if there we think of localization of momentum eigenstates.}.

For a quantum spin-$j$ system, if $\vec{J}=(J_1,J_2,J_3)$ is the operator of angular momentum, 
a {\it pure $J_3$-invariant state} is a vector  $\boldsymbol{u}(j, m)\in\mathcal{H}_j\simeq \mathbb{C}^{n+1}$ satisfying $J_3\boldsymbol{u}(j, m)=m\boldsymbol{u}(j, m)$, $-j\leq m\leq j$, and these comprise the standard basis of the spin-$j$ system, indexed from highest ($m=j$) to lowest ($m=-j$) weight, so that a pure $J_3$-state can also be seen as a projector onto the $k^{th}$ subspace, $\Pi_{k}\in M_{\mathbb C}(n+1)$, 
$1\leq k\leq n+1$,  with $k=j-m+1$, and the convex hull of all projectors $\Pi_{k}$ constitute  the set of all $J_3$-invariant states (pure and mixed).

Considering the sequence of quantum spin-$j$ systems, for a natural sequence 
$$(k_n)_{n\in\mathbb N} \ \ \mbox{satisfying} \ \ k_n\in\mathbb N  , \ 1\leq k_n\leq n+1  ,$$ 
the sequence $(\Pi_{k_n})_{n\in\mathbb N}$ of projectors is said to be {\it $r$-convergent} if $$k_n/n\to r\in[0,1] \ , \  \mbox{as} \ \ n\to\infty \ ,$$ and in this case, given a sequence of symbol correspondences $\boldsymbol W=(W^j)_{n\in\mathbb N}$, 
 the sequence $(\rho^j_{k_n})_{n\in\mathbb N}$ of quasi-probability distributions on $[-1,1]$, given by  $$\rho^j_{k_n}=\frac{n+1}{2}W^j_{\Pi_{k_n}} \ \  \mbox{restricted to $z$-axis} \ ,$$  
 is said to be an {\it $r$-convergent $\Pi$-distribution sequence}, {\it cf.} Definition \ref{pi-sym}.  
Then, a symbol correspondence sequence $\boldsymbol W=(W^j)_{n\in\mathbb N}$ {\it localizes} (resp. {\it anti-localizes}) {\it classically}  if  every  $r$-convergent  $\Pi$-distribution sequence localizes classically at
$$z_0=1-2r \ \ \mbox{(resp.} \ \ z_0=2r-1\mbox{)} \ \in \ [-1,1] \ . $$ 

In the above sentence, the word ``every'' is important and by {\it classical localization at} $z_0$ we mean that the sequence of quasi-probability distributions converges, as distribution, to Dirac's $\delta(z-z_0)$ distribution on $C^{\infty}_{\mathbb C}([-1,1])$, in the sense that   
\begin{equation}\label{limloc}
\lim_{n\to\infty}\int_{-1}^1f(z)\rho^j_{k_n}(z)dz=f(z_0) \ , \ \forall f\in C^{\infty}_{\mathbb C}([-1,1]) \ ,
\end{equation} 
{\it cf.} Definitions \ref{loc}-\ref{cloc}. Hence, for all practical purposes, if a symbol correspondence sequence (anti-)localizes classically,  
for every $r\in[0,1]$ we  can ``see'' the symbols of any $r$-convergent  $\Pi$-distribution sequence ``concentrating'' asymptotically on the parallel of colatitude $\varphi=\arccos(z_0)$ on the sphere (the ``classical picture'').

Then, Corollaries \ref{l_p}, \ref{final} and \ref{SWP=AL} state that a sequence of mapping-positive symbol correspondences,  or else a sequence of isometric symbol correspondences,   is of (anti-)Poisson type if and only if the given symbol correspondence sequence (anti-)localizes  classically. In fact, Corollary \ref{l_p} presents one of the main results of this paper, providing a partial answer to question (\ref{mainQ}) by stating that, for any symbol correspondence sequence $\boldsymbol W$,
\begin{equation}\label{stronger}
\mbox{classical (anti-)localization} \ \Rightarrow \ \mbox{(anti-)Poisson type.}
\end{equation}

However, the converse is not true in general and thus such a classical localization condition is a stronger condition than asymptotic Poisson emergence, for general symbol correspondence sequences. This is another main result of the paper, which is presented in  Theorem \ref{P<loc}. 
Before, Theorem \ref{p_l} presented yet another important result, providing sufficient conditions for classical  (anti-)localization of a symbol correspondence sequence of (anti-)Poisson type, as:  
\begin{equation}\label{polb}
\mbox{direction} \ (\Leftarrow) \ \mbox{in (\ref{stronger})} \ \mbox{holds for a polynomial growth with} \ l \ \mbox{of} \ |c_l^n| \ ,
\end{equation}  
{\it cf.} (\ref{c<d}) for the more explicit bounds. Accordingly, the characteristic numbers of mapping-positive symbol correspondence sequences and of isometric symbol correspondence sequences, both satisfy the polynomial bounds in (\ref{c<d}).

Hence, the standard (resp. alternate) Berezin symbol correspondence sequence, being mapping-positive, and the standard (resp. alternate) Stratonovich-Weyl symbol correspondence sequence, being isometric, both being of Poisson (resp. anti-Poisson) type, both localize (resp. anti-localize)  classically.

Now, for general symbol correspondences there is a well defined notion of duality, {\it cf.} Definition \ref{dualdef}, and the dual of a mapping-positive symbol correspondence is a \emph{positive-dual} symbol correspondence, {\it cf.} Definition \ref{posdualdef}. 
It turns out that, in general, the characteristic numbers of positive-dual symbol correspondence sequences of (anti-)Poisson type do not satisfy the polynomial bounds in  (\ref{c<d}). 

However, there are important examples of positive-dual symbol correspondence sequences of (anti-)Poisson type, specially the standard (resp. alternate) Toeplitz corresspondence sequence, dual of the standard (resp. alternate) Berezin correspondence sequence,
for which their $|c_l^n|$ have  well behaved exponential growth with $l$, allowing us to consider some weaker notions of asymptotic localization of a symbol correspondence sequence, encompassing these special examples.

For such a generalization, called {\it $\mu$-analitical (anti-)localization} of $\boldsymbol W=(W^j)_{n\in\mathbb N}$, the $r$-convergent sequences of quasi-probability distributions $(\rho^j_{k_n})_{n\in\mathbb N}$ converge to   Dirac's $\delta(z-z_0)$ distribution on $\mathcal A_{\mu}([-1,1])\subset C^{\infty}_{\mathbb C}([-1,1])$, the space of complex-analytic extensions to the interior of the Bernstein ellipse $\partial\mathcal E_{\mu}$, with foci $\pm 1$ and sum of semi-axis equal to $\mu$, for some $\mu>1$, {\it cf.} (\ref{Bellmu}) and Definition \ref{locanal}.

This paper is organized as follows. 

In section 2 we present a summary of the main constructions and results of \cite{prios} which are necessary for understanding the questions addressed in this paper. Thus, after reviewing the main definitions and properties of quantum and classical spin systems, we review the main definitions and results on symbol correspondences for spin systems and their associated twisted products of symbols. In order to make this paper minimally self-contained, this section is not too short.

In section 3 we prove the splitting Theorem \ref{propgood}, which states that the subsets of isometric (Stratonovich-Weyl) correspondences,  mapping-positive (coherent-state) correspondences, and positive-dual correspondences, are mutually disjoint.

Then, section 4 comprises the bulk of the first part of this paper. There we address the questions on localization, presenting the main definitions of asymptotic localization and proving the main results alluded to above. In order to develop a better understanding of the issues involved, 
before moving on to the general case, 
we start by studying the questions of asymptotic localization in the case of mapping-positive correspondences, and this leads to another interesting result of the paper, which states that a mapping-positive symbol correspondence sequence localizes (resp. anti-localizes) classically if and only if 
\begin{equation}\label{c12}
\lim_{n\to\infty} c_1^n=1 \ \mbox{(resp.} \ =-1\mbox{)} \ , \ \lim_{n\to\infty} c_2^n=1 \ , 
\end{equation}
{\it cf.} Theorem \ref{v_prop}.
Hence, from (\ref{stronger}) we have that, for mapping-positive symbol correspondence sequences, the Poisson algebra of smooth functions emerges as an asymptotic limit of the twisted algebra of symbols if and only if (\ref{c12}) is satisfied, a much weaker condition than  (\ref{RS}) for general correspondences. In addition we have  that, for mapping-positive and for positive-dual symbol correspondence sequences, (\ref{c12}) implies the r.h.s. ~of (\ref{RS}), a rather strong implication, {\it cf.} Corollary \ref{surprising}.

Up to section 4, we shall have been investigating the localization property for spin systems from the approach which is  inverse of quantization, namely, dequantization and  (semi)classical limit (same approach used in \cite{prios}). But it is also possible to investigate asymptotic localization of symbol correspondences for spin systems from the  quantization approach, which is the subject of the second part of this paper. 

Now, there are various approaches to quantization of a Poisson manifold, but the  
most common one,
deformation quantization,  was shown by Rieffel (over three decades ago) to be ill suited for producing a $SO(3)$-invariant ``star product'' on $S^2$ that actually converges \cite{rieffel} (see also \cite{hawk}). This problem disappears when defining a sequence of twisted products of the form  (\ref{prodchar}), for $\boldsymbol W=(W^j)_{n\in\mathbb N}$ of (anti-)Poisson type ({\it cf.} \cite[Chap. 9]{prios}), and now we can also take advantage of having symbol correspondence sequences  in order to define  \emph{sequential quantizations} of smooth functions on $S^2$, as developed in section 5 and outlined below.

Given a symbol correspondence sequence of (anti-)Poisson type $\boldsymbol W=(W^j)_{n\in\mathbb N}$, for any  $f\in C^{\infty}_{\mathbb C}(S^2)$ we  define its {\it $W$-quantization} as the sequence of operators 
\begin{equation}\label{untildevbF}
 \mathbf F^w=(F^w_n)_{n\in\mathbb N}  , \  F^w_n=[W^j]^{-1}(f)  \ ,   
\end{equation}
with a similar definition for  
\begin{equation}\label{tildevbF}
    \widetilde{\mathbf F}^w=(\widetilde{F}^w_n)_{n\in\mathbb N} \ , \ \widetilde{F}^w_n=[\widetilde{W}^j]^{-1}(f) \ ,
\end{equation}
the {\it $\widetilde{W}$-quantization} of $f$, which is  obtained from the dual symbol correspondence sequence $\widetilde{\boldsymbol W}=(\widetilde{W}^j)_{n\in\mathbb N}$, {\it cf.} Definitions \ref{dualdef} and \ref{quantization-def}. 

On the other hand, given an operator sequence $\vb F=(F_n)_{n\in\mathbb N}$, for each $j$ we have an action $F_n:\mathcal H_j\to\mathcal H_j$ and we want to make sense of such an action in the limit $j\to\infty$. For this, we develop the notion of a \emph{ground Hilbert space} and operators on this asymptotic limit of a nested sequence of Hilbert spaces, as follows.

Let $\textswab{H}=(\mathcal H_j,\langle\cdot|\cdot\rangle_j)_{2j=n\in\mathbb N}$ denote 
a sequence of $(n+1)$-dimensional complex Hilbert spaces, where $\langle\cdot|\cdot\rangle_j$ is the inner product on $\mathcal H_j$, so that $\vb F: \textswab{H}\to \textswab{H}$ as above. We say that $\textswab{H}^<$ is a nested sequence of Hilbert spaces if there are \emph{nesting maps} $\iota_j^{j'}:\mathcal H_j\to\mathcal H_{j'}$ whenever $j\leq j'$, which are isometric inclusions. This allows us to define a \emph{nested norm} $||\cdot||$ on $\textswab{H}^<$ and then the notion of a \emph{convergent state sequence} $\Phi\in\textswab{H}^<_{\infty}$, $\Phi=(\phi^j)_{2j\in\mathbb N}$, $\phi^j\in\mathcal H_j$  (convergent in the sense of Cauchy w.r.t. $||\cdot||$), and a convergent operator sequence $\vb F:\textswab{H}^<_{\infty}\to\textswab{H}^<_{\infty}$, {\it cf.} Definitions \ref{nestedHilb}-\ref{convergentPhi}. 

With the notion of a \emph{well-nested basis sequence} $\textswab{E}=(\{\vb e^j_k\}_{1\leq k\leq 2j+1})_{2j\in\mathbb N}$, where each $\{\vb e^j_k\}_{1\leq k\leq 2j+1}$ is a basis for $\mathcal H_j$, {\it cf.}  Definition \ref{wn}, 
we can identify the ground Hilbert space $\mathcal H$ of $\textswab{H}^<_{\infty}$ with the space of complex $\ell^2$-sequences spanned by the \emph{grounding basis} $\mathcal E=\{\vb e_k\}_{k\in\mathbb N}$, where 
$\vb e_k=\lim_{j\to\infty}\vb e^j_k$,  $\forall k\in\mathbb N$,
since $(\vb e^j_k)_{2j\in\mathbb N}\in \textswab{H}^<_{\infty}$, for each $k\in\mathbb N$, {\it cf.} Definition \ref{ghs} and Theorem \ref{H=H}. 

This  latter theorem also shows that $(\textswab{E},\mathcal E)$ provides an isomorphism between $\mathcal H$ and the set of 
equivalence classes of convergent state sequences, where, for any  $\Phi=(\phi_j)_{2j\in\mathbb N}, \Phi'=(\phi_j')_{2j\in\mathbb N}\in \textswab{H}^<_{\infty}$,  
\begin{equation}
\Phi\approx\Phi' \ \iff \ \lim_{j\to\infty}\phi_j=\lim_{j\to\infty}\phi_j' \  \quad \therefore \ \ \mathcal H \ \simeq \  \textswab{H}^<_{\infty}/\!\approx \ .
\end{equation}

Then, Definition \ref{Fresco} imports this  equivalence relation  to the set of  convergent  operator sequences, 
$$\vb F\approx\vb F' \ \iff \ \vb F(\Phi)\approx\vb F'(\Phi) \ , \ \forall\Phi\in\textswab{H}^<_{\infty} \ , $$
so that, finally, Theorem \ref{point-lim-op} identifies an operator on $\mathcal H$ with the equivalence class $[\vb F]=\vb F/\!\approx$ \ of a convergent operator sequence, that is,  
\begin{equation}\label{groundaction}
    \vb F:\textswab{H}^<_{\infty}\to\textswab{H}^<_{\infty} \  \iff \  [\vb F]=\mathcal F:\mathcal H\to\mathcal H \ .
\end{equation}

These definitions and theorems are illustrated in Examples \ref{example-cp1} and \ref{S1cS2}. In the first one, the nested sequence of Hilbert spaces $\mathcal H_j\hookrightarrow \mathcal H_{j'}$ is identified with a sequence of holomorphic polynomials on $\mathbb CP^1\simeq S^2$ of degrees $2j=n\leq n'=2j'$. Then, the ground Hilbert space is a convergent subspace of the ring of formal power series in one complex variable, which can be thought of as a space of modified holomorphic functions on $\mathbb CP^1$. In the second example, we identify $\mathcal H_j\hookrightarrow \mathcal H_{j'}$ with modified complex Fourier polinomials of degrees $j\leq j'$  on $\mathbb R\!\mod 2\pi$, the latter identified with a closed geodesic $\mathcal X\subset S^2$, 
and the ground Hilbert space  is the space of complex functions on $\mathcal X\subset S^2$ defined via modified complex Fourier series.

Now, for some operator sequences $\mathbf F=(F_n)_{n\in\mathbb N}$, $F_n\in M_{\mathbb C}(n+1)$, one has a natural definition of its asymptotic norm $||\mathbf F||_{\infty}$ which is independent of the ground Hilbert space $\mathcal H$ on which $\vb F$ acts asymptotically via (\ref{groundaction}), in contrast with the usual operator norm for $\mathcal F:\mathcal H\to\mathcal H$. And in fact, for any operator sequence $\vb F$, one has a natural definition of $\vb F$ being \emph{upper-bounded}, which is independent of the sequence of Hilbert spaces on which it acts, {\it cf.} Definition \ref{asympopnorm}. 
Then, Proposition \ref{necessary} states that, given a well-nested sequence of Hilbert spaces $(\textswab{H}^<, \textswab{E})$, an operator sequence $\vb F$ is a convergent operator sequence $\vb F:\textswab{H}^<_{\infty}\to\textswab{H}^<_{\infty}$, inducing an operator on the ground Hilbert space, $\mathcal F=[\vb F]:\mathcal H\to\mathcal H$, only if $\vb F$ is upper bounded. 

In this setting, Example \ref{ex2} show that if a symbol correspondence sequence $\boldsymbol W$ is of (anti-)Poisson type, generally their $W$- and $\widetilde{W}$-quantized functions may not define asymptotic operators on a ground Hilbert space.   Then, Theorems \ref{locFmu} and \ref{locF1} present relations between classical (resp.  $\mu$-analytical) localization of $\boldsymbol W$  and equality of the $L^2$-norm of $f$ with the asymptotic norms $||\cdot||_{\infty}$ of  ${\mathbf F}^w$ and $\widetilde{\mathbf F}^w$, {\it cf.} (\ref{untildevbF})-(\ref{tildevbF}), for  ($J_3$-invariant) functions $f\in C^{\infty}_{\mathbb C}([-1,1])$ (resp. $\mathcal A_{\mu}([-1,1])$). Thus, many of the results in section 5.2 provide relations between asymptotic localization of a symbol correspondence sequence $\boldsymbol W$ and the possibility that their $W$- and $\widetilde{W}$-quantized functions define asymptotic operators on a ground Hilbert space.  

Finally, in the quantization setting there is a natural definition of a symbol correspondence sequence of (anti-)Poisson-type    possessing classical (anti-)expectation, which means that, 
for any $r$-convergent $(\Pi_{k_n})_{n\in\mathbb N}$ sequence, 
\begin{equation}\label{clexp}
\exists \lim_{n\to\infty}\langle \Pi_{k_n}|\widetilde{F}^w_n\rangle  \in\mathbb{C} \ ,  \ \forall f \in C^{\infty}_{\mathbb{C}}(S^2) \ , 
\end{equation}
{\it cf.} (\ref{tildevbF}), 
where $\langle\cdot|\cdot\rangle$ is the usual Hilbert-Schmidt inner product of operators. Then, Theorem \ref{Clocexp2} shows the equivalence between classical (anti-)localization of a symbol correspondence sequence of (anti-)Poisson-type $\boldsymbol W$ and its possession of classical (anti-)expectation. But at first sight, it looks surprising that the mere existence of the limit in (\ref{clexp}) should be equivalent to the specific form of the limit in (\ref{limloc}). Thus, showing their equivalence is the last main result of this paper.

We close, in section 6, with some final thoughts on why asymptotic Poisson emergence is generally a weaker property than the classical localization property.  

For completeness, in the appendix (section 7) we provide a pedestrian proof of a well-known result in probability theory which is used for Theorem \ref{v_prop}, and a proof of a well-known formula, Edmonds formula, used to prove Theorems \ref{loc_pol} and \ref{p_l}.

\section{Review of some basic definitions and results}

This section follows closely to the more detailed and extended treatment in \cite{prios}\footnote{Some of the definitions and results summarized in this section are not present in the preliminary version of \cite{prios} that was first  posted in the arXiv (which could not be later updated there due to copyright restrictions). Although \cite{prios} presents a full treatment of general symbol correspondences for spin system, various excellent treatments of some special symbol correspondences can be found elsewhere, {\it cf. e.g.} \cite{ber}, \cite{vgb} and the review \cite{krg}, and references therein.}. 

\subsection{Quantum and classical spin systems}

Let $SU(2)$  be the special unitary subgroup  of $GL_2(\mathbb{C})$, satisfying: $\det g = 1$ and $gg^*=g^*g = e$, $\forall g\in SU(2)$. From these properties,
topologically $SU(2) \simeq S^3=\{z_1, z_2 \in \mathbb{C}: |z_1|^2+|z_2|^2 = 1\}$, 
hence it is compact and simply connected.
The Lie algebra $\mathfrak{su}(2)$ of $SU(2)$ is generated by the Pauli matrices $\{\sigma_1,\sigma_2,\sigma_3\}$ 
satisfying the commutation relations 
\begin{equation}\label{pauli_comut}\nonumber
[\sigma_a, \sigma_b] = 2i\epsilon_{abc}\sigma_c \ ,
\end{equation}
where $\epsilon_{abc}$ is totally antisymmetric, and is isomorphic to the Lie algebra $\mathfrak{so}(3)$ of the special orthogonal group $SO(3)$. For any isomorphism $d\psi: \mathfrak{su}(2)\to \mathfrak{so}(3)$, the  homomorphism $\psi: SU(2) \to SO(3)$ has kernel  $\mathbb Z_2$, thus $SO(3)=SU(2)/\mathbb Z_2$.

Since $SU(2)$ is compact, every one of its irreducible unitary representations is finite dimensional ({\it cf.} \cite{nach}), commonly labeled by a half-integer number $j$, where $2j+1\in\mathbb{N}$ is its dimension, so we will denote it by $\varphi_j$. A representation $\varphi_j$ of $SU(2)$ is also a representation of $SO(3)$ if and only if $j$ is an integer. Now, for $n=2j\in\mathbb N$, let $M_{\mathbb C}(n+1)$ denote the algebra of $(n+1)$-square complex matrices.

\begin{definition}[\cite{prios}]
A \emph{quantum spin-$j$ system} is a complex Hilbert space $\mathcal{H}_j\simeq \mathbb{C}^{n+1}$ with an irreducible unitary representation $\varphi_j : SU(2)\to G\subset U(n+1)$, $G$ isomorphic to $SU(2)$ or $SO(3)$ according to whether $j$ is strictly half-integer or is integer. The \emph{operator algebra} of the spin-$j$ system\footnote{One often finds in the literature the term {\it spin-$j$ system} referring to a lattice of $n$ spin-$1/2$ particles, but such lattices have spatial degrees of freedom which are fully neglected in our definition.} is $\mathcal{B}(\mathcal{H}_j)\simeq M_{\mathbb C}(n+1)$.
\end{definition}

The spin operators  $J_k=J^j_k=d\varphi_j(\sigma_k/2)\in\mathcal{B}(\mathcal{H}_j)$ correspond to the $x$, $y$, and $z$ components of the total angular momentum or spin\footnote{We are taking $\hbar=1$ (equivalent to a rescaling of units).}. The usual approach for a spin-$j$ system is to diagonalize the operator $J_3\equiv J_z$, which has eigenvalues $m = -j, -j+1,..., j-1, j$. We denote the vectors of an orthonormal basis of $\mathcal{H}_j$ comprised of eigenvectors of $J_3$ by $\boldsymbol{u}(j, m)$. Then, we define the ladder operators  $J_\pm = J_1\pm iJ_2$ and the norm-square operator $J^2 = J_1^2+J_2^2+J_3^2 = J_{\mp}J_{\pm}+J_3(J_3\pm I)$, where $I$ is the identity. The commutation relations are
\begin{equation}
[J_a, J_b] = i\epsilon_{abc}J_c \ , \ \ \ 
[J_+, J_-] = 2J_3 \ , \ \ \ 
[J_3, J_\pm] = \pm J_\pm \ , 
\end{equation}
thus $J^2 = j(j+1)I$, and these operators act as
\begin{equation}
\begin{aligned}\label{stansu2act}
J_3(\boldsymbol{u}(j,m)) &= m(\boldsymbol{u}(j,m)) \ , \\
J_+(\boldsymbol{u}(j,m)) &= \alpha_{j,m}\boldsymbol{u}(j,m+1) \ , \\ J_-(\boldsymbol{u}(j, m)) &= \beta_{j,m}\boldsymbol{u}(j,m-1) \ , 
\end{aligned}
\end{equation} 
where $\alpha_{j,m}$ and $\beta_{j,m}$ are non zero constants, except for $\alpha_{j,j}=\beta_{j,-j}=0$. To (almost completely) eliminate the freedom of choosing an individual phase factor for each $\boldsymbol{u}(j,m)$, we choose a highest weight vector $\boldsymbol{u}(j,j)$ and fix all the other phases so that the constants $\beta_{j,m}$ are nonnegative real numbers. Thus, for such a basis, called \emph{standard}, there is just one free phase on the choice of $\boldsymbol{u}(j, j)$ and 
\begin{equation}\label{ab}
\alpha_{j,m} = \sqrt{(j-m)(j+m+1)} \ ,
\ \ \ 
\beta_{j,m} = \sqrt{(j+m)(j-m+1)} \ .
\end{equation}

To extend the representation $\varphi_j(g)$ acting on $\mathcal{H}_j$ to a representation $\Phi_j(g)$ acting on the operator space $M_{\mathbb{C}}(n+1)\simeq \mathcal{B}(\mathcal{H}_j) = Hom(\mathcal{H}_j, \mathcal{H}_j)\simeq \mathcal{H}_j\otimes\mathcal{H}_j^*$, we  use the dual representation of $\varphi_j(g)$  acting on $\mathcal{H}_j^*$, denoted by $\check{\varphi}_j(g)$. Via the inner-product identification $\mathcal{H}_j^*\leftrightarrow\mathcal{H}_j$, we have  that $\check{\varphi}_j(g) \leftrightarrow \varphi_j(g)^{-1}$, $\forall g\in SU(2)$. For the spin operators, $\check{J}_1 \leftrightarrow -J_1$, $\check{J}_2\leftrightarrow J_2$ and $\check{J}_3\leftrightarrow -J_3$, so $\check{J}_+ \leftrightarrow -J_-$ and $\check{J}_-\leftrightarrow -J_+$. Hence, a standard basis of $\mathcal{H}_j^*$ is formed by vectors 
\begin{equation}\label{dualbasis} 
\check{\boldsymbol{u}}(j,m) \leftrightarrow (-1)^{j+m}\boldsymbol{u}(j,-m) \ , 
\end{equation}
where $\leftrightarrow$ denotes that the l.h.s. is identified as the dual of the r.h.s.

Thus, $\boldsymbol{u}(j,m_1)\otimes\check{\boldsymbol{u}}(j,m_2)=(-1)^{j+m_2}\mathcal{E}_{j-m_1+1,j+m_2+1}$ , where $\mathcal{E}_{k,l}\in M_{\mathbb{C}}(n+1)$ is the one-element matrix 
$[\mathcal{E}_{k,l}]_{p,q} = \delta_{k,p}\delta_{l,q}$. Hence, 
$\varphi_j(g)\otimes\check{\varphi}_j(g)\leftrightarrow \varphi_j(g)\otimes\varphi_j(g)^{-1}$,  \begin{equation}\label{P^g}
    \Phi_j(g):\mathcal{B}(\mathcal{H}_j)\to\mathcal{B}(\mathcal{H}_j) \ , \ P\mapsto P^g = \varphi_j(g)P\varphi_j(g)^{-1} \ , \ \forall g\in SU(2),
\end{equation}
 and the spin operators $\boldsymbol{J}_k= d\Phi_j(\sigma_k/2)$ acting on $\mathcal{B}(\mathcal{H}_j)$ can be identified with  $J_k=d\varphi_j(\sigma_k/2)$ acting on $P\in\mathcal{B}(\mathcal{H}_j)$ via the commutator, that is, 
 \begin{equation}\label{adjact}
  P \ \mapsto \  [J_k, P] =: J_k(P) \ . 
 \end{equation}

\begin{theorem}[\textit{cf. e.g.} \cite{prios}] The Clebsch-Gordan series for $\mathcal{B}(\mathcal{H}_j)$ is 
\begin{equation*}\label{CGseries}
\varphi_j\otimes\check\varphi_j = \bigoplus_{l = 0}^{n}\varphi_l \ .
\end{equation*}
\end{theorem}

Thus, the induced action (\ref{P^g}) of $SU(2)$ on $\mathcal{B}(\mathcal{H}_j)$ is effectively an  $SO(3)$ action. 
For each $\varphi_l$, we find a standard  basis of vectors $\boldsymbol{e}^j(l,m)$ as we did for $\mathcal{H}_j$. The orthonormal basis $\{\boldsymbol{u}(j,m_1)\otimes\check{\boldsymbol{u}}(j, m_2)\}$  of  $\mathcal{B}(\mathcal{H}_j)$  is called \emph{uncoupled}, whereas the basis consisting of $\boldsymbol{e}^j(l,m)$, where $\{\boldsymbol{e}^j(l,m),-l\le m\le l\}$ is a basis of the $(2l+1)$-dimensional $SO(3)$-invariant  subspace $\varphi_l$, is called \emph{coupled}\footnote{Taking $\mathcal{H}_j^*$ as another spin-$j$ system, this is equivalent to addition (actually subtraction) of spin. In Dirac's notation, we can write $\boldsymbol{u}(j,m_1)\otimes\check{\boldsymbol{u}}(j, m_2) = \lvert j, m_1, j, m_2\rangle$ and $\boldsymbol{e}^j(l,m) = \lvert (j, j) l, m\rangle$.}. This basis satisfies 
\begin{equation}\label{c1} 
\begin{aligned}
& [J_+, \boldsymbol{e}^j(l,m)] =\alpha_{l,m} \boldsymbol{e}^j(l,m+1) \ , \ [J_-, \boldsymbol{e}^j(l,m)] = \beta_{l,m}  \boldsymbol{e}^j(l,m-1) \ ,  \\
& [J_3, \boldsymbol{e}^j(l,m)] = m \boldsymbol{e}^j(l,m) \ , \ \sum_{k=1}^{3}[J_k,[J_k, \boldsymbol{e}^j(l,m)]] = l(l+1) \boldsymbol{e}^j(l,m) \ , 
\end{aligned}
\end{equation}
being also orthonormal w.r.t. the Hilbert-Schmidt inner product on $M_{\mathbb C}(n+1)$. 

\begin{theorem}[\cite{prios}]\label{ejlm}
	The coupled standard basis vectors $\boldsymbol{e}^j(l,m)$ of $\mathcal{B}(\mathcal{H}_j)$ satisfy 
	\begin{equation} 
	\boldsymbol{e}^j(l,-m) = (-1)^m\boldsymbol{e}^j(l,m)^T
	\end{equation} 
	and are given explicitly, for $0\leq m\leq l$, by 
	\begin{eqnarray} \label{elm}
	 \boldsymbol{e}^j(l,m) &=& \frac{(-1)^l}{\nu^n_{l,m}}\sum_{k=0}^{l-m}(-1)^k\binom{l-m}{k}J_-^{l-m-k}J_+^lJ_-^k \ , \\
	 \nu^n_{l,m} &=& \frac{l!}{\sqrt{2l+1}}\sqrt{\frac{(n+l+1)!}{(n-l)!}}\sqrt{\frac{(l-m)!}{(l+m)!}} \ .
\end{eqnarray}	
\end{theorem}

The coefficients of the change of orthonormal basis
\begin{equation}\label{cg}
\boldsymbol{u}(j,m_1)\otimes\check{\boldsymbol{u}}(j, m_2) = \sum_{l=0}^{n}\sum_{m=-l}^{l}C_{m_1,m_2,m}^{\,\,\,\,j,\,\,\,\,\,\,j,\,\,\,\,l}\boldsymbol{e}^j(l,m)
\end{equation}
are called Clebsch-Gordan coefficients and for $SU(2)$ they have been extensively studied\footnote{In particular, the Clebsch-Gordan coefficients in (\ref{cg}) vanish when $m\neq m_1+m_2$ so that the summation in $m$ in (\ref{cg}) is actually moot, only the term with $m=m_1+m_2$ survives in its r.h.s.}, {\it cf. e.g.} \cite{bied,varsh}.	They are unique up to a phase 
and we adopt the usual convention in which all Clebsch-Gordan coefficients are real.

We now turn to the classical spin system.

\begin{definition}[\cite{prios}]\label{css}
The \textnormal{classical spin system} is the homogeneous $2$-sphere with its Poisson algebra $\{C_\mathbb{C}^\infty(S^2), \omega\}$, where the symplectic form is the usual area form with  local expression $\omega = \sin \varphi\, d\varphi\wedge d\theta$ in spherical coordinates w.r.t. the north pole.
\end{definition}

We refer to \cite{prios} for a detailed justification of the above definition, but here we point out that every quantum spin-$j$ system is a mechanical system with one degree of freedom, which is consistent with its classical phase space being the $2$-dimensional symplectic manifold which is the generic coadjoint orbit of $SU(2)$. 

Now, just as for the quantum operator spaces, we can find an orthonormal basis for $C^\infty_{\mathbb{C}}(S^2)$ by decomposing the action of $SO(3)$ on $C^\infty_{\mathbb{C}}(S^2)$ in $(2l+1)$-dimensional invariant subspaces which are spanned by the standard basis of spherical harmonics:
\begin{equation}\label{sph_har}
Y_l^m(\boldsymbol{n}) = \sqrt{2l+1}\sqrt{\dfrac{(l-m)!}{(l+m)!}}P_l^m(\cos\varphi)e^{im\theta},
\end{equation} 
where $(\varphi,\theta)$ are spherical polar coordinates of $\boldsymbol{n}\in S^2$ w.r.t. the north pole $\boldsymbol{n}_0$, in other words, the colatitude and longitude on $S^2$, and $P_l^m$ are the associated Legendre polynomials on $[-1,1]$.  As in (\ref{c1}), $Y_l^m$ satisfies 
\begin{equation} 
J_+(Y_l^m)=\alpha_{l,m}Y_l^{m+1} \ , \ J_-(Y_l^m)=\beta_{l,m}Y_l^{m-1} \ , \ J_3(Y_l^m)=mY_l^{m} \ ,
\end{equation}
where $J_{\pm}=J_1\pm iJ_2$ and $J_k=iL_k$ , for the generators $L_k$ of the Lie algebra $\mathfrak{so}(3)$ acting on $C_\mathbb{C}^\infty(S^2)$ via \ $L_1 = z\partial_y - y\partial_z \ , \ L_2 = x\partial_z - z\partial_x \ , \ L_3 = y\partial_x-x\partial_y$ , 
with $(x,y,z)$ denoting the cartesian coordinates of the unit sphere $S^2\subset\mathbb R^3$. Accordingly, for $J^2=J_1^2+J_2^2+J_3^2$, $Y_l^m$ also satisfies 
\begin{equation}
J^2(Y_l^m)=l(l+1)Y_l^m \ \ ,   \ \ \langle Y_l^m | Y_{l'}^{m'}\rangle = \delta_{l,l'}\delta_{m,m'} \ ,  
\end{equation}
where $\langle\cdot |\cdot\rangle$ is the normalized inner product on  $C^\infty_{\mathbb{C}}(S^2)$, {\it cf.} (\ref{innS2}) below. 

\subsection{Symbol correspondences}

By a spin-$j$ symbol correspondence, we mean a map that associates to every spin-$j$ operator a unique  function on the $2$-sphere, called its symbol, satisfying very basic and natural properties, as follows.    

\begin{definition}\label{corresp}
A map $W^j: \mathcal{B}(\mathcal{H}_j)\to C_\mathbb{C}^\infty(S^2)  , \ P\mapsto W^j[P]=W^j_P$ , is a \emph{symbol correspondence for a spin-$j$ system} if, $\forall P,Q \in \mathcal{B}(\mathcal{H}_j)$, $\forall g\in SO(3)$, it satisfies
\begin{enumerate}[label = \roman*)]
\item Linearity and injectivity;
\item Equivariance: $W^j_{P^g} = (W^j_P)^g$;
\item Reality: $W^j_{P^*} = \overline{W^j_P}$;
\item Normalization: $\frac{1}{4\pi}\int_{S^2}{W^j_PdS} = \frac{1}{n+1}tr (P)$.
\end{enumerate}
If injectivity fails but all other properties hold, $W^j$ is called a \emph{pre-symbol map}. 
\end{definition}

\begin{remark}
   In (ii), on the r.h.s.~the action on functions  is the one induced by the standard $SO(3)$ action on the unit sphere $S^2\subset\mathbb R^3$, and on the l.h.s. the action on operators  ($P\mapsto P^g$) is given by (\ref{P^g}) for any of the two choices of lifting $g\in SO(3)$ to $\tilde{g}\in SU(2)$. From (iii), hermitian operators are mapped to real functions. Condition (iv) is necessary to assure that $I\mapsto 1$ (constant function $1$).  
\end{remark}

The above properties for a spin-$j$ symbol correspondence were first set out by Stratonovich \cite{strat} who  imposed a more strict property which implies (iv), as follows:  

\begin{definition}\label{SW}
	A \textnormal{Stratonovich-Weyl correspondence} is a symbol correspondence that is an isometry with respect to the normalized inner products, that is, it satisfies 
	\begin{equation} 
v) \ Isometry: \quad\quad	\langle P| Q\rangle_j =\langle W^j_{P}| W^j_{Q}\rangle  \ ,  \nonumber
	\end{equation} 
	where 
\begin{eqnarray}	
	\langle P| Q\rangle_j &=& \frac{1}{n+1}\langle P|Q\rangle=\frac{1}{n+1}tr(P^*Q) \ , \label{norminn} \\ 
	\langle W^j_{P}| W^j_{Q}\rangle &=& \frac{1}{4\pi}\int_{S^2}\overline{W^j_P}W^j_QdS  \ . \label{innS2}
\end{eqnarray}	
\end{definition}

But as can be seen from various examples, the first one set out by Berezin \cite{ber}, this isometry condition is too strict to be imposed on general symbol correspondences.

Schur's lemma implies that a symbol correspondence is an isomorphism between the subspaces spanned by $\boldsymbol{e}^j(l, m)$ and $Y_l^m$ for fixed $l$. So we can turn the injectivity requirement into bijectivity by taking $W^j: \mathcal{B}(\mathcal{H}_j)\to Poly_\mathbb{C}(S^2)_{\le n}$, where $$Poly_{\mathbb{C}}(S^2)_{\le n}$$ is the space of complex polynomials on $S^2$ of proper degree less than or equal to $n$. 

Any symbol correspondence can be characterized as follows:

\begin{theorem}[\cite{prios}]\label{symb_c}
A map $W^j: \mathcal{B}(\mathcal{H}_j)\to C_\mathbb{C}^\infty(S^2)$ is a symbol correspondence if and only if there is a diagonal matrix $K \in M_\mathbb{C}(n+1)$ such that $tr(K)=1$ and
\begin{equation}\label{kermap}
W^j_P(\boldsymbol{n}) = tr(PK(\boldsymbol{n})),
\end{equation}
where $K(\boldsymbol{n}) = K^g$ ({\it cf.} (\ref{P^g})) for $\boldsymbol{n}=g\boldsymbol{n}_0$,  $\boldsymbol{n}_0$ the north pole on $S^2$, 
and $K$ is given by
\begin{equation}\label{kerexp}
K = \dfrac{1}{n+1}I + \sum_{l = 1}^{n}c^n_l\sqrt{\dfrac{2l+1}{n+1}}\boldsymbol{e}^j(l, 0) \ ,
\end{equation}
with $\boldsymbol{e}^j(l, 0)$ given by (\ref{elm}), where $c^n_l\in \mathbb{R}^*$, for $l=1,...,n$. In particular, 
\begin{equation}\label{etoY}
W^j: \sqrt{n+1}\boldsymbol{e}^j(l,m)\mapsto c^n_lY^m_l\in Poly_\mathbb{C}(S^2)_{\le n} \ .   
\end{equation}
\end{theorem}

\begin{definition}[\cite{prios}]
The diagonal matrix $K$ as above is called the \emph{operator kernel} and the 
	$n$ nonzero real numbers $c^n_l$ as above are called the {\em characteristic numbers} of the spin-$j$ symbol correspondence. 
\end{definition}

The following theorem, first obtained in an equivalent form by Gracia-Bondia and Varilly  \cite{vgb}, follows immediately from (\ref{etoY}). 
\begin{theorem}\label{isom}
A symbol correspondence is a Stratonovich-Weyl correspondence if and only if all of its characteristic numbers have unitary norm, that is, 
\begin{equation}\label{SWc}
|c^n_l|=1 \ , \ 1\leq \forall l\leq n \ .
\end{equation}
\end{theorem}

In particular, we have the following more special cases:

\begin{definition}[\cite{prios}]\label{strat}
The \textnormal{standard Stratonovich-Weyl correspondence}  is the symbol correspondence with all characteristic numbers equal to $1$, {\it i.e.}  given by $\varepsilon^n_l = 1$, $1\leq\forall l\leq n$. The \textnormal{alternate Stratonovich-Weyl correspondence} is the symbol correspondence with characteristic numbers given by $\varepsilon^n_{l-} =(-1)^l$, $1\leq\forall l\leq n$.
\end{definition}

\begin{definition}
A \textnormal{Berezin correspondence}\footnote{The original definition by Berezin \cite{ber} uses only the highest-weight projector $\Pi_1$, but this more general form was, as far as we know, set out by Marc Rieffel (lectures at Berkeley, 2002).} is a symbol correspondence defined via (\ref{kermap}) by an operator kernel which is a projector $\Pi_k = \mathcal{E}_{k, k}$.
\end{definition}

Now, the projectors $\Pi_k$ decompose as
\begin{equation}\label{proj_decomp}
\Pi_k = \dfrac{1}{n+1}I + (-1)^{k+1}\sum_{l=1}^{n}C_{m, -m,\,\, 0}^{\,\,j,\,\,\,\,\,\,\,j,\,\,\,l}\boldsymbol{e}^j(l, 0),
\end{equation}
where $m = j-k+1$ (\textit{cf.} \cite{prios}). Since some Clebsch-Gordan coefficient on the decomposition may vanish, not all $\Pi_k$ define symbol correspondences for all $n\in\mathbb N$.

However, let  $h: \mathbb{C}^{n+1}\times\mathbb{C}^{n+1}\to\mathbb{C}$ denote the inner product that is conjugate linear in the first variable and consider the maps:
\begin{equation*}
\begin{aligned}
\Phi_j : \mathbb{C}^2 \to \mathbb{C}^{n+1} , \ (z_1, z_2) \mapsto \ & \big(z_1^n, ..., \sqrt{\binom{n}{k}}z_1^{n-k}z_2^k, ..., z_2^n\big)\ , \\
\sigma : \mathbb{C}^2 \to \mathbb{C}^2 , \  (z_1,z_2)\mapsto (-\overline{z}_2, \overline{z}_1) \ \ ;  \ \ \pi: \ & \mathbb C^2\to \mathbb R^3 , \ (z_1,z_2)\mapsto \boldsymbol{n}=(x,y,z) \ , \nonumber \\ 
&  x+iy=2\bar{z}_1z_2 \ , \ z=|z_1|^2-|z_2|^2  \ .
\end{aligned}
\end{equation*}
On $S^3\subset \mathbb{C}^2$,\ $\pi:S^3\to S^2$ is the {\it Hopf} map and $\Phi_j: S^3\to S^{2n+1}$. Then we have:

\begin{theorem}[\cite{prios}]\label{std_ber}
For every $n\in \mathbb{N}$, consider the maps  
\begin{equation*}
\begin{aligned}
B^j, B^{j-} : M_\mathbb{C}(n+&1) \to Poly_\mathbb{C}(S^2)_{\le n} \\
B^j: \ & P \mapsto B^j_P(\textbf{n}) = h(\Phi_j(z_1, z_2), P\Phi_j(z_1, z_2)), \\ 
B^{j-}: \ & P \mapsto B^{j-}_P(\textbf{n}) = h(\Phi^{-}_j(z_1, z_2), P\Phi^{-}_j(z_1, z_2)),
\end{aligned}
\end{equation*}
where $\Phi^{-}_j = \Phi_j\circ\sigma$, $\boldsymbol{n}=\pi(z_1, z_2)$, $(z_1,z_2)\in S^3$. Then, each of the maps $B^j, B^{j-}$ satisfies (i)-(iv) in Definition \ref{corresp} and is therefore a symbol correspondence. The operator kernel for $B^j$ is $\Pi_1$ and the one for $B^{j-}$ is $\Pi_{n+1}$. Their characteristic numbers are denoted by $b^n_l$ and $b^n_{l-}$, respectively, and are given by   
\begin{equation}\label{bln}
b^n_l = \sqrt{\frac{n+1}{2l+1}}C_{j, -j,\, 0}^{\,j,\,\,\,j,\,\,l} =  \dfrac{n!\sqrt{n+1}}{\sqrt{(n+l+1)!(n-l)!}} \ , \ b^n_{l-} = (-1)^lb^n_l \ .
\end{equation}
\end{theorem}

\begin{definition}[\cite{prios}]
The map $B^j$ defines the \textnormal{standard Berezin correspondence} ({\it cf.} \cite{ber}) and the map $B^{j-}$ defines the \textnormal{alternate Berezin correspondence}.
\end{definition}

\begin{definition}\label{dualdef}
If $W^j$ is a spin-$j$ symbol correspondence, its \emph{dual} is the spin-$j$ symbol correspondence $\widetilde{W}^j$ satisfying, $\forall P,Q\in   \mathcal{B}(\mathcal{H}_j)$, 
\begin{equation}\label{dualinner}
\langle P|Q\rangle_j=\langle \widetilde{W}^j_P|W^j_Q\rangle=\langle W^j_P|\widetilde{W}^j_Q\rangle \ .
\end{equation}
\end{definition}

\begin{theorem}[\cite{prios}]\label{dualc}
If $\forall P\in\mathcal{B}(\mathcal{H}^j)$,  $W^j:P\mapsto W^j_P$ is the spin-$j$ symbol correspondence defined by the operator kernel $K$ via (\ref{kermap}), with characteristic numbers $c_l^n$, 
then its dual is the symbol correspondence $\widetilde{W}^j:P\mapsto \widetilde{W}^j_P$ defined implicitly by 
\begin{equation*}
P = \dfrac{n+1}{4\pi}\int_{S^2}\widetilde{W}^j_P(\boldsymbol{n})K(\boldsymbol{n})dS \ 
\end{equation*}
and its characteristic numbers are 
\begin{equation}\label{dualch}
 \tilde{c}^n_l=1/c^n_l \ .   
\end{equation}
\end{theorem}

\begin{definition}[\cite{prios}]
The \emph{standard Toeplitz correspondence}\footnote{The standard Berezin and standard Toeplitz correspondences are also commonly known in the literature as Berezin's covariant and contravariant correspondences, respectively ({\it cf.} \cite{ber}).} is the symbol correspondence dual to the standard Berezin correspondence. Likewise, the \emph{alternate Toeplitz correspondence} is dual to the alternate Berezin correspondence. 
\end{definition}

\begin{corollary}\label{toe}
The standard (resp. alternate) Toeplitz correspondence is well defined $\forall n\in\mathbb N$ and has characteristic numbers $t^n_l = 1/b^n_l$ (resp. $t^n_{l-} = (-1)^l/b^n_l$).
\end{corollary}

\begin{remark}
It follows from Definitions \ref{SW} and \ref{dualdef} that  
Stratonovich-Weyl (isometric) correspondences are self-dual, and vice-versa. 
But from (\ref{bln}), standard and alternate Berezin and Toeplitz correspondences are not isometric. 

Note also from (\ref{bln}) that all $b_l^n>0$. Symbol correspondences with all $c_l^n>0$  are called \emph{characteristic-positive}. In particular, the standard Stratonovich-Weyl correspondence is the only characteristic-positive isometric correspondence. 
\end{remark}

Berezin correspondences are particular cases of a more general class: 

\begin{definition}\label{mp_d}
	A symbol correspondence is   \emph{mapping-positive} if it maps positive (resp. positive-definite) operators to positive (resp. strictly-positive) functions.
\end{definition}

\begin{theorem}[\cite{prios}]\label{mp_t}
	A symbol correspondence is mapping-positive  if and only if its operator kernel is a convex combination of projectors $\Pi_k$, for $k=1,...,n+1$.
\end{theorem}

As explained in \cite{prios}, from the above theorem a synonym for mapping-positive correspondence is \emph{coherent-state} correspondence. Like for Berezin correspondences, in general not all convex combinations of projectors define an operator kernel for which the pre-symbol map is injective. But we have the following: 

\begin{theorem}[\cite{prios}]\label{um}
	For every $n\in\mathbb N$, the positive pre-symbol map $M_\mathbb{C}(n+1) \to Poly_\mathbb{C}(S^2)_{\le n} $ defined via (\ref{kermap}) by the operator kernel 
	\begin{equation}\label{umid} 
	S_{1/2}=\frac{1}{2}\left(\Pi_{\lfloor j+1/2\rfloor}+\Pi_{\lfloor j+1\rfloor}\right) \ ,
	\end{equation}
	where $\lfloor x\rfloor$ denotes integer part of $x$, is bijective, {\it i.e.}, the operator kernel $K=S_{1/2}$ as above defines a mapping-positive symbol correspondence for every $n\in\mathbb N$. 
\end{theorem}
\begin{definition}[\cite{prios}]\label{dois}
For every $n=2j\in\mathbb N$, the \emph{upper-middle-state} symbol correspondence is the one 
defined by the operator kernel $S_{1/2}$ in (\ref{umid}).
\end{definition}

The expressions for the characteristic numbers $p_l^n$ of this symbol correspondence are rather long and can be found in \cite[Proposition 6.2.54]{prios}. 
The correspondence with characteristic numbers $p^n_{l-}=(-1)^lp_l^n$, also well defined $\forall n\in\mathbb N$, is called the \emph{lower-middle-state} correspondence. Neither of these two is characteristic-positive.

\subsection{Twisted products and symbol correspondence sequences}

The operator space of a spin-$j$ system has a natural algebraic structure defined by the usual matrix product. 
Via any given symbol correspondence, $Poly_{\mathbb{C}}(S^2)_{\le n}\subset C^{\infty}_{\mathbb C}(S^2)$ imports this algebraic structure. 

\begin{definition}
Given a symbol correspondence $W^j$ such that $\vec{c}=(c_1^n,\cdots,c^n_n)$ is the $n$-tuple of its characteristic numbers,  the \emph{twisted product of symbols}  
$$\star^n_{\vec{c}}:Poly_{\mathbb{C}}(S^2)_{\le n}\times Poly_{\mathbb{C}}(S^2)_{\le n} \to Poly_{\mathbb{C}}(S^2)_{\le n}$$ 
is defined $\forall P,Q\in\mathcal{B}(\mathcal{H}_j)$ by $$W^j_P\star^n_{\vec{c}} W^j_Q = W^j_{PQ} \ .$$ 
\end{definition}

\begin{theorem}
	For any symbol correspondence $W^j$, the twisted product of symbols defines a $SO(3)$-invariant associative unital star algebra on $Poly_{\mathbb{C}}(S^2)_{\le n}$.
\end{theorem}

We refer to \cite{prios} for detailed extensive formulas for twisted products of general and some special symbol correspondences, as well as various of their properties,

For each correspondence $W^j$, the space of symbols $Poly_{\mathbb{C}}(S^2)_{\le n}\subset C^{\infty}_{\mathbb C}(S^2)$ with usual addition and induced twisted product $\star^n_{\vec{c}}$ is called the \emph{$\vec{c}$-twisted $j$-algebra} of smooth functions on the sphere.  
For a given spin number $j$, all {$\vec{c}$-twisted $j$-algebras} are isomorphic, since they are all isomorphic to the operator algebra $M_{\mathbb C}(n+1)$. 

But now we introduce the following:

\begin{definition}[\cite{prios}]\label{seq_c}
Let $\Delta^+(\mathbb{N}^2) = \{(n,l)\in \mathbb{N}^2: n\ge l > 0\}$. For any function $\mathcal{C}: \Delta^+(\mathbb{N}^2)\to\mathbb{R}^*$, we denote by \  $\boldsymbol{W}_{\!\mathcal{C}}=(W^j)_{2j=n\in\mathbb{N}}$ the 
\emph{sequence of symbol correspondences} with characteristic numbers $c^n_l = \mathcal{C}(n,l)$, $c_0^n=1 , \forall n\in\mathbb N$. And we denote by 
$$\boldsymbol{W}_{\!\mathcal{C}}(S^2,\star)=((Poly_{\mathbb{C}}(S^2)_{\le n},\star^n_{\vec{c}}))_{n\in\mathbb N}$$ 
the associated \emph{sequence of twisted algebras} of smooth functions on the sphere. 
In particular, we say that $\boldsymbol{W}_{\!\mathcal{C}}$ is of \emph{limiting type} if $\exists \displaystyle{\lim_{n\to\infty}}\mathcal{C}(n,l) = c^\infty_l \in \mathbb{R}$, $\forall l\in\mathbb{N}$.
\end{definition}

Therefore, we are interested in whether and how  the Poisson algebra of smooth functions on the sphere emerges asymptotically from a 
sequence of twisted algebras associated to a given sequence of symbol correspondences. 

\begin{definition}[\cite{prios}]\label{p_ap}
A sequence of symbol correspondences $\boldsymbol{W}_{\!\mathcal{C}}$  is of \emph{Poisson type} if, $\forall l_1, l_2\in\mathbb{N}$, its twisted products satisfy
\begin{enumerate}[label = \roman*)]
\item $\lim_{n\to\infty}(Y^{m_1}_{l_1}\star^n_{\vec{c}}Y^{m_2}_{l_2}-Y^{m_2}_{l_2}\star^n_{\vec{c}}Y^{m_1}_{l_1}) = 0$,

\item $\lim_{n\to\infty}(Y^{m_1}_{l_1}\star^n_{\vec{c}}Y^{m_2}_{l_2}+Y^{m_2}_{l_2}\star^n_{\vec{c}}Y^{m_1}_{l_1}) = 2Y^{m_1}_{l_1}Y^{m_2}_{l_2}$,

\item $\lim_{n\to\infty}(n[Y^{m_1}_{l_1}\star^n_{\vec{c}}Y^{m_2}_{l_2}-Y^{m_2}_{l_2}\star^n_{\vec{c}}Y^{m_1}_{l_1}]) =2i\{Y^{m_1}_{l_1},Y^{m_2}_{l_2}\}$.
\end{enumerate}
And it is of \emph{anti-Poisson type} if the third property is replaced by
\begin{enumerate}[label = iii')]
\item $\lim_{n\to\infty}(n[Y^{m_1}_{l_1}\star^n_{\vec{c}}Y^{m_2}_{l_2}-Y^{m_2}_{l_2}\star^n_{\vec{c}}Y^{m_1}_{l_1}]) = -2i\{Y^{m_1}_{l_1}, Y^{m_2}_{l_2}\}$.
\end{enumerate}
The convergences taken uniformly.
\end{definition}

Now, there is a simple numerical criterion to know when the above definition is satisfied by a sequence of symbol correspondences.

\begin{theorem}[\cite{prios}]\label{conviso}
A sequence of symbol correspondences $\boldsymbol{W}_{\!\mathcal{C}}$ is of Poisson (resp. anti-Poisson) type if and only if it is of limiting type and its characteristic numbers satisfy $c^\infty_l = 1$ (resp. $c^\infty_l = (-1)^{l}$), $\forall l\in\mathbb{N}$.
\end{theorem}

Thus, generic sequences of symbol correspondences are not of Poisson or anti-Poisson type, nor even of limiting type, and this is also the case for generic 
sequences of Stratonovich-Weyl  (isometric) symbol correspondences, {\it cf.} \cite{prios}.

\begin{corollary}[\cite{prios}]
The sequence of standard Stratonovich-Weyl correspondences and the sequences of standard Berezin and standard Toeplitz correspondences are of Poisson type. The sequences of alternate Stratonovich-Weyl, alternate Berezin and alternate Toeplitz correspondences are all  of anti-Poisson type.
The sequence of upper-middle-state  correspondences is of limiting type but not of Poisson or anti-Poisson type. Likewise for the sequence of lower-middle-state correspondences. The sequences of their dual correspondences are not even of limiting type.
\end{corollary}

The symbol correspondence sequences of Poisson or anti-Poisson type are the ones for which Poisson dynamics emerges asymptotically from quantum dynamics in the limit of high spin number $j\to\infty$. In this paper, we are interested in obtaining a more intuitive criterion which can replace the numerical criterion of Theorem \ref{conviso}. 
This will be the subject of section \ref{locsec}, further  below.

\section{A splitting of the set of spin-$j$ symbol correspondences}

The mapping-positive property for a correspondence implies in many nice analytical properties for the symbols and, in particular, the corresponding symbols of (pure or mixed) states are nonnegative functions which, upon suitable renormalization, can be seen as probability densities on $S^2$.
However, no Stratonovich-Weyl (\textit{i.e.}, isometric) correspondence  possesses this nice mapping-positive property. This statement was stated as a conjecture in \cite{prios}. Here we shall give its proof. In fact, we present below  a more general result, which follows from introducing: 

\begin{definition}\label{posdualdef}
	If \ $W^j$ is mapping-positive symbol correspondence, {\it cf.} Definition  \ref{mp_d}, then its dual \  $\widetilde{W}^j$ is called a \emph{positive-dual}   symbol correspondence. 
\end{definition}

Thus, the standard and alternate Toeplitz correspondences and the duals of  upper and lower-middle state correspondences are  positive-dual correspondences. 

Just like mapping-positive correspondences map  positive(-definite) operators to (strictly-)positive functions, the positive-dual correspondences do the reverse: 

\begin{proposition}\label{p+}
If \ $W^j$ is positive-dual, then $[W^j]^{-1}: Poly_{\mathbb C}(S^2)_{\le n}\to M_{\mathbb C}(n+1)$ maps (strictly-)positive functions to positive(-definite) operators.
\end{proposition}
\begin{proof}
This follows from (\ref{dualinner}) and the positivity of $\widetilde{W}^j$.  
Let $f \in Poly_{\mathbb R^+}(S^2)$ be positive and $F = [W^j]^{-1}(f)$. Since $f$ is real, $F$ is Hermitian and  there is a basis $\{e_1,...,e_{n+1}\}$ that diagonalizes $F$. For $\Pi_{e_k}$ the projector onto the $e_k$ subspace, the eigenvalues of $F$ are $\langle \Pi_{e_k}|F \rangle$. Now, $\langle \Pi_{e_k}|F \rangle = (n+1)\langle \widetilde W^j_{\Pi_{e_k}}|f \rangle$, \textit{cf}. (\ref{dualinner}). But $\Pi_{e_k}$ is a positive operator, so $\widetilde W^j_{\Pi_{e_k}}$ is a (nontrivial) positive function, since $\widetilde W^j$ is mapping-positive. Therefore, $\langle \widetilde W^j_{\Pi_{e_k}}|f \rangle \ge 0$ and the eigenvalues of $F$ are all nonnegative. If $f$ is strictly-positive, every $\langle \widetilde W^j_{\Pi_{e_k}}|f \rangle > 0$ because for each $k$ there is an open $\mathcal B_k \subset S^2$ s.t. $\widetilde W^j_{\Pi_{e_k}}$ and $f$ are both strictly positive on $\mathcal B_k$.
\end{proof}

Now, let us denote by $\mathcal S^j$ the set of all spin-$j$ symbol  correspondences and by 
$\mathcal S^j_=$ , $\mathcal S^j_<$ and $\mathcal S^j_>$ the subsets of isometric, mapping-positive and positive-dual spin-$j$ symbol  correspondences, respectively. We then have the following splitting:

\begin{theorem}\label{propgood}
	For any $n=2j\in\mathbb N$, the
	subsets $\mathcal S^j_=$ , $\mathcal S^j_<$ and $\mathcal S^j_>$ are mutually disjoint. 
\end{theorem}
\begin{proof}
Let $c^n_l$, $l=1,...,n$, be the characteristic numbers of a mapping-positive correspondence $W^j\in \mathcal S^j_<$. From Theorem \ref{mp_t} and eqs. (\ref{kerexp}) and (\ref{proj_decomp}), we have
\begin{equation}\label{cg+}
c^n_l=\sqrt{\dfrac{n+1}{2l+1}}\sum_{k=1}^{n+1}(-1)^{k+1}a_kC_{m, -m,\,\, 0}^{\,\,\,j,\,\,\,\,\,\,\,j,\,\,\,l} \ \ , 
\end{equation}
where $m = j-k+1$, $a_k\ge 0$, for $1 \le k \le n+1$, and $\sum_{k=1}^{n+1}a_k = 1$. Since the Clebsch-Gordan coefficients are coefficients of a unitary transformation of basis, they satisfy $|C_{m, -m,\,\, 0}^{\,\,\,j,\,\,\,\,\,\,\,j,\,\,\,l}| \le 1$. Hence,
\begin{equation*}
\begin{aligned}
|c^n_l| & = \left|\sqrt{\dfrac{n+1}{2l+1}}\sum_{k=1}^{n+1}(-1)^{k+1}a_kC_{m, -m,\,\, 0}^{\,\,\,j,\,\,\,\,\,\,\,j,\,\,\,l}\right| \leq \sqrt{\dfrac{n+1}{2l+1}}\sum_{k=1}^{n+1}a_k\left|C_{m, -m,\,\, 0}^{\,\,\,j,\,\,\,\,\,\,\,j,\,\,\,l}\right|\le \sqrt{\dfrac{n+1}{2l+1}}.
\end{aligned}
\end{equation*}
That is, 
\begin{equation}\label{c<}
|c^n_l| \leq \sqrt{\dfrac{n+1}{2l+1}} \ , \   \  1 \le \forall l \le n  \ .
\end{equation}
In particular,  
\begin{equation}\label{c<1}
|c^n_l|<1 \ , \  \ j<\forall l\le n=2j \ .
\end{equation}
From Theorem \ref{dualc}, eq. (\ref{dualch}), the characteristic numbers $\tilde{c}^n_l$ of its dual $\widetilde{W}^j$ satisfy 
\begin{equation}\label{c>1}
|\tilde{c}^n_l| >1 \ , \ \ j<\forall l\le n=2j \ .
\end{equation} 
Thus, from (\ref{SWc}), (\ref{c<1}) and (\ref{c>1}), 
the three subsets are mutually disjoint. 
\end{proof}

\begin{remark}
	We must bear in mind, however, that  the union \ $\mathcal U^j=\mathcal S^j_<\cup\mathcal S^j_=\cup\mathcal S^j_>$ is a proper subset\footnote{This is easy to see for $j\ge 3/2$, just take some $k\in\mathbb N$ s.t. $j<k<2j$ and take $c_l^n$ to satisfy (\ref{c<1}) for $j<l\le k$ and satisfy (\ref{c>1}) for $k<l\le 2j$.} of $\mathcal S^j$. Therefore, to complete this splitting of $\mathcal S^j$ we must add the complementary subset $\mathcal S^j\setminus \mathcal U^j$.
\end{remark}	

Now, for any $n=2j\in\mathbb N$,  the set $\mathcal S^j$ of all spin-$j$ symbol  correspondences defines  a groupoid $\mathcal D^j$, the {\it dequantization groupoid} of the spin-$j$ system, {\it cf.} \cite[Definition 7.1.19, Proposition 7.2.20]{prios}, which is isomorphic to the pair groupoid over $\mathcal S^j$. Therefore, each of the pair groupoids over  $\mathcal S^j_=$ , or $\mathcal S^j_<$ , or $\mathcal S^j_>$  is isomorphic to a subgroupoid of the dequantization groupoid, these being mutually disjoint.

\section{Asymptotic localization of  correspondence sequences}\label{locsec}

We now come to a main purpose of this paper: establishing a more intuitive or geometric criterion for a given sequence of symbol correspondences to be of Poisson or anti-Poisson type, complementing the algebraic criterion of Theorem \ref{conviso}.

Motivated by a physical view of the projectors $\Pi_k$,  $k=1,...,n+1$, which are $J_3$-invariant pure states of a spin-$j$ system,  given a symbol correspondence sequence we ask whether, or in what way, some kind of asymptotic ($n\to\infty$) localization of the symbols of projectors could be related to the asymptotic emergence of the Poisson algebra from the given sequence of twisted algebras.

Let $W^j$ be a symbol correspondence with characteristic numbers $c^n_l$. From Theorem \ref{symb_c} and expressions (\ref{sph_har}) and (\ref{proj_decomp}), we have
\begin{equation}\label{proj_symb}
W^j_{\Pi_k}(\boldsymbol{n}) = \dfrac{1}{n+1} + \dfrac{(-1)^{k-1}}{\sqrt{n+1}}\sum_{l = 1}^{n}{c_{l}^{n}C_{m, -m,\,\, 0}^{\,\,\,j,\,\,\,\,\,\,\,j,\,\,\,l}\sqrt{2l+1}P_l(\cos\varphi)} \ ,
\end{equation}
where $P_l = P_l^0$ are the Legendre polynomials and $m = j-k+1$. 

Note that these symbols are invariant by rotations around the $z$-axis, $z=\cos\varphi$, according to our general convention of diagonalizing the $z$-spin operator $J_3$.

 \begin{remark}\label{reflection}
   Let $W^{j-}$ be the alternate correspondence of $W^j$, that is, if $c^n_l$ are the characteristic numbers of $W^j$, then $c^n_{l-} = (-1)^lc^n_l$ are the characteristic numbers of $W^{j-}$. Then, {\it cf.} \cite[equations (4.31) and (6.12)]{prios}, 
   $W^j_{\Pi_k}(z) = W^{j-}_{\Pi_k}(-z)$, and thus they are the reflection of each other with respect to the equator.

   But the same reflection relates $(m,-m)$-pairs of symbols of projectors  under the same correspondence, as follows:  recalling that $\Pi_k$ is related to the $J_3$-eigenvalue $m$ by $k=j-m+1$, let $k^{-}$ correspond to the eigenvalue $-m$, \textit{i.e.}, $k^{-} = j+m+1$. Then,
   because Clebsch-Gordan coefficients satisfy $C_{m_1,m_2,m}^{\,\,\,\,\,\,j,\,\,\,\,\,\,j,\,\,\,\,l} = (-1)^{2j-l}C_{-m_1,-m_2,-m}^{\,\,\,\,\,\,\,\,\,\,j,\,\,\,\,\,\,\,\,\,\,j,\,\,\,\,\,\,\,\,l}$ (\textit{cf. e.g.} \cite{bied}), from (\ref{proj_symb}) we have that 
    $W^j_{\Pi_k}(z) = W^j_{\Pi_{k^{-}}}(-z)$. That is, $\forall z\in[-1,1]$, 
\begin{equation}\label{reflpik} 
W^j_{\Pi_{k^{-}}}(z)=W^{j-}_{\Pi_{k}}(z)=W^j_{\Pi_{k}}(-z) \ .
\end{equation}
\end{remark}

We now introduce the main definitions that will be useful for our purposes. First:

\begin{definition}\label{pi-sym}
Let $\boldsymbol{W}_{\!\mathcal{C}}=(W^j)_{n\in\mathbb{N}}$ be a sequence of symbol correspondences (cf. Definition \ref{seq_c}) and let $(k_n)_{n\in\mathbb{N}}$ be a sequence of natural numbers satisfying $1\le k_n\le n+1$, so that \ $\Pi_{k_n}\in \mathcal{B}(\mathcal{H}_j)$ for every $n\in\mathbb N$. 

A \emph{$\Pi$ sequence} is a sequence of projectors $(\Pi_{k_n})_{n\in\mathbb{N}}$. Its corresponding \emph{$\Pi$-symbol sequence} is the sequence of their symbols $(W^j_{\Pi_{k_n}})_{n\in\mathbb{N}}$. Its associated \emph{$\Pi$-distribution sequence} is the sequence $(\rho^j_{k_n})_{n\in\mathbb{N}}$ of quasiprobability distributions on $[-1,1]$, where  
$\rho^j_{k_n}=\frac{n+1}{2}W^j_{\Pi_{k_n}}$ (restricted to the $z$-axis by $J_3$-invariance) so that 
$\int_{-1}^1\rho^j_{k_n}(z)dz=1$.

These sequences  
are said to be \emph{$r$-convergent} if \ $k_n/n \to r\in [0, 1]$ , as $n\to\infty$.
\end{definition}

 The characteristic numbers of the sequence of symbol correspondences $\boldsymbol{W}_{\!\mathcal{C}}$ will also be referred to as the characteristic numbers of its $\Pi$-symbol and $\Pi$-distribution sequences. 
If $\rho^j_{k}$ is an element of a $\Pi$-distribution sequence with characteristic numbers $c^n_l$, from (\ref{proj_symb}) we get explicitly that
\begin{equation}\label{rho}
\rho^j_{k}(z) = \dfrac{1}{2} + \dfrac{(-1)^{k-1}\sqrt{n+1}}{2}\sum_{l = 1}^{n}c_{l}^{n}{C_{m, -m,\,\, 0}^{\,\,j,\,\,\,\,\,\,\,j,\,\,\,l}\sqrt{2l+1}P_l(z)} \ .
\end{equation}

Now, if $\Pi_{k_n}=\Pi_1$, $\forall n\in\mathbb N$,  then $r=0$. Similarly, if $\Pi_{k_n}=\Pi_{n+1}$, $\forall n\in\mathbb N$, then $r=1$. Classically, we could imagine that the symbols for the first sequence would ``localize'' at the north pole $z_0=1$ and the latter at the south pole $z_0=-1$. More generally, if $k_n/n\to r$, the ``standard classical picture'' would be a symbol ``localized'' at the parallel of colatitude $\varphi=\arccos(1-2r)$, or equivalently, a $z$-axis quasiprobabilty distribution ``localized'' at $z_0=1-2r$. 

In order to be more precise, we present our second main definition:  

\begin{definition}\label{loc}
A $\Pi$-distribution sequence $(\rho^j_{k_n})_{n\in\mathbb{N}}$ is  
said to \emph{localize classically} at $z_0\in[-1,1]$ if  
it converges, as distribution, to Dirac's $\delta(z-z_0)$  distribution on $C^{\infty}_{\mathbb C}([-1,1])$, that is, 
\begin{equation}\label{loc-eq}
    \lim_{n\to\infty}\int_{-1}^1f(z)\rho^j_{k_n}(z)dz=f(z_0) \ , \ \forall f\in C^{\infty}_{\mathbb C}([-1,1]) \ .
\end{equation}
\end{definition}
\begin{remark}
    The term \emph{classical} in the above definition refers not only to the asymptotic limit $n\to\infty$, but also to the fact that  $C^{\infty}_{\mathbb C}([-1,1])$ is isomorphic to the space of $J_3$-invariant functions of the classical spin system, {\it cf.} Definition \ref{css}.   
    
    On the other hand, if $f\in C^{\infty}_{\mathbb C}(S^2)$ is not $J_3$-invariant, then its $S^1$-average 
    \begin{equation}\label{3av}
    \bar{f}\in C^{\infty}_{\mathbb C}([-1,1]) \ , \quad   \bar{f}(z)=\frac{1}{2\pi}\int_{-\pi}^{\pi}f(z,\theta)d\theta \ , 
    \end{equation} 
    is $J_3$-invariant, so we can extend Definition \ref{loc} to general classical functions via\footnote{Of course, (\ref{loc-eq}) and (\ref{loc-eq-gen}) are equivalent because $\frac{1}{2\pi}\int_{S^2}f(z,\theta)\rho^j_{k_n}(z)dzd\theta=\int_{-1}^1\bar{f}(z)\rho^j_{k_n}(z)dz$.}  
    \begin{equation}\label{loc-eq-gen}
    \lim_{n\to\infty}\frac{1}{2\pi}\int_{S^2}f(z,\theta)\rho^j_{k_n}(z)dzd\theta=\bar{f}(z_0) \ , \ \forall f\in C^{\infty}_{\mathbb C}(S^2) \ .
\end{equation}
\end{remark}
Thus, ``thinking classically'', we could expect or imagine that every $r$-convergent $\Pi$-distribution sequence would localize classically at $z_0=1-2r$, $\forall r\in [0,1]$. But in view of Remark \ref{reflection}, the ``classical picture'' could also be reflected on the equator. 

This leads to our third main definition: 

\begin{definition}\label{cloc}
   A sequence of symbol correspondences  $\boldsymbol{W}_{\!\mathcal{C}}=(W^j)_{n\in\mathbb{N}}$ is said to \emph{localize classically} if every $r$-convergent $\Pi$-distribution sequence localizes classically at $z_0=1-2r$, $\forall r\in [0,1]$. Likewise, $\boldsymbol{W}_{\!\mathcal{C}}$ is said to \emph{anti-localize classically} if every $r$-convergent $\Pi$-distribution sequence localizes classically at $z_0=2r-1$, $\forall r\in [0,1]$.
\end{definition}

\subsection{Classical localization of mapping-positive  correspondence sequences}

In order to first  bypass the more complicated general case, we start by taking a special look at $\Pi$-distribution sequences that are, in fact, sequences of probability distributions. By Definition \ref{mp_d},  this is assured by considering:

\begin{definition}
A $\Pi$-symbol sequence is \emph{positive} if it is constructed from a sequence of mapping-positive correspondences. Likewise for its $\Pi$-distribution sequence.
\end{definition}

As examples, we present the following explicit expressions (very useful for numerical computations) for the standard and alternate Berezin $\Pi$-symbol  sequences: 

\begin{proposition}\label{b_k}
	The standard/alternate Berezin $\Pi$-symbol sequences, $(B^j_{\Pi_{k_n}})_{n\in\mathbb N}$ and $(B^{j-}_{\Pi_{k_n}})_{n\in\mathbb N}$, are given by
\begin{equation*}
B^j_{\Pi_{k_n}}(z) = {\binom{n}{{k_n}-1}}\dfrac{(1+z)^{n-{k_n}+1}(1-z)^{{k_n}-1}}{2^n} \ , \ B^{j-}_{\Pi_{k_n}}(z)=B^{j}_{\Pi_{k_n}}(-z) \ .
\end{equation*}
\end{proposition}
\begin{proof}
From Theorem \ref{std_ber}, taking $P=\Pi_{k_n}$, we obtain an expression in terms of $|z_1|^2$ and $|z_2|^2$. Using that $|z_1|^2+|z_2|^2 = 1$ and the Hopf map, $z = |z_1|^2-|z_2|^2$, we get the expression for $B^j_{\Pi_{k_n}}$. The second expression is a particular case of (\ref{reflpik}).
\end{proof}

The lemma below follows straightforwardly from Chebyshev's inequality. It is well-known but not so easy to find in exactly these terms in the literature, thus, for completeness, we present a proof of this lemma in the appendix, section \ref{Chebproof}. 

For $(\rho_n)_{n\in\mathbb{N}}$ a sequence of probability distributions on $[-1,1]$, denote by $E_n$ the expected value operator defined by $\rho_n$ and let  $\mu_n = E_n(z)$ and $\sigma^2_n = E_n((z-\mu_n)^2) = E_n(z^2)-\mu_n^2$ denote the mean and variance of $\rho_n$, respectively. 

\begin{lemma}\label{delta_c}
Let $(\rho_n)_{n\in\mathbb{N}}$ be a sequence of probability distributions on $[-1,1]$ with mean values $(\mu_n)_{n\in\mathbb{N}}$ and variances $(\sigma^2_n)_{n\in\mathbb{N}}$. Then $(\rho_n)$ converges, as distribution, to Dirac's $\delta(z-\mu)$ distribution  on $C^0_{\mathbb C}([-1,1])$, that is, 
\begin{equation}\label{locC0}
 \lim_{n\to\infty}\int_{-1}^1f(z)\rho_{n}(z)dz=f(\mu) \ , \ \forall f\in C^{0}_{\mathbb C}([-1,1]) \ , 
\end{equation}
if and only if \ $\mu_n\to \mu$ and $\sigma^2_n\to 0$.
\end{lemma}

Now, let $(\rho^j_{k_n})_{n\in\mathbb{N}}$ be a positive $\Pi$-distribution sequence with characteristic numbers $c^n_l$. To compute $\mu_n$ and $\sigma^2_n$, 
we must integrate $\int_{-1}^{1}{z\rho^j_{k_n}dz}$ and $\int_{-1}^{1}{z^2\rho^j_{k_n}dz}$. Therefore, we need expressions for Clebsh-Gordan coefficients of the form $C_{m, -m,\,\, 0}^{\,\,\,j,\,\,\,\,\,\,\,j,\,\,1}$ and $C_{m, -m,\,\, 0}^{\,\,\,j,\,\,\,\,\,\,j,\,\,2}$. From \cite{varsh}, 
\begin{equation*}
C_{m, -m,\,\, 0}^{\,\,\,j,\,\,\,\,\,\,\,j,\,\,1} = (-1)^{k+1}2(j-k+1)\sqrt{\dfrac{3}{(n+2)(n+1)n}},
\end{equation*}
\begin{equation*}
C_{m, -m,\,\, 0}^{\,\,\,j,\,\,\,\,\,\,j,\,\,2} = (-1)^{k-1}\sqrt{5}\dfrac{(n-k+1)(n-k)-4(k-1)(n-k+1)+(k-1)(k-2)}{\sqrt{(n-1)n(n+1)(n+2)(n+3)}},
\end{equation*}
where, as usual, $m = j-k+1$. Hence, we easily get:  

\begin{lemma}\label{musigma2}
For any positive $\Pi$-distribution sequence $(\rho^j_{k_n})_{n\in\mathbb{N}}$, we have that 
\begin{equation}\label{mean}
\mu_n = c^n_1\dfrac{n-2(k_n-1)}{\sqrt{n(n+2)}},
\end{equation}
\begin{equation}\label{variance}
\begin{aligned}
\sigma^2_n & = \dfrac{2c^n_2[(n-k_n+1)(n-k_n)-4(k_n-1)(n-k_n+1)+(k_n-1)(k_n-2)]}{3\sqrt{(n-1)n(n+2)(n+3)}} \\
& \,\,\,\,\,\,\, +\dfrac{1}{3} - \dfrac{(c^n_1)^2(n-2(k_n-1))^2}{n(n+2)}.
\end{aligned}
\end{equation}
\end{lemma}

From Lemmas \ref{delta_c}-\ref{musigma2} we obtain in a rather straightforward way:

\begin{theorem}
\label{v_prop}
A  mapping-positive symbol correspondence sequence localizes (resp. anti-localizes) classically if and only if 
\begin{equation}\label{c12b}
\lim_{n\to\infty} c_1^n=1 \ \mbox{(resp.} \ =-1\mbox{)} \ , \ \lim_{n\to\infty} c_2^n=1 \ . 
\end{equation}
\end{theorem}

Combined with Theorem \ref{conviso} we immediately have: 

\begin{corollary}\label{loc1}
If a sequence of mapping-positive symbol correspondences is of Poisson  (resp. anti-Poisson) type, then it localizes (resp. anti-localizes) classically. 
\end{corollary}

In particular, the standard (resp. alternate) Berezin symbol correspondence sequence localizes (resp. anti-localizes) classically. But from \cite[Proposition 6.2.54]{prios}, the upper-middle-state and lower-middle-state symbol correspondence sequences do not satisfy the conditions of Theorem \ref{v_prop} and in fact they do not localize or anti-localize classically, 
in accordance with the fact that they are not of Poisson or anti-Poisson type. 
	The failure of these mapping-positive symbol correspondence sequences to (anti-)localize classically can be inferred from numerical computations for finite but growing values of $n$. It also follows from Corollary \ref{l_p} below.

\begin{remark}
As seen from above, for mapping-positive symbol correspondence sequences the condition for their classical (anti-)localization is a condition on the asymptotic limit of their first two characteristic numbers, only. Thus, for mapping-positive symbol correspondence sequences, (anti-)Poisson emergence seems to be a  stronger condition than classical (anti-)localization, {\it cf.} Theorem \ref{conviso}. But we shall see below that this is not so ({\it cf.} Corollaries \ref{final}-\ref{surprising}) and that, in fact, the opposite
is the case for general symbol correspondence sequences.  
\end{remark}

\subsection{Asymptotic localization of general symbol correspondence sequences}

In order to expand our study to the general case, we now use Edmonds formula. Proofs of this formula are found in the literature, but we have not found a proof treating the fully general case, so we present one in the appendix, section \ref{Edproof}. 

\begin{lemma}[Edmonds formula, {\it cf.} \cite{edm}]\label{edm}
Let $(k_n)_{n\in\mathbb{N}}$ be as in Definition \ref{pi-sym} and let \ $m = j - k_n + 1$ (depending implicitly on $n$). If $k_n/n\to r\in[0,1]$, then
\begin{equation}\label{edf}
\lim_{n\to\infty}(-1)^{k_n-1}C_{m, -m,\,\, 0}^{\,\,j,\,\,\,\,\,\,\,j,\,\,\,l}\sqrt{\dfrac{n+1}{2l+1}} = P_l(1-2r) \ , \,\, \forall l\in\mathbb{N} \ ,
\end{equation}
where $P_l$ is the  $l^{th}$ Legendre polynomial. 
\end{lemma}

We then have:

\begin{theorem}\label{loc_pol}
A sequence of symbol correspondences $\boldsymbol{W}_{\!\mathcal{C}}$ is of Poisson type if and only if \ $\forall r \in [0,1]$ its  $r$-convergent $\Pi$-distribution sequences satisfy  
\begin{equation}\label{loc_pol_p}
    \lim_{n\to \infty}\int_{-1}^{1}P_l(z)\rho^j_{k_n}(z)dz = P_l(1-2r), \quad \forall l\in \mathbb{N}.
\end{equation}
And $\boldsymbol{W}_{\!\mathcal{C}}$ is of anti-Poisson type if and only if  \ $\forall r \in [0,1]$ its  $r$-convergent $\Pi$-distribution sequences satisfy
\begin{equation}\label{loc_pol_ap}
    \lim_{n\to \infty}\int_{-1}^{1}P_l(z)\rho^j_{k_n}(z)dz = P_l(2r-1), \quad \forall l\in \mathbb{N}.
\end{equation}
\end{theorem}
\begin{proof}
From (\ref{rho}) and Lemma \ref{edm}, we have
\begin{equation}\label{lim1}
\lim_{n\to\infty}\int_{-1}^{1}P_l(z)\rho^j_{k_n}(z)dz  = \lim_{n\to\infty}(-1)^{k_n-1}c^n_lC_{m, -m,\,\, 0}^{\,\,j,\,\,\,\,\,\,\,j,\,\,\,l}\sqrt{\dfrac{n+1}{2l+1}} \ ,\ \forall l \in \mathbb{N} \ .
\end{equation}
Assuming Poisson, $\lim_{n\to\infty}c^n_l=1, \forall l\in\mathbb N$, then (\ref{loc_pol_p}) follows immediately from Edmonds formula. On the other hand, for all $l\in\mathbb{N}$ there exists $r\in[0,1]$ such that $P_l(1-2r)\ne 0$. So from Edmonds formula, (\ref{loc_pol_p}) holds only if $\lim_{n\to\infty}c^n_l=1, \forall l\in\mathbb N$. Similarly for the anti-Poisson case, using that  
$P_l(-z)=(-1)^lP_l(z)$. 
\end{proof}

In view of the above theorem, we introduce the following definition:  

\begin{definition}\label{polloc}
A $\Pi$-distribution sequence $(\rho^j_{k_n})_{n\in\mathbb N}$ is said to \emph{localize polynomially} at $z_0\in[-1,1]$ if 
\begin{equation}\label{polloc-eq}
    \lim_{n\to\infty}\int_{-1}^1f(z)\rho^j_{k_n}(z)dz=f(z_0) \ , \ \forall f\in Poly_{\mathbb{C}}([-1,1]) \ .
\end{equation}
    Then, a sequence of symbol correspondences  $\boldsymbol{W}_{\!\mathcal{C}}$ is said to \emph{localize polynomially}, resp. \emph{anti-localize polynomially}, if every $r$-convergent $\Pi$-distribution sequence localizes polynomially at $1-2r$, resp. at $2r-1$,
    $\forall r\in [0,1]$. 
\end{definition}

And then Theorem \ref{loc_pol} can be rewritten as 

\begin{corollary}\label{pollocc}
A sequence of symbol correspondences is of Poisson (resp. anti-Poisson) type if and only if it localizes (resp. anti-localizes) polynomially. 
\end{corollary}

And because classical localization ({\it cf.} Definitions \ref{loc} and \ref{cloc}) implies polynomial localization ({\it cf.} Definition \ref{polloc}), we immediately have

\begin{corollary}\label{l_p}
A sequence of symbol correspondences is of Poisson (resp. anti-Poisson) type if it localizes (resp. anti-localizes) classically. 
\end{corollary}

From Theorem \ref{v_prop} and Corollary \ref{l_p} we have a new criterion to complement Theorem \ref{conviso} in the case of mapping-positive symbol correspondence sequences: 

\begin{corollary}\label{final}
A mapping-positive symbol correspondence sequence is of Poisson (resp. anti-Poisson) type if and only if it localizes 
(resp. anti-localizes) classically. 
\end{corollary}

And in addition, we have the  following strong implication\footnote{In the simplest cases, standard/alternate Berezin and Toeplitz correspondence sequences, it is not too difficult to see that this implication also follows from (\ref{cg+}) and some recurrence relations for Clebsch-Gordan coefficients, {\it cf. e.g.} \cite{varsh}.}: 
\begin{corollary}\label{surprising}
For mapping-positive symbol correspondence sequences and for positive-dual symbol correspondence sequences,
\begin{equation*}
\lim_{n\to\infty} c_1^n=1 \ \mbox{(resp.} \ =-1\mbox{)} \ , \ \lim_{n\to\infty} c_2^n=1 \ , 
\end{equation*}
implies 
$$
\lim_{n\to\infty} c_l^n=1 \ \mbox{(resp.} \ =(-1)^l\mbox{)} \ , \ \forall l\in \mathbb N \ .
$$
\end{corollary}
\begin{proof}
For the mapping-positive case, this follows straightforwardly  from Corollary \ref{final} and Theorems \ref{conviso} and \ref{v_prop}. For the positive-dual case, we just recall that every positive-dual correspondence is the dual of a mapping-positive correspondence whose characteristic numbers are related by (\ref{dualch}). 
\end{proof}

Since polynomials are dense in $C^\infty_{\mathbb{C}}([-1,1])$, one might expect that (\ref{loc_pol_p}), or (\ref{loc_pol_ap}), could imply classical (anti-)localization in the sense of Definition \ref{loc}. But there is no such implication, so we now investigate the converse of Corollary \ref{l_p}.  

\begin{theorem}\label{p_l}
If a sequence of symbol correspondences $\boldsymbol{W}_{\!\mathcal{C}}=(W^j)_{n\in\mathbb{N}}$ is of (anti-)Poisson type
and if there exist $d\in\mathbb{N}_0$ and $K_d>0$ such that
\begin{equation}\label{c<d}
|c^n_l|\le K_d\prod_{t=1}^{d}(2(l-t)+1) \ , \  n\ge \forall l> d+1 \ ,
\end{equation}
with the product assumed to be $1$ if $d = 0$, then   $\boldsymbol{W}_{\!\mathcal{C}}$ (anti-)localizes  classically. 
\end{theorem}

\begin{proof}
For any $f\in C^{\infty}_{\mathbb{C}}([-1,1])$, we have that
\begin{equation}\label{f}
f = \lim_{p\to\infty}\sum_{l = 0}^{p}a_lP_l\,\,,
\end{equation}
where the convergence is absolute and uniform. Then, from (\ref{rho}) we have
\begin{equation}\label{rhof}
\begin{aligned}
\lim_{n\to\infty}\int_{-1}^{1}f(z)\rho^j_{k_n}(z)dz & = \lim_{n\to\infty}\lim_{p\to\infty}\sum_{l=0}^{p}\int_{-1}^{1}a_lP_l(z)\rho^j_{k_n}(z)dz\\
& = a_0 + \lim_{n\to\infty}\lim_{p\to\infty}\sum_{l=1}^{p}a_lc^n_l(-1)^{k_n-1}C_{m,-m,0}^{j,j,l}\sqrt{\dfrac{n+1}{2l+1}}\\
& = a_0 + \lim_{n\to\infty}\sum_{l=1}^{n}a_lc^n_l(-1)^{k_n-1}C_{m,-m,0}^{j,j,l}\sqrt{\dfrac{n+1}{2l+1}}\,\,.
\end{aligned}
\end{equation}

In order to apply Edmonds formula, we must be able to take the limit $n\to\infty$ inside the summation in the last line of (\ref{rhof}). 
For simplicity, we now write 
\begin{equation}\label{alphanl}
\alpha^n_l = a_lc^n_l(-1)^{k_n-1}C_{m,-m,0}^{j,j,l}\sqrt{\dfrac{n+1}{2l+1}} \ .
\end{equation}

From the C-G symmetry  
$C_{m,-m,0}^{j,j,l} = \sqrt{\dfrac{2l+1}{n+1}}C_{0,-m,-m}^{l,j,j}$ plus unitarity, we have 
\begin{equation}\label{cg<sqrt}
  |C_{m,-m,0}^{j,j,l}|\le\sqrt{\dfrac{2l+1}{n+1}}  \ ,
\end{equation}
as well as $|C_{m,-m,0}^{j,j,l}|\le 1$. In any case, 
$$|\alpha^n_l|\le|a_lc^n_l| \ .$$ 

We remove the dependence on $n$ using the (anti-)Poisson hypothesis, which implies $|c^\infty_l|=1$. Thus, for any fixed $l$, the sequence $\{|c^n_l|\}_{n\ge l}$ is convergent. 

Hence, for every $l\in\mathbb N$,  there exists $B_l>0$ s.t.
\begin{equation}\label{|alpha|}
    |c^n_l|\le B_l \ , \ \forall n\ge  l \ \  \Rightarrow \ \ |\alpha^n_l|\le B_l|a_l| \ , \ \forall n\ge  l \ . 
\end{equation}
But by hypothesis, there exist $d\in\mathbb{N}_0$ and $K_d>0$ such that
\begin{equation}\label{Kl}
B_l \le K_d\prod_{t=1}^{d}(2(l-t)+1)  \ , \ \forall l> d + 1\ .
\end{equation}

We now use the following lemma, which is a corollary of \cite[Theorem 2.2]{wang}.
\begin{lemma}[\cite{wang}]\label{wanglemma}
For any $f\in C^{k+1}_{\mathbb{C}}([-1,1])$, $k\in \mathbb N_0$, there exists $A_k>0$ \ s.t. 
\begin{equation}\label{al}
|a_l|\le\dfrac{A_k}{\sqrt{2(l-k)-1}}\prod_{t=1}^{k}\dfrac{1}{2(l-t)+1} \ , \ \forall l \geq k+1 \ .
\end{equation}
\end{lemma}

Thus, given $d$ as in the hypothesis of the theorem, take $k= d+1$ in Lemma \ref{wanglemma}, since $f\in C^{\infty}_{\mathbb{C}}([-1,1])\subset C^{d+2}_{\mathbb{C}}([-1,1])$. 
From (\ref{|alpha|})-(\ref{al}), 
\begin{equation}\label{ac}
|\alpha^n_l|\le \dfrac{A_{d+1}K_d}{(2(l-d)-1)^{3/2}} = M^d_l  \ , \ \forall n\ge l\geq d+2 \ .
\end{equation}

Returning to (\ref{rhof}), we write
\begin{equation*}\label{Tan}
\lim_{n\to\infty}\int_{-1}^{1}f(z)\rho^j_{k_n}(z)dz = a_0 + \lim_{n\to\infty}\sum_{l=1}^{d+1}\alpha^n_l + \lim_{n\to\infty}\sum_{l=d+2}^{n}\alpha^n_l \ .
\end{equation*}
The first limit is trivially interchanged with the finite sum. For the second limit, we apply Tannery's Theorem, since $|\alpha^n_l|\le M^d_l$, $\forall n\ge l\geq d+2$, and $\sum_{l=d+2}^{\infty} M^d_l < \infty$, {\it cf.} (\ref{ac}). Then, from (\ref{f})-(\ref{alphanl}), using Edmonds formula (\ref{edf}) and the (anti-) Poisson hypothesis, we conclude the thesis.
\end{proof}

As a direct corollary, we have the important special case: 

\begin{corollary}\label{SWP=AL}
The sequence of standard (resp. alternate) Stratonovich-Weyl symbol correspondences localizes (resp. anti-localizes) classically. 
\end{corollary}

\begin{remark}
	Note from (\ref{c<}) that mapping-positive symbol correspondence sequences satisfy equation (\ref{c<d}), thus Corollary \ref{loc1} of Theorem \ref{v_prop} can also be seen as a direct corollary of Theorem \ref{p_l}.
\end{remark}

In Theorem \ref{p_l}, the bounds (\ref{c<d}) on the characteristic numbers are 
necessary for applying Tannery's theorem and Edmonds formula to (\ref{rhof}), but these are sufficient, in principle not necessary means of assuring classical (anti-)localization ({\it cf.} also Theorem \ref{Clocexp2} further below and its discussion in the concluding section). 

Therefore, one could still ask whether (anti-)Poisson condition implies classical (anti-)localization in general. But in this respect we have the following:

\begin{theorem}\label{P<loc}
For general sequences of symbol correspondences, the classical (anti-)localization property is in fact stronger than the (anti-)Poisson property.
\end{theorem}
\begin{proof} The proof consists in exhibiting a set of symbol correspondence sequences of (anti-)Poisson type that fail to (anti-)localize classically.

For any $f\in C^\infty_{\mathbb{C}}([-1,1])$ with Legendre series (\ref{f}) such that $a_l\ne 0$, $\forall l\in \mathbb{N}$, consider the  sequence of symbol correspondences $\boldsymbol{W}_{\!\mathcal{C}}$ with characteristic numbers
\begin{equation}\label{gex}
    c^n_l = \begin{cases}
    (2l+1)/a_l, \ \ \forall n = l \\
    1, \ \ \textnormal{otherwise}
    \end{cases} \ .
\end{equation}

For every $l\in\mathbb N$ and $n>l$, we have $c^n_l =1$. Thus, $c^\infty_l=1$, $\forall l \in \mathbb N$, which means that the symbol correspondence sequence $\boldsymbol{W}_{\!\mathcal{C}}$ is of Poisson type. Its $r$-convergent $\Pi$-distribution sequences satisfy
\begin{equation}
\begin{aligned}\nonumber
    \int_{-1}^{1}f(z)\rho^j_{k_n}(z)dz - f(1-2r) & = (-1)^{k_n-1}C^{j,j,n}_{m,-m,0}\sqrt{n+1}\left(\sqrt{2n+1}-\frac{a_n}{\sqrt{2n+1}}\right)\\
    &  + a_0+\sum_{l=1}^{n}a_l(-1)^{k_n-1}C^{j,j,l}_{m,-m,0}\sqrt{\dfrac{n+1}{2l+1}} - f(1-2r) \ .
\end{aligned}
\end{equation}
Hence, 
\begin{equation*}
    \left|\int_{-1}^{1}f(z)\rho^j_{k_n}(z)dz - f(1-2r)\right| \ge \Bigg| \left|C^{j,j,n}_{m,-m, 0}\sqrt{n+1}\left(\sqrt{2n+1}-\frac{a_n}{\sqrt{2n+1}}\right)\right| - R_n \Bigg|\ ,
\end{equation*}
where
\begin{equation*} 
    R_n= \left|a_0+\sum_{l=1}^{n}a_l(-1)^{k_n-1}C^{j,j,l}_{m,-m,0}\sqrt{\dfrac{n+1}{2l+1}}- f(1-2r)\right| \ .
\end{equation*}

From \cite{varsh},  we have that
\begin{equation}\label{cjjn}
    C^{j,j,n}_{m,-m,0} = \dfrac{(n!)^2}{(j+m)!(j-m)!\sqrt{(2n)!}} 
\end{equation}
and from (\ref{rhof}) and Corollary \ref{SWP=AL}, we have that $R_n\to 0$ as $n\to \infty$, $\forall r\in[0,1]$. But for $r = 1/2$ we can choose $m=0$ whenever $j$ is integer, thus, under these assumptions, by Stirling approximation (\ref{stir}), we have from (\ref{cjjn}) and (\ref{al}) that
\begin{equation*}
     \left|C^{j,j,n}_{m,-m, 0}\sqrt{n+1}\left(\sqrt{2n+1}-\frac{a_n}{\sqrt{2n+1}}\right)\right| \sim \dfrac{2n}{(\pi n)^{1/4}} 
\end{equation*}
for large integers $j$. Hence, such $r$-convergent $\Pi$-distribution sequence, for $r=1/2$, constructed from a symbol correspondence sequence of Poisson type, fails to converge pointwise on $f$ at $0$. Therefore, $\boldsymbol W_{\mathcal C}$ does not localize classically. 
By considering the symbol correspondence sequence $\boldsymbol{W}_{\!\mathcal{C}}$ with characteristic numbers 
\begin{equation*}
    c^n_l = \begin{cases}
    (2l+1)/a_l, \ \ \forall n = l \\
    (-1)^l, \ \ \textnormal{otherwise}
    \end{cases} \ ,
\end{equation*}
we have that $\boldsymbol{W}_{\!\mathcal{C}}$ is of anti-Poisson type but does not anti-localize classically\footnote{Note that $\boldsymbol{W}_{\!\mathcal{C}}$ built as above is just one of infinite possibilities and, in particular, it would have been sufficient to define $c^l_l = \sqrt{2l+1}/a_l$, for instance. However, the construction we chose above will be useful for defining an example that clarifies another property, later on ({\it cf.} Example \ref{ex2}).}.
\end{proof}

On the other hand, one might think that the more well-known and amply-used symbol correspondence sequences  would satisfy the conditions of Theorem \ref{p_l}. However, although the bounds (\ref{c<d}) do not seem too strong, we actually have:

\begin{proposition}\label{toepprop}
    The polynomial bounds (\ref{c<d}) are not satisfied for the standard and the alternate Toeplitz correspondence sequences. 
\end{proposition}
\begin{proof}
Recall that the characteristic numbers of the standard and alternate Toeplitz correspondences are $t^n_l=1/b^n_l$, with $b^n_l$ given by (\ref{bln}), and $t^n_{l-}=(-1)^lt^n_l$. 
We first verify that for any fixed $l\in\mathbb{N}$ the sequence $\{t^n_l\}_{n\ge l}$ is decreasing:
\begin{equation*}
\begin{aligned}
\dfrac{t^{n+1}_l}{t^n_l} & = \dfrac{\dfrac{1}{(n+1)!}\sqrt{\dfrac{(n+l+2)!(n-l+1)!}{n+2}}}{\dfrac{1}{n!}\sqrt{\dfrac{(n+l+1)!(n-l)!}{n+1}}}\\
& = \sqrt{\dfrac{(n+l+2)(n-l+1)}{(n+2)(n+1)}} = \sqrt{1-\dfrac{l(l+1)}{(n+2)(n+1)}} \ < \ 1 \ .
\end{aligned}
\end{equation*}
Therefore, 
\begin{equation}\label{t<}
t^n_l\le t^l_l \ , \ \forall n\ge l \ . 
\end{equation} 
However, computing explicitly the limiting value of
\begin{equation}
t^l_l = \dfrac{1}{l!}\sqrt{\dfrac{(2l+1)!}{l+1}} = \sqrt{\dfrac{(2l+1)!}{l!(l+1)!}} \ ,
\end{equation}
using Stirling formula (\ref{stir}), we obtain 
\begin{equation}\label{TnL}
t^l_l=|t^l_{l-}|=T_l \sim \dfrac{2^{l+1/2}}{(l\pi)^{1/4}} \ , \ \mbox{as} \ l\to\infty \ .
\end{equation}
Thus, $T_l$ increases exponentially and cannot be bounded by any polynomial. 
\end{proof}

Motivated by these examples, in particular equations (\ref{t<})-(\ref{TnL}), we introduce a one-parameter family of weaker asymptotic localizations, as follows. 

For every $\mu >1$, let $\mathcal E_{\mu}\subset \mathbb C$ denote the closed interior of the Bernstein ellipse 
\begin{equation}\label{Bellmu} 
 \partial\mathcal E_{\mu}=\left\{z\in\mathbb C \ | \ z=(u+u^{-1})/2  , \ u=\mu e^{i\phi} , \ \phi\in[-\pi,\pi] \right\} \ ,
\end{equation}
 with foci at $\pm 1$ and sum of major and minor semi-axis equal to $\mu$.
Denote by 
$$\mathcal A_{\mu}([-1,1])\subset C^{\infty}_{\mathbb{C}}([-1,1]) $$
the subspace of all smooth complex functions on $[-1,1]$ which admit holomorphic extensions to $\mathcal E_{\mu}\subset \mathbb C$.  Note that  
\begin{equation}\label{muorder}
1<\mu_1<\mu_2 \ \Rightarrow \ \mathcal E_{\mu_1}\subset \mathcal E_{\mu_2} \ \Rightarrow \ \mathcal A_{\mu_1}([-1,1])\supset \mathcal A_{\mu_2}([-1,1]) \ ,
\end{equation}
thus $f\in\mathcal A_{\infty}([-1,1])$ if $f$ is the restriction to $[-1,1]\subset\mathbb C$ of an entire function. 
In particular, 
$$Poly_{\mathbb C}([-1,1])\subsetneq \mathcal A_{\infty}([-1,1])= \bigcap_{\mu >1}\mathcal A_{\mu}([-1,1]) \ .$$
\begin{definition}\label{bern}
	We shall say that $f$ is \emph{$\mu$-analytic on} $[-1,1]$  if $f\in\mathcal A_{\mu}([-1,1])$. 
\end{definition}

We now introduce: 

\begin{definition}\label{partialrho}\label{locanal}
A $\Pi$-distribution sequence $(\rho^j_{k_n})_{n\in\mathbb N}$ is said to \emph{localize $\mu$-analytically} at $z_0\in[-1,1]$ if 
\begin{equation}\label{partialrho-eq}
    \lim_{n\to\infty}\int_{-1}^1f(z)\rho^j_{k_n}(z)dz=f(z_0) \ , \ \forall f\in \mathcal A_{\mu}([-1,1]) \ .
\end{equation}
    Then, a symbol correspondence sequence  $\boldsymbol{W}_{\!\mathcal{C}}$ is said to \emph{localize $\mu$-analytically}, resp. \emph{anti-localize $\mu$-analytically}, if every $r$-convergent $\Pi$-distribution sequence localizes $\mu$-analytically at $1-2r$, resp. at $2r-1$,
    $\forall r\in [0,1]$.  
\end{definition}

From (\ref{muorder}), if $\boldsymbol{W}_{\!\mathcal{C}}$ localizes $\mu_1$-analytically, then it  localizes $\mu_2$-analytically,  $\forall \mu_2>\mu_1$. But the converse does not necessarily hold.  Then we have: 

\begin{theorem}\label{propToe}
The sequence of standard (resp. alternate) Toeplitz symbol correspondences localizes (resp. anti-localizes) $\mu$-analytically, $\forall \mu >2$.
\end{theorem}

\begin{proof}
The proof follows similarly to the proof of Theorem \ref{p_l}, using (\ref{t<})-(\ref{TnL}) and the fact that, for any $f\in \mathcal A_{\mu}([-1,1])$, we have from \cite[Corollary 2.1]{xia} that $f$ can be written as (\ref{f}) with
\begin{equation}\label{alrho}
|a_l|\leq \frac{2M\sqrt{l}}{\mu^{l-1}(\mu^2-1)} \ , \ \forall l\geq 1 \ , 
\end{equation}
where $|f(z)|\leq M$, for $z\in \mathcal E_{\mu}$. 
\end{proof}

On the other hand, inspired by the previous considerations, we can think in the opposite direction and ask whether a symbol correspondence sequence can localize in a broader sense than the classical one. This leads to the following:

\begin{definition}\label{genlocdef}
Assume that a $\Pi$-distribution sequence $(\rho^j_{k_n})_{n\in\mathbb N}$ satisfies
\begin{equation}\label{genloceq}
    \lim_{n\to\infty}\int_{-1}^1f(z)\rho^j_{k_n}(z)dz=f(z_0) \ , \ \forall f\in \mathcal F_{\mathbb C}([-1,1]) \ ,
    \end{equation}
 where $\mathcal F_{\mathbb C}([-1,1])$ is a space of complex functions on $[-1,1]$ detailed below. 
 
 In each case, $(\rho^j_{k_n})$ is said to: 
 
  \emph{localize $k$-differentiably} at $z_0$, \  if \  $\mathcal F_{\mathbb C}([-1,1])=C^{k}_{\mathbb C}([-1,1])$. 
  
  (if $k=0$, it \emph{localizes continuously}, and it \emph{localizes differentiably} if $k=1$).  
 
 \emph{localize absolute-continuously} at $z_0$, \ if \  $\mathcal F_{\mathbb C}([-1,1])=AC_{\mathbb C}([-1,1])$.
 
 \emph{localize $\alpha$-H\"{o}lder continuously} at $z_0$, \  if \  $\mathcal F_{\mathbb C}([-1,1])=C^{0,\alpha}_{\mathbb C}([-1,1])$. 
 
 \emph{localize $\alpha$-H\"{o}lder $k$-differentiably} at $z_0$, \  if \  $\mathcal F_{\mathbb C}([-1,1])=C^{k,\alpha}_{\mathbb C}([-1,1])$.

    Then, a symbol correspondence sequence  $\boldsymbol{W}_{\!\mathcal{C}}$ \emph{(anti-)localizes $k$-differentiably} \\(\emph{continuously, differentiably}), if every $r$-convergent $\Pi$-distribution sequence localizes $k$-differentiably (\emph{continuously, differentiably}) at  $1-2r$ (resp. $2r-1$), 
    $\forall r\in [0,1]$. 
    
    And similarly for when $\boldsymbol{W}_{\!\mathcal{C}}$ is said to \emph{(anti-)localize absolute-continuously}, or   \emph{$\alpha$-H\"{o}lder continuously}, or  
    \emph{$\alpha$-H\"{o}lder $k$-differentiably}.
\end{definition}

Now, using Lemma \ref{delta_c}, we immediately generalize Theorem \ref{v_prop} to continuous (anti-)localization, and then, using Theorem \ref{conviso} and Corollary \ref{surprising}, we obtain:

\begin{corollary}\label{continlocpoisson}
A mapping-positive symbol correspondence sequence (anti-)localizes continuously if and only if it is of (anti-)Poisson type.
\end{corollary}

But for general symbol correspondence sequences of (anti-)Poisson type we need additional conditions on their characteristic numbers to guarantee (anti-)localization in each of the senses above, as  already seen for classical (anti-)localization.

We are not aware of sufficient bounds which apply for every case, but   the corollary below follows from Lemma \ref{wanglemma} and the fact that the Legendre series of $f$ converges uniformly to $f$ if $f\in C^{k}_{\mathbb C}([-1,1])$ for $k\geq 1$, {\it cf.} \cite{atk}. 
\begin{corollary}\label{kdiffloc}
Under the same hypothesis of Theorem \ref{p_l}, $\boldsymbol{W}_{\!\mathcal{C}}$ (anti-)localizes $k$-differentiably, for $k= d+2$. 
\end{corollary}

Then, taking $d=0$ in (\ref{c<d}), we also have:
\begin{corollary}\label{SWloc2diff}
The standard (resp. alternate) Stratonovich-Weyl symbol correspondence sequence localizes (resp. anti-localizes) $2$-differentiably.
\end{corollary}

When trying to extend Corollary \ref{SWloc2diff} beyond $2$-differential localization, for the standard and alternate S-W symbol correspondence sequences, we encounter two difficulties. The first one is that for a general $f\in C^{0}_{\mathbb C}([-1,1])$, its Legendre series may not converge to $f$ uniformly and this is used in the proof of Theorem \ref{p_l}. This problem can be avoided if we restrict to $f\in C^{0,\alpha}_{\mathbb C}([-1,1])$, $1/2<\alpha\leq 1$, or to $f\in C^{1}_{\mathbb C}([-1,1])$. But then, upper bounds for the Legendre coefficients of functions in these spaces are not known to us.

\section{Sequential quantizations and asymptotic localization} 

\subsection{Sequential quantizations of the $2$-sphere} 
So far we have been 
investigating the (semi)classical limit of quantum spin systems via their dequantizations. In this section we start investigations in the opposite direction, {\it i.e.}~from the classical spin system to quantum spin systems, often referred to as quantization. One of our main purposes, at this point,  is to find relations between  properties of quantized functions and asymptotic localization of symbol correspondence sequences. But first we need to define what we mean by a quantization of the $2$-sphere. 

We start with some basic definitions.   
Denote by $$\textswab{M} \ = \ \{\mathbf F=(F_n)_{n\in\mathbb N}, \ F_n\in M_{\mathbb C}(n+1)\}$$ the set of all sequences of spin-$j$ operators. In this set, we can define the natural operations of \emph{sum}, \emph{product} and \emph{multiplication by a scalar}, as follows:

\begin{definition}
For any  $\mathbf F, \mathbf G\in\textswab{M}$, their \emph{sum} is given by 
\begin{equation}
\mathbf F + \mathbf G \ = \ (F_n + G_n)_{n\in\mathbb N} \ \in \  \textswab{M}
\end{equation}  
and their \emph{product} is given by 
\begin{equation}
\mathbf F\mathbf G \ = \ (F_nG_n)_{n\in\mathbb N} \ \in \  \textswab{M} \ ,
\end{equation}
where $F_nG_n$ is the product in $M_{\mathbb C}(n+1)$, with \  $\mathbf I=(I_n)_{n\in\mathbb N}\in \textswab{M}$ \ as the unit. 

The multiplication of an operator sequence $\mathbf F$ by  $a\in\mathbb C$,  \
 $a\mathbf F=(a F_n)_{n\in\mathbb N}$,   extends naturally to \ \emph{multiplication by a scalar} \ as being the multiplication of $\mathbf F$ by a sequence of complex numbers \ $\mathbf a=(a_n)_{n\in\mathbb N}$,  $a_n\in\mathbb C$, given by 
\begin{equation}
  \mathbf a\mathbf F \ = \ (a_n F_n)_{n\in\mathbb N} \ \in \  \textswab{M} \ , 
  \end{equation}
  with the set of scalars  $$\textswab{C} \ = \ \{\mathbf a=(a_n)_{n\in\mathbb N}, \ a_n\in\mathbb C\}$$ forming a commutative ring under the natural operations \ 
  \begin{equation} 
  \mathbf a +\mathbf b=(a_n+b_n)_{n\in\mathbb N} \ \ , \ \ \mathbf a\mathbf b=(a_nb_n)_{n\in\mathbb N} \ .
  \end{equation}
\end{definition}
\begin{remark} 
Therefore, the set of operator sequences $\textswab{M}$ has the structure of a unital associative (noncommutative) algebra over the commutative ring $\textswab{C}$ of scalars. However, $\textswab{C}$ is not a field because there are nontrivial zero divisors.
\end{remark}

Recall  the normalized inner product   $\langle\cdot|\cdot\rangle_j: M_{\mathbb C}(n+1)\times M_{\mathbb C}(n+1) \to\mathbb C$, 
\begin{equation}\label{ninn}
\quad\quad\quad\quad \langle F_n|G_n \rangle_j=\frac{1}{n+1}tr(F_n^*G_n) \ , \quad\quad cf. \ (\ref{norminn}),
\end{equation}
satisfying  $\langle I_n,I_n \rangle_j=1$ and let the norm $||\cdot|| : M_{\mathbb C}(n+1)\to\mathbb R^+$ be defined as  
	\begin{equation}\label{normj} 
	||F_n||=\sqrt{\langle F_n|F_n\rangle_j} \ .  
	\end{equation}

\begin{definition}\label{asympopnorm}
For any $\mathbf F, \mathbf G \in\textswab{M}$, their \emph{normalized inner product} is given by 
\begin{equation}\label{<F,G>}
 \langle\mathbf F|\mathbf G\rangle = (\langle F_n|G_n\rangle_j)_{n\in\mathbb N} \in\textswab{C} \ .   \end{equation}

For any  $\mathbf F \in\textswab{M}$,  its  \emph{norm} $||\mathbf F||\in \textswab{R}^+$, where  $\textswab{R}^+=\{\mathbf a=(a_n)_{n\in\mathbb N}, \ a_n\in\mathbb R^+\}\subset \textswab{C}$, is given by
	\begin{equation}\label{Fnorm}
	  ||\mathbf F|| = \left(||F_n||\right)_{n\in\mathbb N} \ , \ ||\mathbf F||^2=\langle\mathbf F|\mathbf F\rangle \ . 
	\end{equation}
	Its \emph{lower asymptotic norm} and \emph{upper asymptotic norm} are given by
	\begin{equation}\label{F<>}
	 ||\mathbf F||_{<}=\liminf_{n\to\infty}||F_n||   \in  \mathbb R^+\cup\{\infty\} \ , \  ||\mathbf F||_{>}=\limsup_{n\to\infty}||F_n||   \in  \mathbb R^+\cup\{\infty\}.
	\end{equation}
	When these are equal, the \emph{asymptotic norm} is denoted by \begin{equation}\label{Finfty}
	 ||\mathbf F||_{\infty}=\lim_{n\to\infty}||F_n|| \  \in \ \mathbb R^+\cup\{\infty\} \ . 
	\end{equation}
	$\mathbf F \in\textswab{M}$ 
	is  \emph{upper bounded} if \ $\exists \  ||\mathbf F||_{>}\in\mathbb R$, and is  \emph{properly bounded} if \ $\exists \  ||\mathbf F||_{\infty}\in\mathbb R$.

For any  $\mathbf F \in\textswab{M}$, its \emph{trace} $\tr(\mathbf F)$ is given by 
\begin{equation}\label{seqtrace}
  \tr(\mathbf F)= (\tr(F_n))_{n\in\mathbb N} \in\textswab{C} \ .
\end{equation}
Its \emph{asymptotic trace}, when it exists, is given/denoted by 
\begin{equation}\label{asymptraceb}
 \tr_{\infty}(\mathbf F)=\lim_{n\to\infty} \tr(F_n) \in\mathbb C \ .  
\end{equation}
$\mathbf F \in\textswab{M}$ is \emph{trace-class} if \  $\exists \ tr_{\infty}(|\mathbf F|)\in\mathbb R$, for $|\mathbf F| = (|F_n| = \sqrt{F_n^\ast F_n})$. 
\end{definition}
	
In the previous section, we were particularly interested in investigating sequences of operators for which, given a symbol correspondence sequence $\boldsymbol{W}_{\!\mathcal{C}}=(W^j)_{n\in\mathbb N}$, its sequence of symbols converge to a $J_3$-invariant classical function $f\in C^{\infty}([-1,1])$ or to a $J_3$-invariant $\mu$-analytic function $f\in\mathcal A_{\mu}([-1,1])$. In this case, if $f$ is the (uniform) limit of a sequence $(f_n)_{n\in\mathbb N}$ of $J_3$-invariant symbols of the form
\begin{equation}\label{genfn}
f_n= \sum_{l=0}^{n}\chi_l^nP_l \ , \ \chi_l^n\in\mathbb C \ ,
\end{equation}
then for simplicity we can consider sequences of $J_3$-invariant operators which 
can be written in the form $\mathbf F^W=(F_n^W)_{n\in\mathbb N}$, for $F_n^W\in M_{\mathbb C}(n+1)$  given by
\begin{equation}\label{FWn}
F_n^W=\sum_{l=0}^n\frac{\chi_l^n}{c^n_l\sqrt{2l+1}}\widehat{\mathbf{e}}^j(l,0) \ ,  
\end{equation}
where
\begin{equation}\label{Ee}
\widehat{\mathbf{e}}^j(l,m)=\sqrt{n+1}\mathbf{e}^j(l,m)
\end{equation}
are orthonormal basis vectors of $M_{\mathbb C}(n+1)$ w.r.t. the normalized inner product $\langle \cdot|\cdot \rangle_j$ on $M_{\mathbb C}(n+1)$, with  
$I_n=\widehat{\mathbf{e}}^j(0,0)$ , that is, 
\begin{equation}\label{EE}
    \langle \widehat{\mathbf{e}}^j(l,m)| \widehat{\mathbf{e}}^j(l',m')\rangle_j = \delta_{l,l'}\delta_{m,m'} \ .
    \end{equation}
Furthermore, from the Legendre series for $f$,
\begin{equation}\label{Lf}
f=\lim_{n\to\infty}\sum_{l=0}^n a_lP_l \ , 
\end{equation}
we have that
\begin{equation}\label{Lc}
\lim_{n\to\infty}\chi_l^n=a_l=\frac{2l+1}{2}\int_{-1}^{1}f(z)P_l(z)dz \ , \ \forall l\in\mathbb N.
\end{equation} 

In view of all this, we now introduce the following general definition: 

\begin{definition}\label{quantization-def}
For any classical function $f\in C^{\infty}_{\mathbb C}(S^2)$, with harmonic series
	\begin{equation}\label{hf} f=\lim_{n\to\infty}\sum_{l=0}^{n}\sum_{m=-l}^l a_l^mY_l^m \ , \ \ a_l^m=\langle Y_l^m|f\rangle  \in\mathbb C \ ,
	\end{equation}
	given a  symbol correspondence sequence $\boldsymbol{W}_{\!\mathcal{C}}$ with characteristic numbers $c_l^n$, the \emph{$W$-pseudoquantization} of $f$ is the operator sequence $\mathbf F^{w}=(F_n^{w})_{n\in\mathbb N}\in\textswab{M}$ given by 
	\begin{equation}\label{FWnlm}
	F_n^{w}=[W^j]^{-1}(f)= \sum_{l=0}^n\sum_{m=-l}^l \frac{a_l^m}{c^n_l }\widehat{\mathbf{e}}^j(l,m) \  , 
	\end{equation}
	and the \emph{$\widetilde{W}$-pseudoquantization} of $f$ is the operator sequence $\widetilde{\mathbf{F}}^{w}=(\widetilde{F}_n^{w})_{n\in\mathbb N}\in\textswab{M}$, 
	\begin{equation}\label{dFWnlm}
	\widetilde{F}_n^{w}=[\widetilde{W}^j]^{-1}(f)= \sum_{l=0}^n\sum_{m=-l}^l a_l^mc^n_l \widehat{\mathbf{e}}^j(l,m) \  . 
	\end{equation}
	
	If \ $\boldsymbol{W}_{\!\mathcal{C}}$ is of (anti-)Poisson type, then $\mathbf F^{w}$ and $\widetilde{\mathbf{F}}^{w}$ are called, respectively, the  \emph{$W$-quantization} and the \emph{$\widetilde{W}$-quantization} of $f$. 
	\end{definition}
	\begin{remark}
	As discussed in \cite[Chapter 9]{prios}, applying the term \emph{quantization} to the Poisson algebra  $\{C^{\infty}_{\mathbb C}(S^2), \omega\}$ requires that $\boldsymbol{W}_{\!\mathcal{C}}$ be of (anti-)Poisson type. \end{remark}
	 
	 Still, nothing has been explicitly said about the sequence of spaces which are acted upon by the operator sequences $\mathbf F^{w}$ and $\widetilde{\mathbf{F}}^{w}$ defined above for $f\in C^{\infty}_{\mathbb C}(S^2)$. 
	 In order to explore concrete representations of the sequential quantizations $\mathbf F^{w}, \widetilde{\mathbf{F}}^{w}\in\textswab{M}$, we start with the 
	following proposition, which was set out by Bargmann \cite{barg}. 
	 
	 \begin{definition}[\cite{barg1, barg, seg}]\label{bar1}
	 Let $\mathcal H_{ol2}$ denote the space of holomorphic functions in $2$ complex variables. We shall denote by $\mathcal H_{ol2}^{2\mu}$ the Hilbert space of 
	 holomorphic functions that are  $L^2$-integrable with respect to the inner product
	 \begin{equation}\label{invinnp}
	     \langle \phi|\psi\rangle_{{\mathcal H}_{ol2}^{2\mu}}= \frac{1}{(2\pi )^2}\int_{\mathbb C^2}\overline{\phi(u)}\psi(u)d\mu(u) \  ,
\end{equation}
where $u=(u_1,u_2)\in\mathbb C^2$ and 
\begin{equation}\label{muu}
   d\mu(u)= e^{-u\overline{u}}dud\overline{u}=d\mu(u_1,u_2) \ , 
\end{equation}
with \ $u\overline{u}=u_1\overline{u}_1+u_2\overline{u}_2$ , $dud\overline{u}= du_1\wedge du_2\wedge d\overline{u}_1\wedge d\overline{u}_2$ .  
	 	 \end{definition}
	 
	 \begin{proposition}[\cite{barg}]\label{bar2}
The inner product defined by (\ref{invinnp})-(\ref{muu}) is invariant under the standard action of $SU(2)$ on $\mathbb C^2$. Under this action, we have the splitting: 
\begin{equation}\label{directsplit}
    \mathcal H_{ol2}^{2\mu}=\bigoplus_{2j=0}^{\infty}\mathcal H^j_{om2} \ , 
\end{equation}
where $\mathcal H_{om2}^j$ is the space of homogeneous polynomials of degree $n=2j$ in $2$ complex variables, so that each $\mathcal H_{om2}^j$, of dimension $n+1$, defines a spin-$j$ system, in other words, the $SU(2)$ action on $\mathcal H_{ol2}^{2\mu}$ splits into irreducible unitary representations $$\varphi_j:SU(2)\to U(n+1) \ \mbox{acting on each} \   \mathcal H_{om2}^j \ .$$   

Furthermore, in each $\mathcal H_{om2}^j$ we define the restricted inner product  $\langle\cdot|\cdot\rangle_{\mathcal H_{om2}^j}$ induced from (\ref{invinnp})-(\ref{muu}), that is, $\forall \phi_j,\psi_j\in {\mathcal H}_{om}^j$, 
\begin{equation}\label{innrest}
  \langle \phi_j|\psi_j\rangle_{{\mathcal H}_{om}^j}= \  \frac{1}{(2\pi )^2}\int_{\mathbb C^2}\overline{\phi_j(u)}\psi_j(u)d\mu(u) \ = \  \langle \phi_j|\psi_j\rangle_{{\mathcal H}_{ol2}^2} \  ,   
\end{equation}
with $d\mu(u)$ still given by (\ref{muu}). 
It follows that the set 
\begin{equation}\label{sbhom2}
    \left\{\mathbf{u}(j,m)= \frac{u_1^{j-m}u_2^{j+m}}{\sqrt{(j-m)!(j+m)!}}\right\}_{-j\leq m\leq j} 
\end{equation}
forms a standard basis for the respective spin-$j$ system, that is, 
$\langle \mathbf{u}(j,m)|\mathbf{u}(j,m')\rangle = \delta_{m,m'}$ , where $\langle\cdot|\cdot\rangle=\langle\cdot|\cdot\rangle_{\mathcal H_{om2}^j}$ , and \ $\mathbf{u}(j,m)$ satisfies (\ref{stansu2act})-(\ref{ab}). 
	 \end{proposition}

\begin{example}\label{bargmann}
In view of the above, any operator $T_{\alpha}^j: \mathcal H_{om2}^j\to\mathcal H_{om2}^j$, $\psi_j\mapsto 	\phi_j$  is determined by an integral kernel \ $\mathcal K_{\alpha}^j$ via 
	 	\begin{equation}\label{kjzh}
	 	\phi_j(u) =\frac{1}{(2\pi)^2}\int_{\mathbb C^2}\mathcal K_{\alpha}^j(u,v)\psi_j(v)d\mu(v)
	 	\end{equation} 	
 which from (\ref{innrest})-(\ref{stansu2act}) has the form 
 \begin{equation}\label{kj2z} 
	 	\mathcal K_{\alpha}^j(u,v) \ =\!\sum_{m,m'=-j}^j \frac{\kappa(\alpha)^j_{m,m'}u_1^{j-m}u_2^{j+m}\overline{v}_1^{j-m'}\overline{v}_2^{j+m'}}
	 		 	{\sqrt{(j-m)!(j+m)!(j-m')!(j+m')!}} \ , 
	 	\end{equation} 
	 	\begin{equation}\label{kamuz}
	 	  \kappa(\alpha)^j_{m,m'}=(-1)^{j-m'} M(\alpha)^j_{m,-m'} \ ,   
	 	\end{equation}
 {\it cf.} (\ref{dualbasis}), where $M(\alpha)^j_{m,m'}$ are the entries in the matrix 
 \begin{equation}\label{MAjz}
	 	\sum_{m,m'=-j}^j  M(\alpha)^j_{m,m'}\boldsymbol{u}(j, m)\otimes\check{\boldsymbol{u}}(j, m') \ \in \ M_{\mathbb C}(n+1) \ . 
	 	\end{equation}
		
		Then, for  $f\in C^{\infty}_{\mathbb C}(S^2)$ and  $\boldsymbol{W}_{\!\mathcal{C}}$ of (anti-)Poisson type, the 
	 	coefficients $\widetilde{\kappa}^{\!w}\!(f)^j_{m,m'}$ in (\ref{kj2z}) of the integral kernel $\widetilde{\mathcal K}_{w}^j[f]$ in the $\widetilde{W}$-quantization of $f$ are given by 
	 	\begin{equation}\label{quantizefzh}
	 	\widetilde{\kappa}^{\!w}\!(f)^j_{m,m'}=\sqrt{n+1}(-1)^{j-m'}\sum_{l=0}^n\sum_{\bar{m}=-l}^l c_l^n\langle Y_l^{\bar{m}}|f\rangle C_{m,-m',\bar{m}}^{\,\,\,j,\,\,\,\,\,\,j,\,\,\,\,\,l}  \ , 
	 	\end{equation}
	 {\it cf.} (\ref{Ee}), (\ref{hf}), (\ref{dFWnlm}), (\ref{cg}), with $\langle \cdot|\cdot\rangle$ given by (\ref{innS2}). 
	 Similarly, from (\ref{FWnlm}), for the coefficients   ${\kappa}^{\!w}\!(f)^j_{m,m'}$ and the integral kernel ${\mathcal K}_{w}^j[f]$ in the ${W}$-quantization sequence of $f$,  substituting $c_l^n\leftrightarrow 1/c_l^n$ in (\ref{quantizefzh}). Thus, ${\mathbf{F}}^{w}$ and $\widetilde{\mathbf{F}}^{w}$ define sequences of kernels $\mathbf{K}_w[f]=(\mathcal K_{w}^j[f])_{n\in\mathbb N}$ and $\widetilde{\mathbf{K}}_w[f]=(\widetilde{\mathcal K}_{w}^j[f])_{n\in\mathbb N}$, determining sequences of integral operators acting via (\ref{kjzh}) on the sequence of Hilbert spaces $\mathcal H_{om2}^j$.  
	 \end{example}

 Now, because $\langle\cdot|\cdot\rangle_{\mathcal H_{om2}^j}$ is the restriction to $\mathcal H_{om2}^j\subset \mathcal H_{ol2}^{2\mu}$ of the inner product (\ref{invinnp}) on ${\mathcal H}_{ol2}^{2\mu}$, it follows that a countable orthonormal basis for $\mathcal H_{ol2}^{2\mu}$ is given by
\begin{equation*}\label{sbhol2}
    \left\{\mathbf{u}(j,m)= \frac{u_1^{j-m}u_2^{j+m}}{\sqrt{(j-m)!(j+m)!}}\right\}_{-j\leq m\leq j, \ 2j\in\mathbb N_0} \ .
\end{equation*}

However, in this Bargmann representation it is not so easy to picture the limit $j\to\infty$. From one perspective, the one described above, the Hilbert space $\mathcal H_{om2}^j$ tends, as $j\to\infty$, to the space of homogeneous holomorphic polynomials of infinite degree in $2$ complex variables, which is meaningless.
More meaningful would be to consider the direct sum (\ref{directsplit}) and consider the Hilbert space $ \mathcal H_{ol2}^{2\mu}$ of $L^2$-bi-holomorphic functions, which could also be seen as a sequence of ``partial sums'' 
\begin{equation*}\label{directsplitk}
  (\mathcal H_{ol2}^{2,k})_{k\in\mathbb N_0} \ , \  \   \mathcal H_{ol2}^{2,k}=\bigoplus_{2j=0}^{k}\mathcal H^j_{om2} \ . 
\end{equation*}
However, for each $k\in\mathbb N_0$, $\dim(\mathcal H_{ol2}^{2,k})=(k+1)(k+2)/2$ thus, for $k>0$, each $\mathcal H_{ol2}^{2,k}$ is ``too big'' for a spin-$j$ system, defining instead a reducible unitary representation of $SU(2)$ consisting of the direct sum of spin-$j$ systems for all $2j\leq k$, including the $1$-dimensional trivial spin system $2j=0$. In other words, the separable Hilbert space $\mathcal H_{ol2}^{2\mu}$ is ``too big'' to be the $j\to\infty$ limit Hilbert space of spin-$j$ systems. 

Therefore, we need a more tailored approach, to be developed below.

\subsection{Ground Hilbert spaces and asymptotic operators}\label{gentheory}

Let  $\textswab{H}$  be a sequence of complex Hilbert spaces, \begin{equation}
	  \textswab{H} = (\mathcal H^j)_{2j=n\in\mathbb N}  \ , \ \ \dim_{\mathbb C}(\mathcal H^j)=n+1 \  \Leftrightarrow \  \mathcal H^j  \simeq \mathbb C^{n+1} \ , 
	\end{equation}   
each $\mathcal H^j$ with its inner product \ $\langle \cdot | \cdot \rangle_j$ ,
conjugate linear in the first entry so that, if $\{\mathbf e^j_k\}_{1\leq k\leq n+1}$ is an orthonormal basis w.r.t. $\langle \cdot | \cdot \rangle_j$ on $\mathcal H^j$, for $\phi^j,\psi^j\in\mathcal H^j$ we have: 
\begin{equation}\label{stdecj}
 \phi^j=\sum_{k=1}^{n+1} \alpha^j_k \mathbf e^j_k \ , \  \psi^j=\sum_{k=1}^{n+1} \beta^j_k \mathbf e^j_k \ \ \Rightarrow \ \  \langle \phi^j | \psi^j \rangle_j =
 \sum_{k=1}^{n+1} \overline{\alpha}^j_k\beta^j_k \ . 
\end{equation}

\begin{definition}\label{MH}
For any $\Phi=(\phi^j)_{2j=n\in\mathbb N}$ and $\textswab{H} = (\mathcal H^j, \langle \cdot | \cdot \rangle_j)_{2j=n\in\mathbb N} $ , we denote 
\begin{equation}\label{notationH}
    \Phi=(\phi^j)_{2j=n\in\mathbb N}\in \textswab{H} \  \iff \  \phi^j\in\mathcal H^j \ , \ \forall n=2j\in\mathbb N \ .  
\end{equation}
For any $\Phi=(\phi^j)_{2j=n\in\mathbb N} , \Psi=(\psi^j)_{2j=n\in\mathbb N} \in\textswab{H}$ , their \emph{sum} is defined naturally by 
\begin{equation}
    \Phi+\Psi= (\phi^j+\psi^j)_{2j=n\in\mathbb N} \ \in \ \textswab{H} \ .
\end{equation}
Also, $\textswab{H}$ is a bi-module of the commutative ring of scalars $\textswab{C}$, for the \emph{multiplication by scalar} defined naturally, for every $\mathbf a\in\textswab{C}$, $\Phi\in\textswab{H}$ , by 
\begin{equation}
    \mathbf a\Phi=\Phi\mathbf a = (a_n\phi^j)_{2j=n\in\mathbb N} \ \in \ \textswab{H}
\end{equation}
and $\textswab{H}$ is a left module for the algebra of operator sequences $\textswab{M}$ with \emph{sequential action} \  $\textswab{M}\times\textswab{H}\to\textswab{H}$ \ given by
\begin{equation}\label{seqaction}
(\mathbf F , \Phi) \ \mapsto \  \  \mathbf F(\Phi)=  (F_n(\phi^j))_{2j=n\in\mathbb N} \ , 
\end{equation}
where $F_n(\phi^j)$ is the usual action $(\mathcal B^j\times\mathcal H^j\to\mathcal H^j)$ $\simeq$ $(M_{\mathbb C}(n+1)\times \mathbb C^{n+1}\to\mathbb C^{n+1})$.

Furthermore, 
we define the \emph{inner product} and \emph{norm} on $\textswab{H}$ , respectively by 
\begin{eqnarray}
\quad\quad  \langle \Phi | \Psi \rangle  &=& (\langle \phi^j | \psi^j \rangle_j)_{2j=n\in\mathbb N} \ \  \in \ \textswab{C} \ , \\
  ||\Phi||^2 &=& \langle \Phi | \Phi \rangle  =  (\langle \phi^j | \phi^j \rangle_j)_{2j=n\in\mathbb N}  =  (||\phi^j||^2)_{2j=n\in\mathbb N} \  \  \in \  \textswab{R}^+ .
\end{eqnarray}

Then, for any $\vb F\in \textswab M$,  the \emph{$\textswab H$-operator norm} $||{\vb F}||_{op}\in \textswab R^+$ is given by
	\begin{equation}
	    ||{\vb F}||_{op} = (||{F_n}||_{op}) \ , \ \ ||{F_n}||_{op} = \sup_{\phi_j \in \mathcal H_j\backslash\{0\}} ||{F_n(\phi_j)}||/||{\phi_j}|| \ , 
	\end{equation}
and its \emph{asymptotic $\textswab{H}$-operator norm}, when it makes sense, is given/denoted by 
\begin{equation}
    ||{\vb F}||_{op}^{\infty}=\lim_{n\to\infty}||F_n||_{op} \in\mathbb R^+\cup\{\infty\} \ .  
\end{equation}
If $\norm{\vb F}_{op}$ is bounded, $\vb F$ is called \emph{$\textswab H$-bounded}, otherwise it is called \emph{$\textswab H$-unbounded}.
\end{definition}

\begin{definition}\label{nestedHilb}
We say that $\textswab{H}^< = (\mathcal H^j,\langle\cdot|\cdot\rangle_j)_{2j=n\in\mathbb N}$ is a \emph{nested sequence of Hilbert spaces} if, for every $j\le j'$, there exists an injective linear \emph{nesting map} $\iota_j^{j'}:\mathcal H^j\to\mathcal H^{j'}$ which is also an isometry, {\it i.e.} such that \ 
$\langle\cdot|\cdot\rangle_j = (\iota_j^{j'})^*\langle\cdot|\cdot\rangle_{j'}$ . 
\end{definition}

It follows that, for any $j\le j'$, $\mathcal H^{j'}=\iota_j^{j'}(\mathcal H^j)\oplus (\iota_j^{j'}(\mathcal H^j))^{\perp}$, where $\perp$ is with respect to $\langle\cdot|\cdot\rangle_{j'}$. 
Thus, $\forall \phi^{j'}\in\mathcal H^{j'}$ we define 
\begin{equation}
\phi^{j'} = \hat{\phi}_{j}^{j'} + \check{\phi}_{j}^{j'} \ , \  \ \hat{\phi}_{j}^{j'} = P_{j,j'}(\phi^{j'}) \ , \  \check{\phi}_{j}^{j'} = P_{j,j'}^{\perp}(\phi^{j'}) \ ,
\end{equation}
where $P_{j,j'}, P_{j,j'}^{\perp}$ are the projectors  $P_{j,j'}:\mathcal H^{j'}\to \operatorname{Im}(\iota_j^{j'})$, $P_{j,j'}^{\perp}:\mathcal H^{j'}\to \operatorname{Im}(\iota_j^{j'})^{\perp}$, and also,  $\forall \psi^{j}\in\mathcal H^{j}$ we define 
\begin{equation}
    {\psi}_j^{j'}=\iota_j^{j'}(\psi^j)\in\mathcal H^{j'}  \implies \ \hat{{\psi}}_j^{j'}={\psi}_j^{j'} \ , \ \check{{\psi}}_j^{j'}= 0 \ . 
\end{equation}

\begin{definition}\label{convergentPhi}
Let $\textswab{H}^< = (\mathcal H^j,\langle\cdot|\cdot\rangle_j)_{2j=n\in\mathbb N}$ be a nested sequence of Hilbert spaces. 
We say that $\vb F=(F_n)_{n\in\mathbb N}\in \textswab{M}$ is \emph{rigid} if \ $\vb F:\textswab{H}^<\to\textswab{H}^<$ given by (\ref{seqaction}) also satisfies 
\begin{equation}
    F_{n'}\circ\iota_{j}^{j'}= \iota_{j}^{j'}\circ F_n \ , \ \forall n, n' \ \mbox{with} \ \ n=2j\leq 2j'=n'\ , 
\end{equation}
and we denote by \ $\textswab{M}^<$ the set of all rigid  operator sequences, $\textswab{M}^<\subset \textswab{M}$.

Furthermore, on $\textswab{H}^< = (\mathcal H^j,\langle\cdot|\cdot\rangle_j)_{2j=n\in\mathbb N}$, the \emph{nested norm} $\norm{\cdot}$ is defined for any \ $\psi^j\in\mathcal H^j$, $\phi^{j'}\in\mathcal H^{j'}$, with $j\le j'$, by 
\begin{equation}\label{cauchyn}
  \norm{\phi^{j'}-\psi^j} := \norm{\phi^{j'}- {\psi}_j^{j'}} \ , 
\end{equation}
where the norm on the r.h.s. of (\ref{cauchyn})  is taken w.r.t. $\langle\cdot|\cdot\rangle_{j'}$, {\it cf.} (\ref{normj}). Similarly for $j'\leq j$.
Then, the \emph{space of convergent state sequences} is
\begin{equation}
    \textswab H_\infty^< = \left\{\Phi = (\phi_j)_{2j=n\in\mathbb N}\in \textswab H^< \ \big| \ \exists \lim_{j\to\infty}\phi_j \right\} \ ,
\end{equation}
where convergence of \ $\Phi = (\phi_j)_{2j=n\in\mathbb N}\in \textswab H^<$  is in the sense of Cauchy w.r.t.~the nested norm on \  $\textswab H^<$ given by (\ref{cauchyn}).

Accordingly, the set of \emph{convergent operator sequences} is defined as
\begin{equation}
    \textswab M_\infty = \left\{\vb F \in \textswab M \ \big| \ \vb F:\textswab H_\infty^< \to \textswab H_\infty^< \right\} \ .
\end{equation}

Here we use notation (\ref{notationH}) and redefine all operations in Definition \ref{MH} as natural restrictions to $\textswab{H}^<_\infty$ and $\textswab M_\infty$, so that, in particular, \  $\textswab{M}_\infty\times\textswab{H}^<_\infty\to\textswab{H}^<_\infty$ is still given by (\ref{seqaction}), $\forall \vb F\in\textswab{M}_\infty$.
\end{definition}

\begin{remark}\label{rig}
The sets $\textswab H_\infty^<$ and  $\textswab M_\infty$ are the relevant ones for studying the   asymptotic limit $j\to\infty$. On the other hand, apart from the identity and some projector sequences, in general not many operator sequences are rigid. But in Example \ref{S1cS2}, the sequence $(J_3^j)_{j\in\mathbb N}$ is rigid, where $J_3^j=d\varphi_j(iL_3)$, $L_3\in\mathfrak{so}(3)$, for $(\varphi_j)_{j\in\mathbb Z}$ a sequence of representations $\varphi_j:SO(3)\to U(n+1)$. Hence, in this case, it is also rigid the sequence $(\varphi_j(g))_{j\in\mathbb N}$, $\forall g=g(\theta)\in S^1=\{\exp(\theta L_3):\theta\in\mathbb R\!\!\!\mod \!2\pi\}\subset SO(3)$.
\end{remark}

Now, given an orthonormal basis  $\{\vb e^j_k\}_{1\leq k\leq 2j+1}$ for $\mathcal H^j\subset \textswab H_\infty^<$ there is a canonical way to choose an orthonormal basis for $\mathcal H^{j+1/2}\subset \textswab H_\infty^<$ , up to a phase, by taking 
\begin{equation}\label{almostcan}
\begin{aligned}
   \vb e^{j+1/2}_k &= \iota_j^{j+1/2}(\vb e^j_k) \ , \ 1\leq k\leq 2j+1 \ , \\ 
   \vb e^{j+1/2}_{2j+2} &\in (\iota_j^{j+1/2}(\mathcal H^j))^{\perp} \ , \ \norm{\vb e^{j+1/2}_{2j+2}}=1 \ .
   \end{aligned}
\end{equation}
Thus, starting with a standard basis for $\mathcal H^{1/2}\subset\textswab H_\infty^<$, we can inductively define an orthonormal basis for every $\mathcal H^j\subset \textswab H_\infty^<$, up to ${2j-1}$ choices of phases.   

\begin{definition}\label{wn}
We say that $\textswab H_\infty^<$ is \emph{well-nested} if there exists a canonical choice of phase for $\vb e^{j+1/2}_{2j+2}$ in (\ref{almostcan}), for every $2j=n\in \mathbb N$.

The unique sequence of orthonormal basis $\textswab E=(\mathcal E^j)_{2j\in\mathbb N}=(\{\vb e^j_k\}_{1\leq k\leq 2j+1})_{2j\in\mathbb N}$ for 
$\textswab H_\infty^<$ which is obtained inductively from a standard basis $\mathcal E^{1/2}$ for $\mathcal H^{1/2}\subset\textswab H_\infty^<$ using the canonical phase choice is called a \emph{well-nested} basis sequence\footnote{Alternatively, given a sequence of representations $\varphi_j:SU(2)\to U(n+1)$ defining a sequence of standard basis $\mathcal E^j$ for $\mathcal H^j$, the well-nested basis sequence is defined by specifying the nesting maps $\iota_j^{j'}:\mathcal E^j\to\mathcal E^{j'}$ plus a unique choice of ``overall constant'' for each $j$, that is, a function $\eta:\mathbb N\to\mathbb C^*$, {\it cf. e.g.} (\ref{cc}) for $\eta(2j)=\nu_j$ in Example \ref{example-cp1} and (\ref{newrho}) for $\eta(2j)=\rho_j$ in Example \ref{S1cS2}.} for $\textswab H_\infty^<$. 
\end{definition}

From now on, we shall assume that $\textswab H_\infty^<$ is {well-nested}, with a well-nested basis sequence $\textswab E=(\mathcal E^j)_{2j\in\mathbb N}=(\{\vb e^j_k\}_{1\leq k\leq 2j+1})_{2j\in\mathbb N}$.

\begin{definition}\label{ghs}
The \emph{ground Hilbert space} $\mathcal H$ for well-nested $(\textswab H_\infty^<,\textswab E)$ is the space of complex $\ell^2$-sequences spanned by the  
 \emph{grounding basis}\footnote{From $\mathbb N\leftrightarrow\mathbb Z$, sometimes it may be useful to describe this countable basis as $\mathcal E=\{\mathbf e_m\}_{m\in\mathbb Z}$ or as $\mathcal E=\{\mathbf e_m\}_{2m\in\mathbb Z}$, with $\vb e_m=\lim_{j\to\infty}\vb e^j_m$ for $\{\vb e^j_m\}_{-j\leq m\leq j}$ orthonormal basis of $\mathcal H^j$.} $\mathcal E=\{\vb e_k\}_{k\in\mathbb N}=\mathcal E^{\infty}$,  
\begin{equation}\label{infbasis}
 \vb e_k := \lim_{j\to\infty} \vb e^j_k \ , \ \forall k\in\mathbb N \ ,     
\end{equation}
where each limit in (\ref{infbasis}) is taken in the sense of Definition \ref{convergentPhi}, 
with inner product $\langle\cdot |\cdot\rangle$ on $\mathcal H$ being conjugate-linear in the first entry and satisfying  
\begin{equation}\label{kk'}
    \langle \vb e_k | \vb e_{k'}\rangle = \delta_{k,k'} \ , \ \forall k,k'\in\mathbb N \ .
\end{equation}
That is, $\mathcal E$ provides the identification $\mathcal H\ni\phi\iff(\alpha_k)_{k\in\mathbb N}\in \ell^2$, $\alpha_k\in\mathbb C$, via  
\begin{equation}
    \phi\in\mathcal H \ \iff \  \phi=\sum_{k=1}^{\infty}\alpha_k \vb e_k \ , \ \langle\phi|\phi\rangle=\sum_{k=1}^{\infty}|\alpha_k|^2 < \infty \ . 
\end{equation}
\end{definition}

Noting that,  $\forall j\in\mathbb N$, $\langle\cdot|\cdot\rangle_j$ is conjugate-linear in the first entry and satisfies 
\begin{equation}\label{kk'j}
    \langle \vb e_k^j | \vb e_{k'}^j\rangle_j = \delta_{k,k'} \ , \ 1\leq\forall k,k'\leq 2j+1 \ , 
\end{equation}
we can take (\ref{kk'}) to be the consistency condition between (\ref{infbasis}) and (\ref{kk'j}). 

\begin{theorem}\label{H=H}
Let $(\textswab H_\infty^<,\textswab E)$ be well-nested and let $[\Phi]$ denote the equivalence class of $\Phi\in\textswab{H}_{\infty}^<$, under the equivalence relation $(\phi^j)=\Phi\approx \widetilde \Phi=(\widetilde \phi^j)\in\textswab{H}_{\infty}^<$ given by
\begin{equation}\label{er1}
  \Phi\approx\widetilde\Phi \  \iff \ 
  \lim_{j\to\infty}\phi^j = \lim_{j\to\infty}\widetilde \phi^j \ .
\end{equation} 
Then, $\mathcal H$ is isomorphic to \  $\textswab{H}_{\infty}^</\approx$ and \ 
$\textswab E$ \ provides a canonical isomorphism 
\begin{equation}\label{Hidentif}
    \mathcal H \ = \ \{\phi=\lim_{j\to\infty}\phi^j\, \equiv [\Phi] \ | \ \Phi = (\phi^j)\in[\Phi]\} \ . 
\end{equation}

Furthermore, $\textswab E$ determines the sequence of 
isometries $\Gamma: (\textswab H_\infty^<,\textswab E)\to(\mathcal H,\mathcal E)$, with  $\Gamma=(\gamma_j)_{2j\in\mathbb N}$ given by 
\begin{equation}\label{isoj}
\gamma_j:(\mathcal H^j,\mathcal E^j)\to (\mathcal H,\mathcal E) \ , \ \  \phi^j=\sum_{k=1}^{2j+1}\alpha^j_k\vb e^j_k \  \mapsto \  \sum_{k=1}^{2j+1}\alpha^j_k\vb e_k = \gamma_j(\phi^j) \ .
\end{equation}
\end{theorem}

\begin{proof}
First, for any $k\in\mathbb N$, let $(\hat{\vb e}_k^j)_{2j\in\mathbb N}\in \textswab H^<$ be such that $\hat{\vb e}_k^j=0$, for $2j+1<k$, $\hat{\vb e}_k^j={\vb e}_k^j$ , otherwise. Then, clearly, 
$(\hat{\vb e}_k^j)_{2j\in\mathbb N}\in \textswab H_\infty^<$, hence, for any $k\in\mathbb N$, we have the canonical identification $\vb e_k=[(\hat{\vb e}_k^j)_{2j\in\mathbb N}]\in \textswab H_\infty^</\approx$ .

Now, define $L: \mathcal H\to\textswab H_\infty^<$ and $\Lambda: \mathcal H\to\textswab H_\infty^</\approx$ , by 
\begin{eqnarray}\label{phi-Phi}
    \phi = \sum_{k=1}^\infty \alpha_k \, \vb e_k &  \mapsto &  L\phi = (\phi^j) \ , \ \phi^j = \sum_{k=1}^{n+1}\alpha_k \, \vb e_k^j \in \mathcal H^j \\
    \Lambda\phi & = & [L\phi] \ . \label{Lambdaphi}
\end{eqnarray}
It is clear that $\Lambda$ is linear, we will show it is a bijection. 

For $\phi, \psi \in \mathcal H$, let $L\phi= (\phi^j)$ and $L\psi= (\psi^j)$ be as in  (\ref{phi-Phi}). If $\phi \neq \psi$, then $\norm{\phi-\psi} > 0$, where the norm $\norm{\cdot}$ in $\mathcal H$ is the $\langle\cdot|\cdot\rangle$ norm, and because $\langle\vb e^j_k|\vb e_{k'}^{j}\rangle=\langle\vb e_k|\vb e_{k'}\rangle$, $\forall 2j+1\geq k,k'$, there exists $j_0$ such that $\norm{\phi^{j'}-\psi^j}>0$, $\forall j',j\ge j_0$. Thus, $\Lambda\phi \neq \Lambda\psi$ and $\Lambda$ is an injection.

Now, let $\widetilde\Phi = (\widetilde\phi^j)\in \textswab H_\infty^<$ be a general element, with
\begin{equation}\label{tilPhi}
    \widetilde \phi^j = \sum_{k=1}^{n+1}\alpha_k^j \, \vb e_k^j \ .
\end{equation}
For any $\epsilon> 0$ there is $2j_0 = n_0\in \mathbb N$ such that
\begin{equation}\label{cauchy-phi}
\norm{\widetilde\phi^{j'} - \widetilde\phi^j}^2 = \sum_{k=1}^{2j+1} \left|\alpha_k^{j'}-\alpha_k^j\right|^2 + \sum_{k=2j+2}^{2j'+1}\left|\alpha^{j'}_k\right|^2 < \epsilon    
\end{equation}
for every $j' > j \ge j_0$. Thus, for any $k\in \mathbb N$, $(\alpha^j_k)_{2j\in\mathbb N}$ is a Cauchy sequence of complex numbers and, hence, there exists $(\alpha_k)_{k\in\mathbb N}\in \textswab C$ satisfying
\begin{equation}\label{ak>a}
    \lim_{j\to\infty} \alpha^j_k = \alpha_k \ . 
\end{equation}

From (\ref{cauchy-phi}), we get
\begin{equation*}
    \sum_{k=2j_0+2}^{2j+1}\left|\alpha_k\right|^2 = \lim_{j'\to\infty}\sum_{k=2j_0+2}^{2j+1}\left|\alpha_k^{j'}\right|^2 \le \limsup_{j'\to\infty}\sum_{k=2j_0+2}^{2j'+1}\left|\alpha_k^{j'}\right|^2 \le \epsilon \ ,
\end{equation*}
for any $j> j_0$, thus
\begin{equation}\label{tail-series}
    \sum_{k=2j_0+2}^\infty|\alpha_k|^2 \le \epsilon \ .
\end{equation}
This implies that $(s_n)_{n\in\mathbb N}\in \textswab R^+$, with $s_n = \sum_{k=1}^{n+1}|\alpha_k|^2$, is a Cauchy sequence, which means that $(\alpha_k)_{k\in\mathbb N}$ is a complex $\ell^2$-sequence, and thus
\begin{equation*}
    \phi = \sum_{k=1}^\infty \alpha_k \, \vb e_k \in \mathcal H \ .
\end{equation*}

We then need to show that $L\phi$ given by (\ref{phi-Phi})
is equivalent to $\widetilde \Phi$ given by (\ref{tilPhi}), with the relation between $\alpha_k^j$ and $\alpha_k$ given by (\ref{ak>a}). Again, from (\ref{cauchy-phi}), 
\begin{equation}\label{phij-phi-conv}
    \norm{\phi^j-\widetilde \phi^j}^2 = \sum_{k=1}^{2j+1}\left|\alpha_k-\alpha^j_k\right|^2 = \lim_{j'\to \infty} \sum_{k=1}^{2j+1}\left|\alpha_k^{j'}-\alpha_k^j\right|^2 \le \epsilon
\end{equation}
for $j\ge j_0$, therefore $L\phi \approx \tilde \Phi$. Thus, $\Lambda$ is a surjection, hence it is a bijection. 

In particular, we have $\Lambda\vb e_k = [(\hat{\vb e}_k^j)_{2j\in\mathbb N}]$, as should be. Also, (\ref{tail-series}) and (\ref{phij-phi-conv}) show that $\lim_{j\to\infty}\widetilde \phi^j = \phi$, so (\ref{Hidentif}) holds.

To finish, we show that the family of nesting maps $\{\iota^{j'}_j\}$ induces the sequence of isometries $\Gamma$. 
Fix $j_0$. For any $\phi^{j_0}\in \mathcal H^{j_0}$, take $\Phi = (\phi^j)$, where
\begin{equation*}
    \phi^j = \begin{cases}
    \iota^j_{j_0}(\phi^{j_0}) \ , \ \ j\ge j_0\\
    0 \ , \ \ j< j_0
    \end{cases} \ .
\end{equation*}
Then, $\norm{\phi^j-\phi^{j_0}} = 0$ for every $j\ge j_0$. So $\tilde \gamma_{j_0}: \mathcal H^{j_0}\to \textswab H_\infty^</\approx \ : \phi^{j_0}\mapsto [\Phi]$ is well defined. So we make $\gamma_{j_0} = \Lambda^{-1}\circ \tilde \gamma_{j_0}$. A straightforward calculation gives $\gamma_{j_0}(\vb e^{j_0}_k) = \vb e_k$ for any $\vb e^{j_0}_k\in \mathcal E^{j_0}$, and this proves that (\ref{isoj}) holds.
\end{proof}

\begin{corollary}\label{cor-conv}
For any $\Phi \in \textswab H^<$, we have that $\Phi \in \textswab H_\infty^<$ if and only if \ $\Gamma(\Phi)$ is convergent in $\mathcal H$. Also, for $\phi = \lim_{n\to\infty}\Gamma(\Phi)$, we have $[\Phi] \equiv \phi$ , {\it cf.} (\ref{phi-Phi})-(\ref{Lambdaphi}).
\end{corollary}

We now investigate the relations between operator sequences $\vb F$ and operators on the ground Hilbert space $\mathcal H$. First we have the following natural definitions.

\begin{definition}\label{cos}
Let $(\mathcal H,\mathcal E)$ be the ground Hilbert space for a well-nested $(\textswab H^<_{\infty}, \textswab E)$.

Given any $\vb F = (F_n)\in \textswab M$, for \  $\textswab M\times\textswab H^<\to\textswab H^<$ as in (\ref{seqaction}), the \emph{$\Gamma$-induced operator sequence} $(\mathcal F_n)_{n\in\mathbb N}$ is the sequence of operators $\mathcal F_n: \mathcal H\to\mathcal H$ defined by
\begin{equation}\label{Frescon}
    \mathcal F_n(\phi) : = \mathcal F_n\circ \gamma_j(\phi^j) = \gamma_j\circ F_n(\phi^j) \ , \ \ \mbox{for} \ \ \phi^j\in L\phi \ , \ \ n=2j  \ .
\end{equation}
\end{definition} 
\begin{definition}\label{op-norm}
For any $\mathcal F:\mathcal H\to\mathcal H$, its \emph{$\mathcal H$-operator-norm}  $||\mathcal F||_{op}$ is given by  
\begin{equation}
    ||\mathcal F||_{op}=\sup_{\phi\in\mathcal H\backslash\{0\}}\norm{\mathcal F(\phi)}/\norm{\phi} \ \in \  \mathbb R^+\cup \{\infty\} \ ,
\end{equation}
and $||\mathcal F||_{op}<\infty$ means that $\mathcal F$ is \emph{bounded}, or $\mathcal H$-\emph{bounded}. 
\end{definition}

\begin{proposition}\label{convFn}
 Let $\vb F=(F_n)_{n\in\mathbb N}\in\textswab{M}$. If its $\Gamma$-induced operator sequence $(\mathcal F_n)_{n\in\mathbb N}$ converges in $\mathcal H$-operator-norm, then \ $\vb F \in \textswab M_\infty$ .
\end{proposition}
\begin{proof}
If $\Phi = (\phi^j)\in \textswab H_\infty^<$, we have
\begin{equation*}
\begin{aligned}
    \norm{\mathcal F_{n'}\circ \gamma_{j'}(\phi^{j'}) - \mathcal F_n\circ \gamma_j(\phi^j)} & \le \norm{\mathcal F_{n'}\circ \gamma_{j'}(\phi^{j'})-\mathcal F_n\circ \gamma_{j'}(\phi^{j'})}\\
    & \ \ \ \ + \norm{\mathcal F_n\circ \gamma_{j'}(\phi^{j'})-\mathcal F_n \circ \gamma_j(\phi^j)}\\
    & \le \norm{\mathcal F_{n'}-\mathcal F_n}_{op}\norm{\phi^{j'}}\\
    & \ \ \ \ + \norm{\mathcal F_n}_{op}\norm{\gamma_{j'}(\phi^{j'})-\gamma_j(\phi^j)}
\end{aligned}
\end{equation*}
for $n'=2j'$ and $n=2j$. Let $\phi \in \mathcal H$ and $\mathcal F: \mathcal H \to \mathcal H$ be such that $\gamma_j(\phi^j)\to \phi$ and $\mathcal F_n \to \mathcal F$. There is $n_1=2j_1 \in \mathbb N$ for which 
$$n=2j> n_1\implies\norm{\phi^j}\le 2\norm{\phi} \ , \ \norm{\mathcal F_n}_{op} \le 2\norm{\mathcal F}_{op} \ .$$ 
In addition, given any $\epsilon >0$ there is $n_2=2j_2 \in \mathbb N$ for which 
\begin{equation*}
n'=2j',n=2j> n_2 \implies 
\norm{\mathcal F_{n'}-\mathcal F_n}_{op}\le \dfrac{\epsilon}{4\norm{\phi}} \ , \  \norm{\gamma_{j'}(\phi^{j'})-\gamma_j(\phi^j)}\le \dfrac{\epsilon}{4\norm{\mathcal F}_{op}} \ .    
\end{equation*}
Then, for $n_0 = 2j_0 = \max\{n_1, n_2\}$, we have
\begin{equation*}
n'=2j',n=2j> n_0\implies 
    \norm{\mathcal F_{n'}\circ \gamma_{j'}(\phi^{j'}) - \mathcal F_n\circ \gamma_j(\phi^j)} \le \dfrac{\epsilon}{2}+\dfrac{\epsilon}{2} = \epsilon \ .
\end{equation*}
Thus, $(\mathcal F_n\circ \gamma_j(\phi^j))$ is a Cauchy sequence and $(F_n(\phi^j))\in \textswab H_\infty^<$ , by Corollary \ref{cor-conv}.
\end{proof}

\begin{proposition}\label{op-norm-seq-b}
If \  $\vb F \in \textswab M_\infty$ , then \  $\vb F$ is $\textswab H$-bounded.
\end{proposition}
\begin{proof}
Suppose that ${\vb F}$ is $\textswab H$-unbounded and assume without loss of generality that $\norm{F_n}_{op}\neq 0$ for all $n\in \mathbb N$ and $\norm{F_n}_{op}\to \infty$. Let $(\phi^j) \in \textswab H^<$ be such that $\norm{\phi^j} = 1$ and $\norm{F_n}_{op} = \norm{F_n(\phi^j)}$. Now,
\begin{equation*}
    \psi^j = \dfrac{1}{\sqrt{\norm{F_n}_{op}}}\phi^j
\end{equation*}
defines a sequence $\Psi = (\psi^j)\in \textswab H_\infty^<$ such that $[\Psi] \equiv 0$, but 
\begin{equation*}
    \norm{F_n(\psi^j)} = \dfrac{\norm{F_n(\phi^j)}}{\sqrt{\norm{F_n}_{op}}} = \sqrt{\norm{F_n}_{op}} \to \infty \ ,
\end{equation*}
in contradiction with the fact that $\vb F \in \textswab M_\infty$.
\end{proof}

\begin{proposition}\label{e=e}
If $\Phi, \Psi \in \textswab H_\infty^<$ and $\vb F \in \textswab M_\infty$, then $\vb F(\Phi)\approx \vb F(\Psi)$ if $\Phi\approx \Psi$.
\end{proposition}
\begin{proof}
It is sufficient to show that if $\Phi \in \textswab H_\infty^<$ satisfies $[\Phi] \equiv 0$ and $\vb F \in \textswab M_\infty$, then $\vb F(\Phi)\approx 0$. Let $\Phi = (\phi^j)$ and $\vb F = (F_n)$. 
Since $\norm{F_n(\phi^j)} \le \norm{F_n}_{op}\norm{\phi^j}$ and $\norm{\phi^j}\to 0$, by Proprosition \ref{op-norm-seq-b}, we get $\norm{F_n(\phi^j)}\to 0$.
\end{proof}

\begin{definition}\label{Fresco}
Two operator sequences $\vb F, \vb F' \in \textswab M_\infty$ are \emph{equivalent}, denoted $\vb F\approx \vb F'$, if \ $\vb F(\Phi) \approx \vb F'(\Phi)$ for every $\Phi \in \textswab H_\infty^<$. Let us denote the space of equivalent classes of convergent operator sequences by \  $\mathcal B = \textswab M_\infty/\approx$ .

Then, given any $[\vb F]\in\mathcal B$, the \emph{induced operator} $\mathcal F:\mathcal H\to\mathcal H$ is defined by
\begin{equation}\label{eqFresco}
   \phi\mapsto \mathcal F(\phi) \equiv [\vb F'(\Phi)] \ , \ \mbox{for} \ \ \phi \equiv [\Phi] \ , \  \Phi \in \textswab H_\infty^< \ ,
\end{equation}
for any $\vb F' \in [\vb F]\in\mathcal B$. 
\end{definition}

The following shows that the operator $\mathcal F:\mathcal H\to\mathcal H$ is well defined, according to Definitions \ref{ghs}, \ref{cos}, \ref{Fresco}, Theorem  \ref{H=H}  and Propositions \ref{convFn}, \ref{op-norm-seq-b}, \ref{e=e}.

\begin{theorem}\label{point-lim-op}
For any $[\vb F]\in \mathcal B$, the induced operator 
$\mathcal F$ on $\mathcal H$, 
given by (\ref{eqFresco}), is bounded and coincides,  for any $\vb F' \in [\vb F]$, with the pointwise limit of the induced sequence of operators $(\mathcal F_n')$ on $\mathcal H$, that is, $(\mathcal F_n'(\phi))\to\mathcal F(\phi)$ for every $\phi \in \mathcal H$, with $\mathcal F_n'$ given by (\ref{Frescon}).
\end{theorem}

\begin{proof}
Given $\phi \in \mathcal H$, let $L\phi=(\phi^j) = \Phi$ be as in (\ref{phi-Phi}), then 
\begin{equation*}
  \phi = \sum_{k=1}^{\infty} \alpha_k \, \vb e_k \implies   \gamma_j(\phi^j) = \sum_{k=1}^{n+1}\alpha_k \, \vb e_k \ ,
\end{equation*}
so $\phi \equiv [\Phi]$ and $\mathcal F_n'(\phi) = \mathcal F_n'(\gamma_j(\phi^j))$. We also have $\mathcal F(\phi)\equiv [\vb F'(\Phi)]$, then
\begin{equation*}
    \mathcal F(\phi) = \lim_{n\to\infty} \gamma_j \circ F'_n(\phi^j) = \lim_{n\to\infty} \mathcal F'_n(\gamma_j(\phi^j)) = \lim_{n\to\infty} \mathcal F_n'(\phi) \ ,
\end{equation*}
{\it cf.} (\ref{er1})-(\ref{Hidentif}) and (\ref{Frescon}). 
That is, $(\mathcal F'_n)$ is a pointwise convergent sequence to $\mathcal F$.

Now, for any $\vb F'\in[\vb F]\in\mathcal B$,  since $\mathcal F$ is the pointwise limit of the operator sequence $(\mathcal F_n')$ induced by $\vb F'$ then, because each $\mathcal F_n'$ is bounded and $\vb F'$ is $\textswab H$-bounded, we get that $\mathcal F$ is also $\mathcal H$-bounded and $\mathcal F(\phi)\in\mathcal H$.  \end{proof}

Thus, from Definition \ref{Fresco} and Theorem \ref{point-lim-op}, we have the identification 
\begin{equation}\label{groundaction2}
    \vb F:\textswab{H}^<_{\infty}\to\textswab{H}^<_{\infty} \  \iff \  [\vb F]=\mathcal F:\mathcal H\to\mathcal H \ ,
\end{equation}
{\it cf.} (\ref{groundaction}). And in view of 
Theorem \ref{point-lim-op} and Proposition \ref{convFn}, we can single out: 

\begin{definition}
We say that $\vb F=(F_n)_{n\in\mathbb N}\in \textswab{M}_{\infty}$ is a \emph{strongly convergent operator sequence} if its $\Gamma$-induced operator sequence $(\mathcal F_n)_{n\in\mathbb N}$ converges in $\mathcal H$-operator-norm. We denote by \  $\hat{\textswab{M}}_{\infty}$ the set of all strongly convergent operator sequences.
\end{definition}

\begin{proposition}
$\hat{\textswab{M}}_{\infty}$ is a proper subset of \  ${\textswab{M}}_{\infty}$. 
\end{proposition}
\begin{proof}
Take, for instance, the operator sequence $\vb F=(F_n)_{n\in\mathbb N}$ determined by 
\begin{equation*}
    F_n(\vb e_k^j)=\sqrt{\frac{k}{n+1}}\vb e_k^j \ , \ \forall \vb e_k^j \in \mathcal E^j \ , \ \forall n=2j\in\mathbb N \ , 
\end{equation*}
where $\textswab{E}=(\mathcal E^j)_{2j\in\mathbb N}$ is a well-nested basis for $\textswab{H}^<$. 
It is straightforward to see that $\vb F\in {\textswab{M}}_{\infty}$ but $\vb F\notin \hat{\textswab{M}}_{\infty}$.
\end{proof}

We end this subsection with the important results stated in the next propositions. 

\begin{proposition}
If $\vb F = (F_n)\in \textswab M_\infty$, then $\vb F^\ast = (F_n^\ast)\in \textswab M_\infty$ and the operator induced by $[\vb F^\ast]$ is the Hermitian conjugate of the operator induced by $[\vb F]$.
\end{proposition}
\begin{proof}
Let $(\mathcal F_n)$ and $(\mathcal F_n^\ast)$ be the $\Gamma$-induced operator sequences of $\vb F$ and $\vb F^\ast$, respectively. Also, let $\mathcal F$ be the  operator induced by $[\vb F]$. Given $\Phi = (\phi^j)\in \textswab H_\infty^<$ and $\psi \in \mathcal H$, let $\phi = \lim_{n\to\infty}\gamma_j(\phi^j)$. Then
\begin{equation}
    \ip{\mathcal F_n^\ast\circ \gamma_j(\phi^j)}{\psi} = \ip{\gamma_j(\phi^j)}{\mathcal F_n (\psi)} \to \ip{\phi}{\mathcal F(\psi)}\ .
\end{equation}
Since $\psi$ is any element of $\mathcal H$, we conclude that $(\mathcal F_n^\ast\circ \gamma_j(\phi^j))$ converges to $\mathcal F^\ast (\phi)$.
\end{proof}

\begin{definition}
For any operator $\mathcal F:\mathcal H\to\mathcal H$, its \emph{trace}, if it exists, is
\begin{equation}
    \tr(\mathcal F) = \sum_{k=1}^{\infty}\langle \mathbf e_k | \mathcal F(\mathbf e_k)\rangle \ , 
\end{equation}
where $\mathcal E=\{\mathbf e_k\}_{k\in\mathbb N}$ is a countable orthonormal basis for $\mathcal H$. And $\mathcal F$ is \emph{trace-class} if \  $\exists \  \tr(|\mathcal F|)\in\mathbb R^+$, for $|\mathcal F| = \sqrt{\mathcal F^\ast \mathcal F}$. 
\end{definition}

We recall that the set of trace-class operators forms an ideal in the algebra of operators on an  infinite-dimensional Hilbert space. Then, we have:

\begin{proposition}
For $[\vb F]\subset \textswab M_\infty$, if $\vb F'$ is trace-class, for any $\vb F'\in [\vb F]$, then the induced operator $\mathcal F: \mathcal H\to \mathcal H$ is trace-class and
\begin{equation}
    \tr(|\mathcal F|) \le \tr_\infty(|\vb F|) \ .
\end{equation}
\end{proposition}
\begin{proof}
As before, for $\vb F'\in [\vb F]$, take the induced operator sequence $(\mathcal F'_n)$. Then, 
\begin{equation}
\begin{aligned}
    \tr(|\mathcal F|) & = \sum_{k=1}^{\infty} \ip{\vb e_k}{|\mathcal F|(\vb e_k)} 
    = \lim_{n\to\infty}\sum_{k=1}^{n+1}\ip{\vb e_k}{|\mathcal F|(\vb e_k)} \\
    & = \lim_{n\to\infty} \lim_{n'\to\infty}\sum_{k=1}^{n+1}\ip{\vb e_k}{|\mathcal F'_{n'}|(\vb e_k)}\\
    & \le  \lim_{n'\to \infty}\tr(|\mathcal F'_{n'}|)
     = \lim_{n'\to\infty}\tr(|F'_{n'}|) \ .
\end{aligned}
\end{equation}
So if $\vb F'$ is trace-class, then $\tr(|\mathcal F|) \le \tr_\infty(|\vb F'|)$.
\end{proof}

The converse of the last proposition is not true. 
Take, for example, 
\begin{equation}\label{cex}
  \vb F = (F_n) \ , \   F_n = \dfrac{I_n}{\sqrt{n+1}}  = |F_n| \ .
\end{equation}
Trivially, $\norm{F_n}_{op} = 1/\sqrt{n+1}$, thus we have $(F_n) \in \textswab M_\infty$ by Proposition \ref{convFn}. Let $\mathcal F$ be the operator induced by $[\vb F]$. For any $\phi \in \mathcal H$, from Proposition \ref{point-lim-op}, we have
\begin{equation*}
    \norm{\mathcal F(\phi)} = \lim_{n\to\infty} \norm{\mathcal F_n(\phi)} \le \lim_{n\to\infty} \norm{\mathcal F_n}_{op}\norm{\phi} = 0 \ .
\end{equation*}
Thus, $\mathcal F$ is identically null. So $|\mathcal F|$ is also identically null and $\tr(|\mathcal F|) = 0$. However, 
from (\ref{cex}), $\tr(|F_n|) = \sqrt{n+1}$, thus  $\tr_\infty(|\vb F|) = \tr_\infty(\vb F) = \infty$.

\subsection{Concrete examples of asymptotic spin quantization}\label{examples-section} 

We now provide two concrete examples that illustrate some of the previous definitions and results.

\begin{example}\label{example-cp1} 
$S^2$ is a K\"ahler manifold, $S^2\simeq\mathbb CP^1$. Then,   
recalling Definition \ref{bar1} and Proposition \ref{bar2}, 
	for each $j$ we can pass from $\mathcal H_{om2}^j$ to $Poly(\mathbb C)_{\leq n}^h$ , the space of holomorphic polynomials of degree $\leq n$ in one complex variable, via identification 
	\begin{equation}\label{identf}
	 id^j_1: \  \mathcal H_{om2}^j\ni u_1^{j-m}u_2^{j+m}\longleftrightarrow \nu_j z^{j-m}\in Poly(\mathbb C)_{\leq n}^h \ , 
	\end{equation}
	where $\nu_j\in\mathbb C$ is a constant, for each $j$, and $z\in\mathbb C$ is a holomorphic coordinate of $\mathbb CP^1$, the latter seen in this coordinate system as the flattened plane, with $SU(2)$ now acting on $z\simeq u_1/u_2\in\mathbb C$ via M\"obius transformations.  
	
	Thus, for each $\phi^j\in Poly(\mathbb C)_{\leq n}^h$ there is a unique ``lift'' $\widetilde{\phi}^j\in \mathcal H_{om2}^j$ satisfying 
	$$\phi^j(z)=\widetilde{\phi}^j(u_1,u_2) \ , 
	$$
	via the identification $id^j_1$ given by (\ref{identf}). From this, it follows that, for each $j$, we can define a $SU(2)$-invariant inner product $\langle\cdot|\cdot\rangle_{Poly(\mathbb C)_{\leq n}^h}^j$ via (\ref{identf}) by 
	\begin{equation}\label{invinncp1}
	   \langle\phi^j|\psi^j\rangle_{Poly(\mathbb C)_{\leq n}^h}^j = \langle\widetilde{\phi^j} |\widetilde{\psi^j}  \rangle_{\mathcal H_{om2}^j} \ . 
	\end{equation}
	
	However, just as the identification (\ref{identf}) is $j$-dependent, so is the inner product on the l.h.s. of (\ref{invinncp1}), therefore this product cannot be written simply as the restriction to $Poly(\mathbb C)_{\leq n}^h$ of a $j$-invariant inner product on the space $\mathcal H_{ol1}$ of holomorphic functions on $\mathbb C$. The inner product (\ref{invinncp1}) can also be written in integral form,  
	\begin{equation}\label{invinnj}
  \langle \phi^j|\psi^j\rangle^j_{Poly(\mathbb C)_{\leq n}^h}= \ \int_{\mathbb C}\overline{\phi^j(z)}\psi^j(z)d\mu^j(z) \  ,   
\end{equation}
but, in contrast with (\ref{innrest}), now $d\mu^j(z)$	depends explicitly on $j$. 

Anyway, from (\ref{invinncp1}), the set 
\begin{equation}\label{deltaz}
\ \left\{\delta^m_j(z)\equiv 
\frac{\nu_j z^{j-m}}{\sqrt{(j-m)!(j+m)!}}\right\}_{-j\leq m\leq j} \ , 
\end{equation}
forms an orthonormal basis for  $$id^j_1(\mathcal H_{om2}^j, \langle\cdot|\cdot\rangle)=:({\mathcal H}^j_z,\langle\cdot|\cdot\rangle_j):=(Poly(\mathbb C)_{\leq n}^h,\langle \cdot|\cdot\rangle^j_{Poly(\mathbb C)_{\leq n}^h}) \ .  $$ 

In principle, for each $j$ the choice of ($j$-dependent) constant $\nu_j\in\mathbb C$ is arbitrary. But there is a canonical way to choose $\nu_j$ for every $j$, by looking at the function $1$. 
Setting $m=j$ in (\ref{deltaz}) we obtain
\begin{equation}\label{cc}
    1\equiv\frac{\nu_j}{\sqrt{(2j)!}} \implies \nu_j=\sqrt{(2j)!} \ \ \mbox{as a canonical choice} \ .
\end{equation}
With the canonical choice (\ref{cc}), identifying \ $p=j-m$, we set 
\begin{equation}\label{u=z}
\begin{aligned}
\mathcal E^j & = \left\{\boldsymbol{u}(j, m)\equiv 
\frac{z^{j-m}\sqrt{(2j)!}}{\sqrt{(j-m)!(j+m)!}}\right\}_{-j\leq m\leq j}\\
& \!\Leftrightarrow \ \ \left\{\boldsymbol{u}(j, m)\equiv \sqrt{\binom{n}{p}}z^{p}=:\boldsymbol{u}_p^n\right\}_{0\leq p\leq n}
\end{aligned}
\end{equation}
as the standard basis  for $(\mathcal H^j_z,\langle\cdot|\cdot\rangle_j)$ satisfying (\ref{stansu2act})-(\ref{ab}), so that 
\begin{equation}\label{coordprod}
   \langle \phi^j|\psi^j\rangle^j_{Poly(\mathbb C)_{\leq n}^h}\equiv \  \langle \phi^j|\psi^j\rangle_j \ = \ \sum_{p=0}^n\overline{\alpha}_p^j\beta_p^{j} \ , 
\end{equation}
\begin{equation}\label{holdec}
    \phi^j(z)= \sum_{p=0}^n \alpha_p^j\sqrt{\binom{n}{p}}z^p \ \ , \  \ \psi^j(z)= \sum_{p=0}^n \beta_p^j\sqrt{\binom{n}{p}}z^p \ ,
\end{equation}
and we have the identification 
\begin{equation}\label{J3ninvj}
    J_3^j \ \longleftrightarrow \ j-z\frac{\partial}{\partial z} : \mathcal H^j_z\to\mathcal H^j_z \ , \ \forall n=2j\in\mathbb N \ .
\end{equation}

We now note that the sequence of Hilbert spaces $\textswab{H}^<=(\mathcal H^j_z,\langle\cdot|\cdot\rangle_j)_{2j\in\mathbb N}$ is a nested sequence, with nesting maps $\iota_j^{j'}:\mathcal H^j_z\to\mathcal H^{j'}_z$ determined for $j\leq j'$ by  
\begin{equation}\label{iotaz}
    \iota_j^{j'}(\boldsymbol{u}_p^n)= \boldsymbol{u}_p^{n'} , 
\end{equation}
and the canonical choice (\ref{cc}) defines a well-nested basis sequence \  $\textswab{E}=(\mathcal E^j)_{2j\in\mathbb N}$, with $\mathcal E^j$ given by (\ref{u=z}).

Thus, from Theorem \ref{H=H}, if $(\phi^j)_{2j\in\mathbb N}\in \textswab H_\infty^<$ is given as in (\ref{holdec}), then $(\alpha_p)_{p \in \mathbb N_0}$, with $\alpha_p = \lim_{j\to\infty}\alpha^j_p\in \mathbb C$, is a $\ell^2$-sequence and $\phi = \lim_{j\to\infty}\phi^j \in \mathcal H_z$, where $\mathcal H_z$ is the ground Hilbert space of $\textswab{H}^<_{\infty}$. In this case, from Theorem  \ref{point-lim-op}, an operator sequence $\vb T\in \textswab{M}_{\infty}$ takes the $\ell^2$-sequence $(\alpha_p)_{p\in\mathbb N_0}$ to an $\ell^2$-sequence $(\beta_p)_{p\in\mathbb N_0}$, with $\beta_p=\lim_{j\to\infty}\beta_p^j\in \mathbb C$, for $\beta_i^j$ as in (\ref{holdec}), and defines an operator $\mathcal T:\mathcal H_z\to\mathcal H_z$.

But the inner product on each $\mathcal H^j_z$ can also be written in the integral form (\ref{invinnj}) and  one obtains straightforwardly that the measure/integrator is given by  
\begin{equation}\label{dmuj} 
d\mu^j(z)= \frac{n+1}{2\pi i}\frac{d\overline{z}\wedge d{z}}{(1+z\overline{z})^{n+2}} \ ,
\end{equation} 
which yields the set (\ref{u=z}) as an orthonormal basis for ${\mathcal H}^j_z$ with respect to the inner product $\langle \cdot|\cdot\rangle_j$ given by  (\ref{invinnj}) and (\ref{coordprod})-(\ref{dmuj}), {\it cf.} also \cite{ber}.  

Thus, from the above and (\ref{phi-Phi})-(\ref{Lambdaphi}), we can also characterize the ground Hilbert space $\mathcal H_z$ as the subspace of  formal power series in $z$, $\mathbb C[[z]]$ , satisfying\footnote{It is necessary to restrict to $\psi^j\in Poly(\mathbb C)_{\leq n}^h\subset\mathbb C[[z]]$ in (\ref{Hint}), with $(\psi^j)_{2j\in\mathbb N} = L\psi$ , because the inner product defined by  (\ref{invinnj}) and (\ref{dmuj}) is ill-defined or divergent on $\mathbb C[[z]]\setminus Poly(\mathbb C)_{\leq n}^h$.}
\begin{equation}\label{Hint}
    \ \psi\in\mathcal H_z\subset\mathbb C[[z]]\iff\exists\lim_{j\to\infty}\int_{\mathbb C}|\psi^j(z)|^2d\mu^j(z) \ , \ (\psi^j)_{2j\in\mathbb N}= L\psi \ , \ cf. \  (\ref{phi-Phi}).
\end{equation}

Then, repeating the steps in Example \ref{bargmann}, using (\ref{invinnj})-(\ref{dmuj}) we can write any operator $T_{\alpha}^j: {\mathcal H}^j_z\to{\mathcal H}^j_z$, $\psi_j\mapsto 	\phi_j$  as determined by an integral kernel \ $\mathcal K_{\alpha}^j$ via 
	 	\begin{equation}\label{kjz}
	 	\phi_j(z_1) =\int_{\mathbb C}\mathcal K_{\alpha}^j(z_1,z_2)\psi_j(z_2)d\mu^j(z_2) \ , 
	 	\end{equation} 	
with $d\mu^j(z)$ given by (\ref{dmuj}), which thus has the form, {\it cf.} (\ref{dualbasis}) and (\ref{MAjz}), 
 \begin{equation*}
	 	\mathcal K_{\alpha}^j(z_1,z_2) \ =\!\sum_{p_1,p_2=0}^n (-1)^{p_2} M(\alpha)^j_{j-p_1,p_2-j}\sqrt{\binom{n}{p_1}}\sqrt{\binom{n}{p_2}} z_1^{p_1}\overline{z}_2^{p_2} \ , 
	 	\end{equation*} 
so that, for  $f\in C^{\infty}_{\mathbb C}(S^2)$, with  $\boldsymbol{W}_{\!\mathcal{C}}$ of (anti-)Poisson type, the ${W}$-quantization of $f$, ${\mathbf{F}}^{w}$, determines a sequence of integral operators $\mathbf T_w[f]=(T_w^j[f])_{2j=n\in\mathbb N}$ acting on the well-nested sequence of Hilbert spaces \ $\textswab{H}^< = ({\mathcal H}^j_z, \langle\cdot|\cdot\rangle_j)_{2j=n\in\mathbb N}$ via (\ref{kjz}) by the sequence of integral kernels $\mathbf{K}_w[f]=(\mathcal K_{w}^j[f])_{n\in\mathbb N}$, where
\begin{equation}\label{Kfwz2}
	    {\mathcal K}_{w}^j[f](z_1,z_2)= \sqrt{n+1} \sum_{p_1,p_2, l=0}^n  (-1)^{p_2}\sqrt{\binom{n}{p_1}}\sqrt{\binom{n}{p_2}} C_{j-p_1,p_2-j,m}^{\,\,\,j,\,\,\,\,\,\,\,\,\,\,\,j,\,\,\,\,\,l}\frac{\langle Y_l^m|f\rangle}{c_l^n} z_1^{p_1}\overline{z}_2^{p_2}
	 \end{equation}
{\it cf.} (\ref{Ee}), (\ref{hf})-(\ref{dFWnlm}), (\ref{cg}), with $\langle \cdot|\cdot\rangle$ given by (\ref{innS2}).
And similarly for $(\widetilde{\mathcal K}_{w}^j[f])_{2j\in\mathbb N}$ from $\widetilde{\mathbf{F}}^{w}$, the $\widetilde{W}$-quantization of $f$, replacing $c_l^n\leftrightarrow 1/c_l^n$ in (\ref{Kfwz2}).

Therefore,  for $\psi\in\mathcal H_z$ as in (\ref{Hint}), 
setting \ $(\psi^j)_{2j\in\mathbb N}=L\psi$ in (\ref{kjz}), 
{\it cf.} (\ref{phi-Phi}),  
 if $\boldsymbol{W}_{\!\mathcal{C}}$ is such that $\mathbf T_w[f]=(T_w^j[f])_{2j\in\mathbb N}\in \textswab{M}_{\infty}$, then $\mathcal T_w[f]:\psi\mapsto \phi\in\mathcal H_z$, where 
\begin{equation}\label{HtoH}
    \phi(z_1)=\lim_{j\to\infty}\int_{\mathbb C}{\mathcal K}_{w}^j[f](z_1,z_2)\psi^j(z_2)d\mu^j(z_2) \ , 
\end{equation}
and similarly for $\widetilde{\mathcal T}_w[f]$ with $\widetilde{\mathcal K}_{w}^j[f]$ in (\ref{HtoH}).  
\end{example}

Na\"{\i}vely, one could have thought of defining integral operators on the space of $L^2$-holomorphic functions on $\mathbb C$,   with respect to the inner product on $\mathcal H_{ol1}$ defined by analogy with the inner product on $\mathcal H_{ol2}$, as 
\begin{equation}\label{innpHol1}
    \langle \phi|\psi\rangle_{\mathcal H^{2\mu}_{ol1}}=\int_{\mathbb C}\overline{\phi(z)}\psi(z)d\mu(z)  \ , \ \ d\mu(z)=e^{-z\overline{z}}\frac{d\overline{z}\wedge dz}{2\pi i} \ . 
\end{equation}
However, this inner product is not $SU(2)$-invariant, for the $SU(2)$-action on $\mathcal H^{2\mu}_{ol1}$ induced from the $SU(2)$-M\"obius-action on $z\in\mathbb C$. 
This can also be seen from the fact that an orthonormal basis for $\mathcal H^{2\mu}_{ol1}$ w.r.t. the above inner product is given by 
\begin{equation}\label{e=z}
    \{\mathbf{v}_p={z^p}/{\sqrt{p!}}\}_{p\in\mathbb N_0} \ . 
\end{equation}
Therefore, for any $2j=n>1$, the set $\{\mathbf{e}_p={z^p}/{\sqrt{p!}}\}_{0\leq p\leq n}$ does not constitute a standard basis for the spin-$j$ system.
In fact, from (\ref{u=z}) and (\ref{e=z}), we have that 
\begin{equation*}\label{changebasis}
    \boldsymbol{u}_p^n=\sqrt{\frac{n!}{(n-p)!}}\mathbf{v}_p
\end{equation*}
so that, from (\ref{stansu2act})-(\ref{ab}), for $J_3^j=d\varphi_j(\sigma_3/2)$,  $J_{\pm}^j=d\varphi_j(\sigma_{\pm}/2)$, where $\varphi_j:SU(2)\to U(n+1)$ is the irreducible representation on $\widetilde{\mathcal H}^j\subset \mathcal H_{ol1}$, 
\begin{equation*}
  \ \ \  J_3^j(\boldsymbol{u}_p^n)= (j-p)\boldsymbol{u}_p^n \iff    J_3^j(\mathbf{v}_p)= (j-p)\mathbf{v}_p
\end{equation*}
but, {\it cf.} (\ref{stansu2act})-(\ref{ab}), with $p=j-m$, 
\begin{equation}
\begin{aligned}\label{notinvbasis}
J_+^j(\boldsymbol{u}_p^n)=\sqrt{p(n-p+1)}\boldsymbol{u}_{p-1}^n \ &\implies  \ J_+^j(\mathbf{v}_p)=\sqrt{p}\mathbf{v}_{p-1} \ , \\
J_-^j(\boldsymbol{u}_p^n)=\sqrt{(n-p)(p+1)}\boldsymbol{u}_{p+1}^n \ &\implies  \  J_-^j(\mathbf{v}_p)=(n-p)\sqrt{(p+1)}\mathbf{v}_{p+1} \ .
\end{aligned}
    \end{equation}
Thus, while both are basis formed by eigenvectors of $J_3^j$, for any $j$, we see from (\ref{notinvbasis}) that the basis   $\{\mathbf{v}_p={z^p}/{\sqrt{p!}}\}_{0\leq p\leq n}$ is not $SU(2)$-equivariant, for any $2j=n>1$.

\begin{example}\label{S1cS2}
  It is common to think of the quantization of a function $f$ on a real symplectic manifold $\mathcal M$ as defining a differential or an integral operator on a Hilbert space of functions on a real Lagrangian submanifold $\mathcal L\subset\mathcal M$. For spin systems, any simple closed curve on $S^2$ is a Lagrangian submanifold, but, by symmetry considerations, it is natural to focus on the closed geodesics. 
  
  Thus, let $\mathcal X\subset S^2$ be a closed geodesic, which we can take to be the equator of the coordinate system $(\varphi,\theta)$ on $S^2$, so that $\theta$ parametrizes $\mathcal X$. We now construct three well-nested sequences of Hilbert spaces adapted to $\mathcal X\subset S^2$, as follows. 
  
  In the first construction, we trivially   ``repeat'' Example \ref{example-cp1}, looking at restrictions of functions on $\mathbb CP^1$ to $\mathcal X\simeq S^1=\{z=e^{i\theta}\}\subset\mathbb CP^1\simeq S^2$ via identification 
  \begin{equation}
     \widetilde{id}_2^j: \mathcal H^j_z\equiv Poly(\mathbb C)_{\leq n}^h\ni \nu_j z^{j-m} \ \longleftrightarrow \  \nu_je^{i(j-m)\theta} \in {\mathcal H}^j_{e^{i\theta}} \ ,  
\end{equation}
so that all expressions are imported directly from Example \ref{example-cp1} and then everything gets well defined on $\mathcal X\subset S^2$, but we don't get anything new and we loose the integral expressions of Example \ref{example-cp1}. This is $SU(2)$-invariant, but uninteresting. 

\ 

The second construction is quite more interesting, but now we shall restrict to the subsequence of $SU(2)$ representations with integer $j$'s, in other words, this second construction is restricted to the sequence of $SO(3)$ representations.

  Then, for each $j\in\mathbb N$,
  let ${\mathcal H}^j$ be the $(2j+1)$-dimensional complex vector space spanned by $\{e^{im\theta}\}_{-j\leq m\leq j}$. To be more explicit, we shall denote it by ${\mathcal H}^j_{\theta}$.  In order to turn $({\mathcal H}^j_{\theta})_{j\in\mathbb N}$ into a well-nested sequence of Hilbert spaces, recalling identification  (\ref{identf}) of Example \ref{example-cp1}, we start by identifying, 
  for each $j\in\mathbb N$,   
 \begin{equation}\label{id2}
     id_2^j: \mathcal H^j_z\equiv Poly(\mathbb C)_{\leq n}^h\ni \nu_j z^{j-m} \ \longleftrightarrow \  \rho_je^{im\theta} \in {\mathcal H}^j_{\theta} \ , 
 \end{equation}
 where $\rho_j\in\mathbb C$ is a constant for each $j$. 
 Then, from (\ref{deltaz}), a $SO(3)$-invariant inner product $\langle\cdot|\cdot\rangle_j$ on $\mathcal H^j_{\theta}$ is determined by setting 
 \begin{equation}\label{gammaj}
     \left\{\omega_j^m=\frac{\rho_je^{im\theta}}{\sqrt{(j-m)!(j+m)!}}\right\}_{-j\leq m\leq j}
 \end{equation}
 as an orthonormal basis for $\mathcal H^j_{\theta}$, so  that \  $\langle \omega_j^m|\omega_j^{m'}\rangle_j=\delta_{m,m'}$.

 In order to complete the construction of a well-nested basis sequence $\textswab{E}$ for the nested sequence $\textswab{H}^<= (\mathcal H^j_{\theta}, \langle\cdot|\cdot\rangle_j)_{2j\in\mathbb N}$, we need to define the nesting maps and specify a canonical choice for all $\rho_j$'s. For the choice of $\rho_j$'s, we can proceed as in Example \ref{example-cp1} and look at the constant function $1$ on $\mathcal X\subset S^2$, 
 \begin{equation}\label{j!}
 (j\in\mathbb N, \ m=0): \ 1\equiv \frac{\rho_j}{j!}\implies \rho_j=j! \ \ , \ \mbox{as a canonical choice} \ . 
 \end{equation}
 Thus, with the canonical choice (\ref{j!}) for $\rho_j$ in (\ref{gammaj}), 
 \begin{equation}\label{wnEX}
 \mathcal E^j =  \left\{\boldsymbol{u}(j,m)=\frac{j!e^{im\theta}}{\sqrt{(j-m)!(j+m)!}}\right\}_{-j\leq m\leq j}  
 \end{equation}
 is a standard basis for $\mathcal H^j_{\theta}$ satisfying (\ref{stansu2act})-(\ref{ab}).

 The nesting maps ${\iota}_j^{j'}: \mathcal H^j_{\theta}\to\mathcal H^{j'}_{\theta}$ must be consistent with $id_2^j$ given by (\ref{id2}) and nesting maps $\check{\iota}_j^{j'}: \mathcal H^j_{z}\to\mathcal H^{j'}_{z}$, that is, $\forall j\leq j'$ they must satisfy  \ ${\iota}_j^{j'}\circ  id_2^j \ = \ id_2^{j'}\circ \check{\iota}_j^{j'}$. The first tentative is to fix $\check{\iota}_j^{j'}$ as determined by (\ref{iotaz}). With this choice, the nesting maps $\widetilde{\iota}_j^{j'}:\mathcal H^j_{\theta}\to\mathcal H^{j'}_{\theta}$ satisfy \ $\widetilde{\iota}_j^{j'}(\boldsymbol{u}(j,m))=\boldsymbol{u}(j,m+j'-j)$. However, these nesting maps are not well suited to the choice (\ref{j!}) for $\rho_j$.
 Because, in Example \ref{example-cp1}, for every $j$ the constant function $1$ was associated to the highest-weight vector, and the nesting maps (\ref{iotaz})  took highest weight to highest weight. Now, the constant function $1$ is associated to the middle-weight vector, for every $j$, but the nesting maps $\widetilde{\iota}_j^{j'}$ induced from (\ref{iotaz}) do not take middle-weight to middle-weight. 
 
 So, we define new nesting maps $\iota_j^{j'} : {\mathcal H}^j_{\theta}\to{\mathcal H}^{j'}_{\theta}$, determined for $j\leq j'$ by  
\begin{equation}\label{iotat}
    \iota_j^{j'}(\boldsymbol{u}(j,m))= \boldsymbol{u}(j',m) \ . 
\end{equation}
Since both $j$ and $j'$ are integers, (\ref{iotat}) is well defined, but it induces new nesting maps $\widetilde{\iota}_j^{j'} : {\mathcal H}^j_{z}\to{\mathcal H}^{j'}_{z}$ on the subsequence $(\mathcal H^j_z)_{j\in\mathbb N}$  via \ 
${\iota}_j^{j'}\circ  id_2^j \ = \ id_2^{j'}\circ \widetilde{\iota}_j^{j'}$.

Thus, with the identification (\ref{id2}), the nesting defined by (\ref{iotat}) and the canonical choice (\ref{j!}), 
 our well-nested basis sequence for $\textswab{H}^<= (\mathcal H^j_{\theta}, \langle\cdot|\cdot\rangle_j)_{2j\in\mathbb N}$ is given by $\textswab{E}=(\mathcal E^j)_{2j\in\mathbb N}$, 
 where each $\mathcal E^j$ is given by (\ref{wnEX}).

 In this way, a sequence $\Phi=(\phi^j)_{j\in\mathbb N}\in \textswab{H}^<$ can be identified with a sequence of $SO(3)$-equivariant Fourier polynomials of degree $j$ on
 $\mathcal X\subset S^2$, with
	 	\begin{equation}\label{psij2}
	 	 \phi^j(\theta)\ = \  j!\!\sum_{m=-j}^{j}\frac{\alpha_m^je^{im\theta}}{\sqrt{(j-m)!(j+m)!}} \ = \
	 	  \phi^j(\theta + 2\pi) \ \in \ \mathcal H^j_{\theta} \ . 
	 	\end{equation}
Accordingly, we shall call $\alpha^j_m$ in (\ref{psij2}) the modified Fourier coefficients of $\phi^j$, so that, if 	 	
	 	$\alpha^j_m, \beta^j_m$ are modified Fourier coefficients of $\phi^j, \psi^j\in\mathcal H^j_{\theta}$ as in (\ref{psij2}), then 	
\begin{equation}\label{prodjX}
    \langle\phi^j|\psi^j\rangle_j=\sum_{m=-j}^j\overline{\alpha}_m^j\beta_m^j \ .
\end{equation}		
Furthermore, it follows from (\ref{wnEX}) and (\ref{psij2}) that the operator $J_3$ can be identified in this representation  as 
\begin{equation}\label{J3invj}
    J_3 \longleftrightarrow -i\frac{\partial}{\partial \theta} :\mathcal H^j_{\theta}\to\mathcal H^j_{\theta} \ , \ \forall j\in\mathbb N \ . 
\end{equation}
Note that $J_3$ is $j$-invariant (compare with (\ref{J3ninvj})), but recall that the sequence $(J_3^j)_{j\in\mathbb N}$ is a rigid operator sequence for the nesting maps (\ref{iotat}), {\it cf.} Remark \ref{rig}.
 
 However, in contrast with Example \ref{example-cp1}, now it is not so easy to write the inner product given by (\ref{psij2})-(\ref{prodjX}) in terms of an integral\footnote{The standard definition $\frac{1}{2\pi}\int_{-\pi}^{\pi}\overline{\phi}_j(\theta)\psi_j(\theta)d\theta$ \  is not $SO(3)$-invariant, similarly to (\ref{innpHol1}).} on ${\mathcal X}\simeq\mathbb R\!\mod 2\pi$.  Thus, we will now restrict ourselves to a purely discrete description. 
 
 Then, repeating the steps in Example \ref{example-cp1}, now without integral descriptions, given $f\in C^{\infty}_{\mathbb C}(S^2)$, with  $\boldsymbol{W}_{\!\mathcal{C}}$ of (anti-)Poisson type, the ${W}$-quantization of $f$, ${\mathbf{F}}^{w}$, determines a sequence of operators $\mathbf T_w[f]=(T_w^j[f])_{j\in\mathbb N}$ acting on the well-nested sequence of Hilbert spaces \ $\textswab{H}^< = ({\mathcal H}^j_{\theta}, \langle\cdot|\cdot\rangle_j)_{j\in\mathbb N}$  as   
$$T_w^j[f]: \mathcal H^j_{\theta}\to\mathcal H^j_{\theta} \ , \  \phi^j\mapsto\psi^j \ , $$
where, for $\phi^j(\theta)$ as in (\ref{psij2}), 
\begin{equation}\label{TjX}
  \  \psi^j(\theta)=j!\sqrt{2j+1} \sum_{l=0}^{2j}\sum_{m,m'=-j}^j \!\!\!{\frac{\langle Y_l^{\bar m}|f\rangle}{c^{2j}_l }}C_{m,-m',\bar m}^{\,\,j,\,\,\,\,\,\,j,\,\,\,\,\,l}
    \frac{(-1)^{j-m'}\alpha^j_{m'}e^{im\theta}}{\sqrt{(j-m)!(j+m)!}} 
\end{equation}	 	
and similarly for the 	$\widetilde{W}$-quantization of $f$, replacing $c_l^n\leftrightarrow 1/c_l^n$ in  (\ref{TjX}). 	
	 	
Put another way, $T_w^j[f]$ takes the modified Fourier coefficients $\alpha_m^j$ of $\phi^j$ to the modified Fourier coefficients $\beta_m^j$ of $\psi^j$, {\it cf} (\ref{psij2}),  where 	 	
	 \begin{equation}\label{betajX}
    \beta_m^j =  \sqrt{n+1} \sum_{l=0}^{n}\sum_{m'=-j}^j \!\!{\frac{\langle Y_l^{\bar m}|f\rangle}{c^n_l}}C_{m,-m',\bar m}^{\,\,j,\,\,\,\,\,\,j,\,\,\,\,\,l}
    (-1)^{j-m'}\alpha_{m'}^j \ . 
 \end{equation}

From Theorem \ref{H=H}, if $(\phi^j)_{j\in\mathbb N}\in \textswab H_\infty^<$ is given as in (\ref{psij2})  and $\mathcal H_{\theta}$ denotes the ground Hilbert space of $\textswab H_\infty^<$,  then we have the identification\footnote{Slightly abusing nomenclature, we also refer to the ordered set  $(\alpha_m)_{m\in\mathbb Z}$ as a sequence.}:   
\begin{equation}\label{p1X}
 \phi=\lim_{j\to\infty}\phi^j\in\mathcal H_{\theta} \ \iff \  (\alpha_m)_{m\in\mathbb Z} \ , \ \sum_{m=-\infty}^{\infty}|\alpha_m|^2 \ < \infty \ ,
\end{equation}
where, {\it cf.} (\ref{iotat}),  
\begin{equation}\label{alphalim2}
 \alpha_m = \lim_{j\to\infty}\alpha^j_{m}\in \mathbb C \ , \ \forall m \in \mathbb Z \ ,    
\end{equation}
 so that \ $\phi\equiv[L\phi=(\widetilde{\phi}^j)_{j\in\mathbb N}] = \lim_{j\to\infty}\widetilde{\phi}^j$ \ is given explicitly by 
\begin{equation}\label{phinfty2}
 \phi \ = \   \lim_{j\to\infty}j!\!\sum_{m=-j}^{j}\frac{\alpha_{m} 
    e^{i m\theta}}{\sqrt{(j-m)!(j+m)!}} \ \in \ \mathcal H_{\theta} \ . 
\end{equation}
 
Then, if $\vb T_w[f]\in \textswab{M}_{\infty}$ , from Theorem  \ref{point-lim-op} we have that 
$$\vb T_w[f]\implies \mathcal T_w[f]: \mathcal H_{\theta}\to\mathcal H_{\theta} \ , \  \phi\mapsto\psi \ , $$
where \ $\psi\equiv[L\psi=(\widetilde{\psi}^j)_{j\in\mathbb N}]=\lim_{j\to\infty}\widetilde{\psi}^j$ \ is given by 
\begin{equation}\label{psinfty2}
 \psi \ = \   \lim_{j\to\infty}j!\!\sum_{m=-j}^{j}\frac{\beta_{m} 
    e^{i m\theta}}{\sqrt{(j-m)!(j+m)!}} \ \in \ \mathcal H_{\theta} \ , 
\end{equation}
with, {\it cf.} (\ref{betajX}), 
 \begin{equation}\label{betajjX2}
    \beta_{m} =  \lim_{j\to\infty}\sqrt{n+1} \sum_{l=0}^{n}\sum_{m'=-j}^j \!\!{\frac{\langle Y_l^{\bar m}|f\rangle}{c^n_l}}C_{m,-m',\bar m}^{\,\,j,\,\,\,\,\,\,j,\,\,\,\,\,l}
    (-1)^{j-m'}\alpha_{m'} \ , 
 \end{equation} 	
satisfying 
\begin{equation}\label{beta<}
    \sum_{m=-\infty}^{\infty}|\beta_m|^2 \ < \infty \ , \ \ \mbox{so that} \quad  (\beta_m)_{m\in\mathbb Z} \  \longleftrightarrow \  \psi \in \mathcal H_{\theta} \ .
\end{equation}
And similarly for $\widetilde{\mathcal T}_w[f]: \mathcal H_{\theta}\to\mathcal H_{\theta}$, by substituting $c_l^n\leftrightarrow 1/c_l^n$ in (\ref{betajjX2}).

Now, $\phi$ and $\psi$ satisfying (\ref{phinfty2})-(\ref{psinfty2})  converge in nested-$\textswab{H}$-norm. Therefore, using Tannery, if $(\alpha_m)_{m\in\mathbb Z}$ satisfies (\ref{p1X}) and $\vb T_w[f]\in \textswab{M}_{\infty}$ then 
\begin{equation}\label{fs2}
\lim_{j\to\infty} \ (j!)^2\!\sum_{m=-j}^{j}\frac{|\beta_{m}|^2}{(j-m)!(j+m)!} \ = \ \lim_{j\to\infty}\sum_{m=-j}^{j} |\beta_m|^2 \ ,
\end{equation}
with a similar equation holding for $(\alpha_m)_{m\in\mathbb Z}$.

On the other hand, functional convergence of $\phi$ and $\psi$ given by (\ref{phinfty2})-(\ref{psinfty2}) is assured if the sequences $(\alpha_m)_{m\in\mathbb Z}$ and $(\beta_m)_{m\in\mathbb Z}$ are both $\ell^1$-sequences, that is, with absolutely convergent series, and then it follows that, 
\begin{equation}
\infty \ > \ \lim_{j\to\infty} \ j!\!\sum_{m=-j}^{j}\frac{|\alpha_{m}|}{\sqrt{(j-m)!(j+m)!}}\iff  \sum_{m=-\infty}^{\infty} |\alpha_m| \ < \ \infty
\end{equation}
and likewise for $(\beta_m)_{m\in\mathbb Z}$, 
in which case we have that, $\forall \theta\in\mathbb R\!\!\mod 2\pi$,   
\begin{equation}\label{fs212}
\lim_{j\to\infty} \ j!\!\sum_{m=-j}^{j}\frac{\gamma_{m}e^{im\theta}}{\sqrt{(j-m)!(j+m)!}} \ = \ \lim_{j\to\infty}\sum_{m=-j}^{j} \gamma_me^{im\theta} \ ,
\end{equation}
where $(\gamma_m)_{m\in\mathbb Z}$ stands for both $(\alpha_m)_{m\in\mathbb Z}$ and $(\beta_m)_{m\in\mathbb Z}$.

\

In the third construction, we extend the second construction, from the sequence of $SO(3)$ representation to the full sequence of $SU(2)$ representations, as follows.

For $j$ half-integer, we adjust  (\ref{id2})-(\ref{gammaj}) in order to accommodate the constant function $1$ so that it is  associated to $m=1/2, -1/2, 1/2, -1/2,\cdots$ Hence, \begin{equation}\label{id3}
   id_3^j: \mathcal H^j_z\equiv Poly(\mathbb C)_{\leq n}^h\ni \nu_j z^{j-m} \ \longleftrightarrow \  \rho_je^{i(m-\epsilon_j)\theta} \in {\mathcal H}^j_{\theta} \ ,  
\end{equation} 
where 
\begin{equation}\label{epsilonj}
    \epsilon_j=\frac{\sin(j\pi)}{2} \ , \ 2j\in\mathbb N \ . 
\end{equation}
Then, 
\begin{equation}\label{wnEX2}
   \mathcal E^j =  \left\{\boldsymbol{u}(j,m)=\frac{\rho_je^{i(m-\epsilon_j)\theta}}{\sqrt{(j-m)!(j+m)!}}\right\}_{-j\leq m\leq j}   
 \end{equation}
is our standard basis for ${\mathcal H}^j_{\theta}$, $2j\in\mathbb N$, where 
\begin{equation}\label{newrho}
    \rho_j=\left\{\begin{array}{ll}
       j!  & , \  j\in\mathbb N \\
       (j-1/2)!\sqrt{j+1/2}  & , \  j=k-1/2 \ , \ k\in\mathbb N 
    \end{array} \right.
\end{equation}
is the canonical choice so that $1\longleftrightarrow(m_j)_{2j\in\mathbb N}=(1/2,0,-1/2,0,1/2,0,-1/2,0,\cdots)$. Accordingly, the nesting maps are now determined by 
\begin{equation}\label{newnest}
    \iota_j^{j'}(\boldsymbol{u}(j,m))=\boldsymbol{u}(j',m+\epsilon_{j'}-\epsilon_j) \ . 
\end{equation}
In this way, our well-nested sequence of Hilbert spaces $(\textswab{H}^<,\textswab{E})$, where   
$\textswab{H}=({\mathcal H}^j_{\theta})_{2j\in\mathbb N}$, $\textswab{E}=(\mathcal E^j)_{2j\in\mathbb N}$, is now defined by (\ref{epsilonj})-(\ref{newnest}). 

Now, the sequence of spin operators $(J_3^j)$ is not rigid anymore, since it has an explicitly dependence on $j$. This can also be seen from (\ref{stansu2act}), which implies
\begin{equation}
\begin{aligned}
    \iota^{j'}_j\circ J_3^j(\vb u & (j, m)) = m\, \vb u(j',m+\epsilon_{j'}-\epsilon_j)\\
    & \ne (m+\epsilon_{j'}-\epsilon_j)\, \vb u(j',m+\epsilon_{j'}-\epsilon_j) = J^{j'}_3\circ \iota^{j'}_j(\vb u(j,m))
\end{aligned}
\end{equation}

And now, a sequence $\Phi=(\phi^j)_{2j\in \mathbb N}\in \textswab H^<$ can be identified with a sequence of complex Fourier polynomials
\begin{equation}\label{phi-j-semint}
    \phi^j  = \ \rho_j\sum_{m=-j}^{j}\dfrac{\alpha^j_{m-\epsilon_j}  e^{i(m-\epsilon_j)\theta}}{\sqrt{(j-m)!(j+m)!}} \in \mathcal H^j_\theta \ ,
\end{equation}
where $\alpha^j_{m-\epsilon_j}$ are the modified Fourier coefficients, and then we identify the $W$-quantization $\vb F^w$ of $f \in C^\infty_{\mathbb C}(S^2)$ via $\boldsymbol W_{\mathcal C}$ of Poisson type with a sequence of operators $\vb T_w[f] = (T^j_w[f])_{2j\in\mathbb N}$ acting on $\textswab H^< = (\mathcal H^j_\theta, \ip{\cdot}{\cdot}_j)_{2j\in\mathbb N}$, so that 
\begin{equation}
\psi^j = \ \rho_j \sum_{m=-j}^{j}\dfrac{\beta^j_{m-\epsilon_j}  e^{i(m-\epsilon_j)\theta}}{\sqrt{(j-m)!(j+m)!}} \in \mathcal H^j_\theta \ ,
\end{equation}
for $\psi^j = T^j_w[f](\phi^j) = \psi^j$ and $\phi^j \in \mathcal H_\theta ^j$ as in (\ref{phi-j-semint}), where
\begin{equation}
    \beta^j_{m-\epsilon_j} = \  \sqrt{n+1} \sum_{l=0}^{n}\sum_{m'=-j}^j \!\!{\frac{\langle Y_l^{\bar m}|f\rangle}{c^n_l}}C_{m,-m',\bar m}^{\,\,j,\,\,\,\,\,\,j,\,\,\,\,\,l}
    (-1)^{j-m'}\alpha_{m'-\epsilon_j}^j \ .
\end{equation}

From Theorem \ref{H=H}, if $\Phi \in \textswab H^<_\infty$, then we have the identification (\ref{p1X}), that is,
\begin{equation*}
    \phi = \lim_{j\to\infty}\phi^j \in \mathcal H_\theta \iff (\alpha_m)_{m\in\mathbb Z} \  ,   \sum_{m=-\infty}^{\infty} |\alpha_m|^2 < \infty \ ,
\end{equation*}
where now
\begin{equation}
    \alpha_m = \lim_{j\to\infty}\alpha^j_{m-\epsilon_j}\in\mathbb C \ , \ \ \forall m \in \mathbb Z \ .
\end{equation}
So $\phi \equiv [L\phi = (\widetilde \phi^j)] = \lim_{j\to\infty} \widetilde \phi^j$ is given by
\begin{equation}\label{phinftyhalf}
    \phi \ = \ \lim_{j\to\infty} \ \rho_j\!\sum_{m=-j-\epsilon_j}^{j-\epsilon_j}\dfrac{\alpha_{m}  e^{im\theta}}{\sqrt{(j-m-\epsilon_j)!(j+m+\epsilon_j)!}} \ .
\end{equation}

If $\vb T_w[f]\in \textswab M_\infty$, the $\Gamma$-induced operator $\mathcal T_w[f]: \mathcal H_\theta\to \mathcal H_\theta$, maps $\phi \mapsto \psi$, for $\psi \equiv [\Psi = (\psi^j)]$. Using again $\psi \equiv [L\psi = (\widetilde \psi^j)] = \lim_{j\to\infty}\widetilde \psi^j$, we get
\begin{equation}\label{psinftyhalf}
\psi \ = \ \lim_{j\to\infty} \ \rho_j\!\sum_{m=-j-\epsilon_j}^{j-\epsilon_j}\dfrac{\beta_{m}  e^{im\theta}}{\sqrt{(j-m-\epsilon_j)!(j+m+\epsilon_j)!}} \ ,    
\end{equation}
where $(\beta_m)_{m\in\mathbb Z}$ satisfying (\ref{beta<}) are given by 
\begin{equation}
    \beta_m = \lim_{j\to\infty}\sqrt{n+1} \sum_{l=0}^{n}\sum_{m'=-j-\epsilon_j}^{j-\epsilon_j} \!\!{\frac{\langle Y_l^{\bar m}|f\rangle}{c^n_l}}C_{m-\epsilon_j,-m'+\epsilon_j,\bar m}^{\,\,j,\,\,\,\,\,\,j,\,\,\,\,\,l}
    (-1)^{j-m'}\alpha_{m'} \ . 
\end{equation}

Again, it is important to emphasize that (\ref{phinftyhalf}) and (\ref{psinftyhalf}) always converge in the nested-$\textswab{H}$-norm. Thus, again from (\ref{p1X}), (\ref{beta<}) and Tannery's theorem,  
\begin{equation}\label{fs3}
\lim_{j\to\infty} \ \rho_j^2\!\sum_{m=-j-\epsilon_j}^{j+\epsilon_j}\frac{|\beta_{m}|^2}{(j-m-\epsilon_j)!(j+m+\epsilon_j)!} \ = \ \lim_{j\to\infty}\sum_{m=-j}^{j} |\beta_m|^2 \ ,
\end{equation}
with a similar equation holding for $(\alpha_m)_{m\in\mathbb Z}$.  

However, now 
the study of functional convergence of (\ref{phinftyhalf}) and (\ref{psinftyhalf}), {\it i.e.} the study of whether $\phi$ and $\psi$  are  actually well defined functions on the whole or {\it a.e.} on $\mathcal X\subset S^2$, can be simplified by taking $\textswab H^<$ as three disjoint subsequences: $$\textswab H^<_1 = (\mathcal H^j_\theta)_{2j\equiv 1 \!\!\!\pmod 4} \ , \ \textswab H^<_2 = (\mathcal H^j_\theta)_{j\in\mathbb N} \ , \ \textswab H^<_3 = (\mathcal H^j_\theta)_{2j\equiv 3 \!\!\!\pmod 4} \ . $$ 

But note that we could have defined the extension $SO(3)\hookrightarrow SU(2)$ of this quantization by choosing $\epsilon_j'=-\epsilon_j$ so that $1\longleftrightarrow (m_j)_{2j\in\mathbb N}=(-1/2,0,1/2,0, \cdots)$, in which case the roles of $\textswab H^<_1$ and $\textswab H^<_3$ would be interchanged. Hence, we could also define the $SU(2)$ quantization by taking the average of these two choices.

Anyway, 
it is not hard to see that all choices are asymptotically equivalent and all subsequences converge equally, as functions on $\mathcal X\subset S^2$, when 
both $(\alpha_m)_{m\in\mathbb Z}$ and $(\beta_m)_{m\in\mathbb Z}$ are complex $\ell^1$-sequences, in which case we have that,  $\forall \theta\in\mathbb R\!\!\mod 2\pi$,  
\begin{equation}\label{fs312}
\lim_{j\to\infty} \ \rho_j\!\sum_{m=-j-\epsilon_j}^{j+\epsilon_j}\frac{\gamma_{m}e^{im\theta}}{\sqrt{(j-m-\epsilon_j)!(j+m+\epsilon_j)!}} \ = \ \lim_{j\to\infty} \sum_{m=-j}^{j} \gamma_me^{im\theta} \ , 
\end{equation}
with $(\gamma_m)_{m\in\mathbb Z}$ standing for both $(\alpha_m)_{m\in\mathbb Z}$ and $(\beta_m)_{m\in\mathbb Z}$.

\

We emphasize that the theory developed in Section \ref{gentheory} asserts that  $(\beta_m)_{m\in\mathbb Z}\in\ell^2$ when $(\alpha_m)_{m\in\mathbb Z}\in\ell^2$, if \ $\mathbf F\in\textswab{M}_{\infty}$ . But when the series of $(\alpha_m)_{m\in\mathbb Z}$ is also absolutely convergent, that is, if $(\alpha_m)_{m\in\mathbb Z}\in\ell^1$, then further hypotheses on  $\mathbf F$ may generally be needed for  $(\beta_m)_{m\in\mathbb Z}$ to also be a complex $\ell^1$-sequence. We shall not investigate these further hypotheses here. We shall also defer further investigations on the functional convergence of (\ref{phinfty2})-(\ref{psinfty2}) and (\ref{phinftyhalf})-(\ref{psinftyhalf}) in the more subtle cases when $(\alpha_m)_{m\in\mathbb Z}$ and $(\beta_m)_{m\in\mathbb Z}$ are complex $\ell^2$-sequences but not complex $\ell^1$-sequences.
\end{example}

\subsection{Quantized functions and asymptotic localization}	

The general theory developed in Section \ref{gentheory} and exemplified in Section \ref{examples-section} asserts that the operator sequences $\vb T\notin\textswab{M}_{\infty}$ are not suited to the limit $j\to\infty$, {\it cf.} Theorem \ref{point-lim-op}.

Here, it is important to clarify that the limit $j\to\infty$ is asymptotic and the existence of a ground Hilbert space and operators therein could be seen as providing consistency requirements for a sequential quantization of $S^2$, but recalling that we only have a quantum spin system when $2j\in\mathbb N$. 
	Thus, for instance, in (\ref{fs2}) of Example \ref{S1cS2}, the equality generally becomes an inequality if the limit $j\to\infty$ is not taken on both sides, that is, we have an inequality for any $j\in\mathbb N$. 
Similarly in (\ref{fs212}), as well as in (\ref{fs3})-(\ref{fs312}) for any $2j\in\mathbb N$. And this is very important to consider when working out expansions in $j^{-1}$ to the asymptotic expressions. 

Nonetheless, the asymptotic consistency is fundamental for quantization, so we now 	 	
 start investigating the conditions for a  function $f\in C^{\infty}_{\mathbb C}(S^2)$ to define operator sequences 
$\vb F^w\simeq \vb T_w[f]\in{\textswab{M}}_{\infty}$  and/or $\widetilde{\vb F}^w\simeq \widetilde{\vb T}_w[f]\in{\textswab{M}}_{\infty}$, as the ones in Examples \ref{example-cp1}-\ref{S1cS2}, in terms of a given  correspondence sequence $\boldsymbol{W}_{\!\mathcal{C}}$. For this, we shall use: 
\begin{proposition}\label{necessary}
Let $(\textswab{H}^<_{\infty},\textswab{E})$ be  well-nested with ground $\mathcal H$. 
If $\vb F\in \textswab M_\infty$, then $\vb F$ is upper bounded ({\it cf.} Definition \ref{asympopnorm}).  
\end{proposition}
\begin{proof}
Let $\textswab{E}=(\vb e_k)_{k\in\mathbb N}$, $\vb e_k=\lim_{j\to\infty}\vb e^j_k$, $\forall k\in\mathbb N$. 
For $F_n\in \vb F$, we have
\begin{equation}\label{opup}
    \norm{F_n}^2 = \dfrac{1}{n+1}\sum_{k=1}^{n+1}\ip{\vb e^j_k}{F_n^\ast F_n(\vb e^j_k)} = \dfrac{1}{n+1}\sum_{k=1}^{n+1}\norm{F_n(\vb e^j_k)}^2 \le \norm{F_n}_{op}^2
\end{equation}
By Proposition \ref{op-norm-seq-b}, $\vb F\in \textswab M_\infty\implies{\vb F}$ is $\textswab H$-bounded, thus ${\vb F}$ is upper bounded.
\end{proof}

 Now, given a symbol correspondence sequence $\boldsymbol{W}_{\!\mathcal{C}}$, from equations (\ref{hf})-(\ref{dFWnlm}) in Definition \ref{quantization-def}, we see that the $W$- and $\widetilde{W}$-(pseudo)quantizations of $f$ are
defined via the spherical harmonic series of $f$. Thus, in a weaker sense, for any $f\in L^2_{\mathbb C}(S^2)$ we can define ${\mathbf{F}}^{w}$ and $\widetilde{\mathbf{F}}^{w}$. This includes all continuous functions and much more.  

However, in a stronger sense, which is tied up with the questions of asymptotic localization, we need uniform convergence to $f$ of the spherical harmonic series of $f$. And this is guaranteed if $f\in C_{\mathbb C}^{k,\alpha}(S^2)$, with $k+\alpha>1/2$, {\it cf. e.g.} \cite{atk}. 
Thus, in this strong sense, Definition \ref{quantization-def} can in principle be extended to differentiable functions or $\alpha$-H\"{o}lder continuous functions,   with $1/2<\alpha\leq 1$.

Anyway, such extensions shall not be studied here and all results will only be stated for classical functions, that is, $f\in C_{\mathbb C}^{\infty}(S^2)$. 
So, we now start investigating the conditions for such  a $W$- or $\widetilde{W}$-quantized function to be upper bounded, in terms of the properties of the symbol correspondence sequence $\boldsymbol{W}_{\!\mathcal{C}}$. 

But for our  investigations related to asymptotic localization, it will often be simpler to restrict our attention to $J_3$-invariant functions and operators (most generalizations to non-$J_3$-invariant cases being rather straightforward).  
	Thus, if $(a_l)_{l\in\mathbb N_0}$ is the sequence of Legendre coefficients of a $J_3$-invariant classical function $f\in C_{\mathbb C}^{\infty}([-1,1])$, 
	with $a_l\in\mathbb C$ as in  (\ref{Lf})-(\ref{Lc}), then ${\mathbf{F}}^{w}=({F}_n^{w})_{n\in\mathbb N}$ and $\widetilde{\mathbf{F}}^{w}=(\widetilde{F}_n^{w})_{n\in\mathbb N}$ are given by ({\it cf.} equations (\ref{genfn})-(\ref{dFWnlm}) and Definition \ref{quantization-def}):
	\begin{eqnarray}
	F_n^{w}&=&[W^j]^{-1}(f)\ = \  \sum_{l=0}^n\frac{a_l}{c^n_l\sqrt{2l+1}}\widehat{\mathbf{e}}^j(l,0) \  , \label{FWn0} \\
	\widetilde{F}_n^{w}&=&[\widetilde{W}^j]^{-1}(f) \ = \ \sum_{l=0}^n\frac{a_lc^n_l}{\sqrt{2l+1}}\widehat{\mathbf{e}}^j(l,0) \ \label{dFWn0} . 
	\end{eqnarray}

Thus, from (\ref{EE}) we have that 
\begin{eqnarray}\label{FJ3Norms} 
\quad\quad\quad\quad||\mathbf F^{w}||^2_{<} = \liminf_{n\to\infty}\sum_{l=0}^n \frac{1}{2l+1}\left|\frac{a_l}{c_l^n}\right|^2  &,& ||\mathbf F^{w}||^2_{>} = \limsup_{n\to\infty}\sum_{l=0}^n \frac{1}{2l+1}\left|\frac{a_l}{c_l^n}\right|^2 \label{L|F|} \\
||\widetilde{\mathbf F}^{w}||^2_{<} = \liminf_{n\to\infty}\sum_{l=0}^n \frac{|a_lc_l^n|^2}{2l+1}  &,& ||\widetilde{\mathbf F}^{w}||^2_{>} = \limsup_{n\to\infty}\sum_{l=0}^n \frac{|a_lc_l^n|^2}{2l+1} \label{L|tF|} \end{eqnarray}
with obvious generalizations of (\ref{L|F|}) and (\ref{L|tF|}) when $\mathbf F^{w}$ and $\widetilde{\mathbf F}^{w}$ are given in the general form (\ref{FWnlm}) and (\ref{dFWnlm}), using
(\ref{EE}).

On the other hand, the $L^2$-norm $||\cdot|| : C_{\mathbb C}^{\infty}([-1,1])\to \mathbb R^+$, $f\mapsto ||f||$, is given by 
	\begin{equation}\label{|f|} 
	||f||^2=\frac{1}{2}\int_{-1}^{1} |f(z)|^2dz \ ,
	\end{equation}
being the $L^2$-norm on $C_{\mathbb C}^{\infty}(S^2)$ restricted to $J_3$-invariant functions.  Thus, because 
$$\frac{1}{2}\int_{-1}^{1}P_l(z)P_{l'}(z)dz = \frac{\delta_{l,l'}}{2l+1} \ , $$
in terms of the Legendre coefficients of $f$, {\it cf.} (\ref{Lf}), we have that
\begin{equation}\label{L|f|} 
||f||^2= \lim_{n\to\infty}\sum_{l=0}^n \frac{|a_l|^2}{2l+1}  \ . 
\end{equation}

For starters, we do not assume $\boldsymbol{W}_{\!\mathcal{C}}$ is of (anti-)Poisson type. We then have: 
\begin{proposition}\label{obvious}
  If $\boldsymbol{W}_{\!\mathcal{C}}$ is an isometric symbol correspondence sequence, then 
	\begin{equation}\label{F=f-iso}
	||\mathbf F^{w}||_{\infty} \ = \ ||f|| \ = \ ||\widetilde{\mathbf F}^{w}||_{\infty} \ , \ \forall f\in C_{\mathbb C}^{\infty}([-1,1]) \ .
	\end{equation}
	If $\boldsymbol{W}_{\!\mathcal{C}}$ is a positive-dual symbol correspondence sequence, then 
	\begin{equation}\label{F<f}
	||\mathbf F^{w}||_{>} \ \leq \ ||f|| \ \leq \ ||\widetilde{\mathbf F}^{w}||_{<} \ , \ \forall f\in C_{\mathbb C}^{\infty}([-1,1]) \ .
	\end{equation}
	Equivalently, $\boldsymbol{W}_{\!\mathcal{C}}$ is a mapping-positive symbol correspondence sequence, then 
	\begin{equation}\label{f<F}
	||\widetilde{\mathbf F}^{w}||_{>} \ \leq \ ||f|| \ \leq \ ||\mathbf F^{w}||_{<} \ , \ \forall f\in C_{\mathbb C}^{\infty}([-1,1]) \ .
	\end{equation}  
\end{proposition}
\begin{proof}
For isometric symbol correspondence sequences, (\ref{F=f-iso}) is obvious from (\ref{L|F|}), (\ref{L|tF|}) and (\ref{L|f|}). For  mapping-positive symbol correspondence sequences, recalling (\ref{cg+}) and (\ref{cg<sqrt}), plus the fact that $\sum_{k=1}^{n+1}a_k=1$ in (\ref{cg+}),  we see that $|c^n_l| \le 1$, $0\leq \forall l\leq n, \ \forall n\in\mathbb N$. So, $\forall f\in C_{\mathbb C}^{\infty}([-1,1])$, from (\ref{L|F|})-(\ref{L|tF|}) and (\ref{L|f|}) we get (\ref{f<F}). Equivalence between (\ref{F<f}) and (\ref{f<F}) is obvious from the definitions. 
\end{proof}

We note that if $\boldsymbol{W}_{\!\mathcal{C}}$ is just mapping-positive (or equivalently just positive-dual), the inequalities in   (\ref{f<F}) or (\ref{F<f}) can be very strict and, in fact, some of the asymptotic operator norms can be $0$ or $\infty$, as shown by the example of the upper-middle-state symbol correspondence sequence. Because, for this correspondence, we recall from  \cite[eq.(6.57)]{prios} that its characteristic numbers $p_l^n$ satisfy  $\lim_{n\to\infty}p_l^n=0$ for every $l=2k+1$ odd. Hence, if $f$ is an odd function, $||\widetilde{\mathbf F}^w||_<=||\widetilde{\mathbf F}^w||_>=0$, and if $f$ is not an even function, $||{\mathbf F}^w||_<=||{\mathbf F}^w||_>=\infty$.

Thus, in order to have more control, we recall:  
\begin{definition}[\cite{prios}]\label{quasi-c}
  A symbol correspondence sequence $\boldsymbol{W}_{\!\mathcal{C}}$ is of \emph{quasi-classical} type if  
  \begin{equation}\label{qcc}
   \lim_{n\to\infty}|c_l^n|=1 \ , \ \forall l\in\mathbb N \ .    
  \end{equation}
\end{definition}

Of course, every symbol correspondence sequence of (anti-)Poisson type is of quasi-classical type, but the converse does not hold in general, {\it cf.} Theorem \ref{conviso}. 

Then we have:

\begin{proposition}\label{quant_norm}
	If $\boldsymbol{W}_{\!\mathcal{C}}$ is positive-dual and also of quasi-classical type, then 
	\begin{equation}\label{F=f}
	||\mathbf F^{w}||_{\infty} = ||f|| \ , \ \forall f\in C_{\mathbb C}^{\infty}([-1,1]) \ .
	\end{equation}
	Equivalently, if $\boldsymbol{W}_{\!\mathcal{C}}$ is mapping-positive and also of quasi-classical type, then 
	\begin{equation}\label{F=f2}
	||\widetilde{\mathbf F}^{w}||_{\infty} = ||f|| \ , \ \forall f\in C_{\mathbb C}^{\infty}([-1,1]) \ .
	\end{equation}
	\end{proposition}
\begin{proof}
Because the two statements are equivalent, we prove the mapping-positive case. Then, from the first inequality in (\ref{f<F}) and the fact that $||f||<\infty$, \textit{cf.} (\ref{|f|})-(\ref{L|f|}), we apply Tannery's theorem to (\ref{L|tF|}) to get from the  quasi-classical condition (\ref{qcc}) that $||\widetilde{\mathbf F}^w||_<=||\widetilde{\mathbf F}^w||_>=||f||$.  
\end{proof}	
\begin{remark}
Propositions \ref{obvious} and \ref{quant_norm} can be written in the general form, with $C_{\mathbb C}^{\infty}(S^2)$ replacing $C_{\mathbb C}^{\infty}([-1,1])$ in their statements.
\end{remark}

 In principle, however, we cannot guarantee equality for the remaining inequality in (\ref{F<f}), resp. (\ref{f<F}), if we restrict to generic positive-dual, resp. mapping-positive symbol correspondence sequences of quasi-classical or even of (anti-)Poisson type. But in this respect, we have the following result for the more special cases: 

\begin{theorem}\label{locFmu}
    For the standard and alternate Berezin correspondence sequences, or equivalently for  the standard and alternate Toeplitz correspondence sequences, 
\begin{equation}\label{F=fmu}
	||\mathbf F^{w}||_{\infty} = ||f|| = ||\widetilde{\mathbf F}^{w}||_{\infty} \ , \ \forall f\in \mathcal A_{\mu}([-1,1]) \ , \ \forall \mu > 2 \ .
	\end{equation}    
 	\end{theorem}
\begin{proof}
As in the proof of Proposition \ref{quant_norm}, here it is sufficient to prove for just one of the correspondence sequences of the statement. Take the standard Toeplitz correspondence sequence. Since $\mathcal{A}_\mu([-1,1])\subset C^\infty_{\mathbb{C}}([-1,1])$ for all $\mu>1$, from Proposition \ref{quant_norm} we have that $||\mathbf{F}^{w}||_{\infty} = ||f||$, $\forall f\in\mathcal{A}_\mu([-1,1])$, $\forall \mu>2$. On the other hand, following the proof of Theorem \ref{propToe} we conclude that, if $f\in\mathcal{A}_\mu([-1,1])$ for any $\mu>2$, then we can apply Tannery's theorem to (\ref{L|tF|}) and  then the quasi-classical condition (\ref{qcc}) implies that  $||\widetilde{\mathbf F}^w||_< = ||\widetilde{\mathbf F}^{w}||_>=||f||$. 
\end{proof}

In the same vein, if $\boldsymbol{W}_{\!\mathcal{C}}$ is a symbol correspondence sequence of (anti-)Poisson type, but without further conditions, it  may happen that, for some $f\in C_{\mathbb C}^{\infty}([-1,1])$, the asymptotic operator norm of its $W$-quantization, or of its $\widetilde{W}$-quantization, blows up  
to infinity, as the following example shows.

\begin{example}\label{ex2}
Take the sequence of symbol correspondences dual to $\boldsymbol W_{\!\mathcal{C}}$ in the proof of Theorem \ref{P<loc} ({\it cf.} (\ref{gex})), that is, for any $f\in C^\infty_{\mathbb{C}}([-1,1])$ with Legendre coefficients $a_l\ne 0$, $\forall l\in \mathbb{N}$, set
\begin{equation}
    c^n_l = \begin{cases}
    a_l/(2l+1), \ \ \forall n = l \\
    1, \ \ \textnormal{otherwise}
    \end{cases} \ .
\end{equation}

This sequence of symbol correspondences is also of Poisson type, but, for $n > m$, the norm of the $W$-quantization of $f$ satisfies
\begin{equation}
    ||F^w_n||^2-||F^w_{m}||^2 =
    2(n-m)+\sum_{l=m+1}^{n-1}\dfrac{|a_l|^2}{2l+1} > 2(n-m) > 2 \ .
\end{equation}
Thus, $||\mathbf{F}^w||$ is not a Cauchy sequence. Being increasing, $||\mathbf{F}^w||_{<}=||\mathbf{F}^w||_{>}=\infty$.

Of course, by taking the symbol correspondence sequence $\boldsymbol W_{\!\mathcal{C}}$ as in the proof of Theorem \ref{P<loc} ({\it cf.} (\ref{gex})), we have that $\boldsymbol W_{\!\mathcal{C}}$ is of Poisson type but $||\widetilde{\mathbf{F}}^w||_{\infty}=\infty$.  
\end{example} 

Therefore, if $\boldsymbol{W}_{\!\mathcal{C}}$ is just of (anti-)Poisson type, equality (\ref{F=f-iso}) does not hold in general. 
But recalling Theorem \ref{p_l} and Lemma \ref{wanglemma}, we have:

\begin{theorem}\label{locF1} 
If a symbol correspondence sequence $\boldsymbol{W}_{\!\mathcal{C}}$ is of quasi-classical type and if there exist $d_1,d_2\in\mathbb{N}_0$ and $K_{d_1},K_{d_2}>0$ such that
\begin{equation}\label{d<c<d}
\frac{1}{K_{d_1}\prod_{t=0}^{d_1}(2(l-t)+1)}\le|c^n_l|\le K_{d_2}\prod_{t=0}^{d_2}(2(l-t)+1) \ , \  n\ge \forall l>d+1 \ ,
\end{equation}
where $d=max\{d_1,d_2\}$, 
then (\ref{F=f-iso}) holds, that is, 
\begin{equation}\label{g=g=g}
||\mathbf F^{w}||_{\infty} = ||f|| = ||\widetilde{\mathbf F}^{w}||_{\infty} \ , \ \forall f\in C_{\mathbb C}^{\infty}([-1,1]) \ .
\end{equation}
\end{theorem}
\begin{proof}
Following the proof of Theorem \ref{p_l},  from Lemma \ref{wanglemma}, the inequality (\ref{d<c<d}) implies that, if $f\in  C_{\mathbb C}^{\infty}([-1,1])$, then we can apply Tannery's theorem to (\ref{L|F|}) and (\ref{L|tF|}), hence the quasi-classical condition (\ref{qcc}) implies (\ref{F=f-iso}). 
\end{proof}

In particular, of course, the above theorem provides  a sufficient condition on a symbol correspondence sequence $\boldsymbol{W}_{\!\mathcal{C}}$ of (anti-)Poisson type to guarantee that the  $L^2$-norm of every classical function agrees with the asymptotic norms of its $W$-quantized and $\widetilde{W}$-quantized operator sequences.  

 Therefore, Theorems \ref{locFmu} and \ref{locF1} provide connections between the conditions for equality of classical and asymptotic operator norms, and the conditions for the asymptotic localization of symbol correspondence sequences, as expressed in  Theorems \ref{p_l} and \ref{propToe}. 
 But the relations between localization of symbol correspondence sequences and properties of quantized functions can be further explored.  
 
 First, we have the following proposition, complementing Propositions \ref{obvious}-\ref{quant_norm}.
 
\begin{proposition}\label{q-d}
Let $\boldsymbol{W}_\mathcal{C} = (W^j)$ and $\widetilde{\boldsymbol{W}}_{{\mathcal{C}}} = (\widetilde{W}^j)$
be a symbol correspondence sequence and its dual. For every operator sequence $\mathbf{F}=(F_n)_{n\in\mathbb{N}}$, $F_n\in M_{\mathbb C}(n+1)$, 
\begin{equation}\label{A}
 ||\mathbf F||_\infty \in \mathbb{R} \ \Rightarrow   \lim_{n\to\infty} ||W^j_{F_n}|| \in \mathbb{R} 
\end{equation}
if and only if, for every sequence of polynomials $(f_n)_{n\in\mathbb{N}}$, 
$f_n\in Poly_{\mathbb{C}}(S^2)_{\le n}$,
\begin{equation}\label{B}
    \lim_{n\to\infty}||f_n||\in\mathbb{R}\Rightarrow ||\widetilde{\mathbf F}^w||_\infty  \in \mathbb{R} \ ,
\end{equation}
where $\widetilde{\mathbf{F}}^w = (\widetilde{F}_n^w)$ is the operator sequence given by $\widetilde{F}_n^w = [\widetilde{W}^j]^{-1}(f_n)$.
\end{proposition}
\begin{proof}
To simplify notation, we show for $J_3$-invariant operators and functions, the generalization being straightforward. 
Given a sequence of $J_3$-invariant operators $\mathbf{F}=(F_n)$, we can write
\begin{equation*}
    F_n = \sum_{l=0}^{n} \dfrac{\chi^n_l}{\sqrt{2l+1}}\widehat{\mathbf{e}}^j(l,0) \ ,
\end{equation*}
where $\chi^n_l \in \mathbb{C}$. Recalling that $c^n_0 = 1$, we get
\begin{equation*}
    W^j_{F_n} = \sum_{l=0}^n\chi^n_lc^n_lP_l \ .
\end{equation*}
Thus,
\begin{equation}\label{norm1}
    ||F_n|| = \sum_{l=0}^n\dfrac{|\chi^n_l|^2}{2l+1} \ \ \textnormal{and} \ \
    ||W^j_{F_n}|| = \sum_{l=0}^n\dfrac{|\chi^n_lc^n_l|^2}{2l+1} \ .
\end{equation}

Analogously, given a sequence of $J_3$-invariant  polynomials $(f_n)$, we can write
\begin{equation*}
    f_n = \sum_{l=0}^{n}\chi^n_l P_l \ ,
\end{equation*}
from which we get
\begin{equation*}
    \widetilde{F}_n^w = \sum_{l=0}^{n}\dfrac{\chi^n_lc^n_l}{\sqrt{2l+1}} \widehat{\mathbf{e}}^j(l,0) \ .
\end{equation*}
So we have
\begin{equation}\label{norm2}
        ||f_n|| = \sum_{l=0}^n\dfrac{|\chi^n_l|^2}{2l+1} \ \ \textnormal{and} \ \
    ||\widetilde{F}_n^w|| = \sum_{l=0}^n\dfrac{|\chi^n_lc^n_l|^2}{2l+1} \ .
\end{equation}

From (\ref{norm1}) and (\ref{norm2}), we have the enunciated equivalence.
\end{proof}

Therefore, in a rather strong sense 
we can say that, if a symbol correspondence sequence $\boldsymbol{W}_{\!\mathcal{C}} = (W^j)_{n\in\mathbb N}$  is ``more appropriate'' for dequantization, then its dual is ``more appropriate'' for (pseudo)quantization, and vice-versa. 
This falls in line with Definition \ref{mp_d} and Proposition \ref{p+}, that is:
\begin{proposition}
If \ $\boldsymbol{W}_{\!\mathcal{C}} = (W^j)_{n\in\mathbb N}$ is a mapping-positive symbol correspondence sequence, then, for every strictly positive classical function $f\in C^{\infty}_{\mathbb R^+}(S^2)$, its $\widetilde{W}$-(pseudo)quantization \  $\mathbf{\widetilde{F}}^w=(\widetilde{F}_n^w)_{n\in\mathbb N}$, $\widetilde{F}_n^w=[\widetilde{W}^j]^{-1}(f)$, is a sequence of positive-definite operators $\forall n\geq n_0$, for some $n_0\in\mathbb N$.
\end{proposition} 
\begin{proof} 
Let $f_n$ be the $n^{th}$ partial sum of the decomposition of $f$ in spherical harmonics, {\it cf.} (\ref{hf}). Because $(f_n)\to f$ in the sup norm and $f$ is strictly positive, $\exists n_0\in\mathbb N$ such that $f_n$ is strictly positive, $\forall n\geq n_0$. But each $f_n$, $n\geq n_0$, is a polynomial function, and because each ${W}^j$ is mapping-positive, $\widetilde{W}^j$ is positive dual, 
thus from Proposition \ref{p+}, $[\widetilde{W}^j]^{-1}(f)=[\widetilde{W}^j]^{-1}(f_n)$ is  positive-definite, {\it cf.} (\ref{dFWnlm}).  
\end{proof}

We thus introduce the following definitions:
 
\begin{definition}\label{dequantiz}
We shall say that an  
operator sequence $\mathbf{F} = (F_n)_{n\in\mathbb N}$ possesses  \textnormal{asymptotic expectation} if, for every $r$-convergent $\Pi$-sequence $(\Pi_{k_n})_{n\in\mathbb N}$, the expectation sequence $(\langle\Pi_{k_n}| F_n\rangle)_{n\in\mathbb N}$ also converges. 
\end{definition}

\begin{remark}
Note that the above definition uses the usual Hilbert-Schmidt inner product $\langle\cdot|\cdot\rangle$ and \emph{not} the normalized product  $\langle\cdot|\cdot\rangle_j=\frac{1}{n+1}\langle\cdot|\cdot\rangle$ given by (\ref{ninn}).
\end{remark}

\begin{definition}\label{expect} 
We say that a symbol correspondence sequence of (anti-)Poisson type
$\boldsymbol{W}_{\!\mathcal{C}}=(W^j)_{n\in\mathbb N}$ 
possesses \emph{classical (anti-)expectation} if the $\widetilde{W}$-quantization $\mathbf{\widetilde{F}}^w$ of any classical  function $f\in C^\infty_{\mathbb{C}}(S^2)$ possesses asymptotic expectation.
\end{definition}

And then we have:

\begin{theorem}\label{Clocexp2}
A symbol correspondence sequence $\boldsymbol{W}_{\!\mathcal{C}}$ (anti-)localizes classically ({\it cf.} Definitions \ref{loc}-\ref{cloc})  if and only if it possesses classical (anti-)expectation.
\end{theorem}
\begin{proof}
To simplify notation, assume $f=\bar{f}$, where $\bar{f}$ denotes the $S^1$-average of $f$, as in (\ref{3av}).  The generalization to $f\neq \bar{f}$ is straightforward from (\ref{3av})-(\ref{loc-eq-gen}).

Suppose that $\boldsymbol{W}_{\!\mathcal{C}}$ possesses classical expectation. Then, it is of Poisson type and, given any $r$-convergent $\Pi$ sequence,
\begin{equation}\label{eq1.1}
    \exists \lim_{n\to\infty}\langle \Pi_{k_n}|\widetilde{F}^w_n\rangle = \lim_{n\to\infty}\int_{-1}^1W^j_{F^w_n}(z)\rho^j_{k_n}(z)dz \in\mathbb{C}  ,  \ \forall f \in C^{\infty}_{\mathbb{C}}([-1,1])  , 
\end{equation}
{\it cf.} (\ref{dualinner}), where $\rho^j_{k_n}=\frac{n+1}{2}W^j_{\Pi_{k_n}}$, with $\mathbf{F}^w = (F^w_n)$ and $\widetilde{\mathbf{F}}^w = (\widetilde{F}^w_n)$ being the $W$-quantization and $\widetilde{W}$-quantization of $f$, respectively. 
We shall denote $W^j_{F_n}\equiv f_n$, which is the $n^{th}$ partial sum of the Legendre series of $f$. 
From (\ref{eq1.1}) we have 
\begin{equation}\label{limrho_n}
    \lim_{n\to\infty}\int_{-1}^{1}f(z)\rho^j_{k_n}(z)dz = \lim_{n\to\infty}\int_{-1}^1f_n(z)\rho^j_{k_n}(z)dz \in\mathbb C  , \ \forall f \in C^{\infty}_{\mathbb{C}}([-1,1]) .
\end{equation}

Now, the space $C^{\infty}_{\mathbb C}(S^2)$ is a Fr\'echet space for the topology $\mathcal E$ generated by the collection of seminorms $\{||\cdot||_m\}_{m\in\mathbb N_0}$, where $||g||_m \ = \sup_{u\in S^2}\{|-\Delta^{m/2} g(u)|\}$,  $\Delta$ the Laplace operator on $S^2$. Likewise for $C^{\infty}_{\mathbb C}([-1,1])$, by restricting to $J_3$-invariant functions, and  we also have 
({\it cf. e.g.} \cite[Proposition 2.47 and Corollary 2.49]{morm}):
\begin{lemma}\label{lem54}
      	For any $f\in C^{\infty}_{\mathbb{C}}([-1,1])$, \  $(f_n)\to f$ in  the topology $\mathcal E$.
\end{lemma}
Furthermore, we have the following lemma ({\it cf. e.g.} \cite[Theorem 2.8]{Rud}):
\begin{lemma}\label{lemcont}
 The  quasi-probability distribution $\rho$ on $C^{\infty}_{\mathbb{C}}([-1,1])$, defined by     
 \begin{equation}\label{defrho}
\int_{-1}^1f(z)\rho(z)dz \ :\!= \lim_{n\to\infty} \int_{-1}^{1} f(z) \rho^j_{k_n} dz \ ,
\end{equation}
is continuous in the topology \ $\mathcal E$. 
\end{lemma}
On the other hand, from the Poisson condition we have that 
\begin{equation}\label{rho_fn}
\int_{-1}^{1}f_n(z)\rho(z)dz = f_n(1-2r)\ , \ \forall n\in\mathbb N \  ,
\end{equation}
{\it cf.} Definition \ref{polloc}, eqs. (\ref{polloc-eq}) and (\ref{defrho}),  and Corollary \ref{pollocc}.

Thus, from (\ref{rho_fn}) and Lemmas  
\ref{lem54}-\ref{lemcont}, 
$$\int_{-1}^{1}f(z)\rho(z)dz=f(1-2r) \ , \ \forall f\in C^{\infty}_{\mathbb{C}}([-1,1]) \ .$$

If we suppose anti-Poisson condition, the r.h.s.~of (\ref{rho_fn}) must be $f_n(2r-1)$, so the result of the expression above changes to $f(2r-1)$, which means that the sequence of correspondences anti-localizes classically. 

Conversely, assuming classical (anti-)localization of $\boldsymbol{W}_{\!\mathcal{C}}$,  asymptotic expectation of the $\widetilde{W}$-quantization of a classical function follows trivially from definitions.      	
\end{proof}
\begin{remark}
For $\mathcal E$ in Lemmas \ref{lem54}-\ref{lemcont}, we could also have used the simpler collection of seminorms  $\{||\cdot||_m\}_{m\in\mathbb N_0}$, where $||g||_m \ = \sup_{z\in [-1,1]}\{|g^{(m)}(z)|\}$. 
\end{remark} 

We end this section with some comments on possible extensions/restrictions of Theorem \ref{Clocexp2}.

First, we should realize that a 
fundamental hypothesis of Theorem \ref{Clocexp2} is, besides (anti-)Poisson, the  convergence of equation (\ref{rhof})  $\forall f\in C^{\infty}_{\mathbb{C}}([-1,1])$. But this hypothesis can be restricted to the case 
$\forall f\in \mathcal A_{\mu}([-1,1])$ or extended to the cases 
$\forall f\in C^{k}_{\mathbb{C}}([-1,1])$, $k\geq 1$, or $\forall f\in C^{0,\alpha}_{\mathbb{C}}([-1,1])$, $1/2<\alpha\leq 1$, because in all these cases we have that the Legendre series of $f$ converges uniformly to $f$.

However, we are not sure if Lemma \ref{lemcont} holds true for $\mathcal A_{\mu}([-1,1])$ with the topology of holomorphic functions, and it is known that Lemma \ref{lem54} does not hold true for $C^{0,\alpha}_{\mathbb{C}}([-1,1])$ with the topology given by its natural norm.

Finally, the difficulty in generalizing Theorem \ref{Clocexp2} to $C^{k}_{\mathbb{C}}(S^2)$ stems from the fact that 
Lemma \ref{lem54} cannot be generalized to the natural topology $\mathcal E_k$ that makes $C^{k}_{\mathbb{C}}([-1,1])$ a Fr\'echet space because, for any $f\in C^{k}_{\mathbb{C}}([-1,1])$,
one can  guarantee $f_n\to f$ in all seminorms  $||\cdot||_{m}$ , $0\leq m\leq k-2$, but $\mathcal E_k$ is the topology on $C^{k}_{\mathbb{C}}([-1,1])$ that is generated by the collection of seminorms $\{||\cdot||_m\}_{0\leq m\leq k}$.

\section{Summary and discussion}

If a symbol correspondence sequence $\boldsymbol{W}_{\!\mathcal{C}}$ localizes (resp. anti-localizes) classically, 
then it is of Poisson (resp. anti-Poisson) type, {\it cf.} Corollary \ref{l_p}, but the converse does not 
hold in general, {\it cf.} Theorem \ref{P<loc}.
A sufficient condition for the converse of Corollary \ref{l_p} to hold is that the characteristic numbers of $\boldsymbol{W}_{\!\mathcal{C}}$ satisfy certain bounds, {\it cf.} (\ref{c<d}), but these polynomial bounds are not satisfied for general symbol correspondence sequences of Poisson (resp. anti-Poisson) type, as per the standard (resp. alternate) Toeplitz correspondence sequence, {\it cf.} Proposition \ref{toepprop}.

Motivated by these examples, we thus defined a notion of $\mu$-analytic localization, $\mu>1$, which  amounts to $\Pi$-distribution sequences converging to Dirac's distribution on an appropriate subspace $\mathcal A_{\mu}([-1,1])\subset C^{\infty}_{\mathbb{C}}([-1,1]) $, {\it cf.} Definition \ref{partialrho}, and in this way we have that the standard (resp. alternate) Toeplitz symbol correspondence sequence localizes (resp. anti-localizes) $\mu$-analytically, for every $\mu>2$.

However, we should have in mind that the specific polynomial bounds on the characteristic numbers stated in Theorem \ref{p_l}, {\it cf.} (\ref{c<d}), were assumed in view of the sharpest known bounds on the coefficients of Legendre expansions of functions in $C^{\infty}_{\mathbb{C}}([-1,1])$, {\it cf.} \cite{wang}. If new sharper bounds for these coefficients can be found, then rougher bounds on  the characteristic numbers could be assumed.  
In the same vein, if new bounds sharper than (\ref{alrho}) can be found for the Legendre coefficients of functions in $\mathcal A_{\mu}([-1,1])$, we might still ask whether the standard (alternate) Toeplitz correspondence sequence (anti-)localizes $\mu$-analytically for some $\mu<2$.

In order to approach 
these questions from the inverse direction, going from the classical system to quantum systems, we had to develop the theory of sequential quantizations of $S^2$, particularly the 
notion of a (well-)nested sequence of Hilbert spaces $\textswab{H}^<$, the notion of a convergent state sequence $\Phi\in\textswab{H}^<_{\infty}$ and the notion of a ground Hilbert space $\mathcal H$, all of which were made explicit in Examples \ref{example-cp1} and \ref{S1cS2}. 

We saw that an operator sequence $\vb F:\textswab{H}^<\to\textswab{H}^<$ defines an asymptotic operator $\mathcal F:\mathcal H\to\mathcal H$  only if $\vb F$ is upper bounded, {\it cf.} Proposition \ref{necessary}. Then, we found that just being of (anti-)Poisson type is not sufficient for a symbol correspondence sequence to define upper bounded quantized functions, \textit{cf}. Example \ref{ex2}. In this setting, Theorems \ref{locFmu} and \ref{locF1} imply 
that the $L^2$-norm of any classical function is equal to the asymptotic norms of its $W$-quantization and/or $\widetilde{W}$-quantization when $\boldsymbol{W}_{\!\mathcal{C}}$ 
satisfies similar sufficient conditions for its  asymptotic (anti-)localization. 

Furthermore, the quantization approach provided an equivalent characterization for classical (anti-)localization of a symbol correspondence sequence in terms of $\boldsymbol{W}_{\!\mathcal{C}}$ possessing the natural property of classical (anti-)expectation, \textit{cf}. Definition \ref{dequantiz} and  Theorem \ref{Clocexp2}. But a closer look at the hypothesis of Theorem  \ref{Clocexp2} reveals that 
classical (anti-)localization of  $\boldsymbol{W}_{\!\mathcal{C}}$ of (anti-)Poisson type is equivalent to having every $r$-convergent sequence $(\rho^j_{k_n})_{n\in\mathbb N}$ converge to an element in the dual of $C^{\infty}_{\mathbb{C}}([-1,1])$.
Hence,  
for Theorem  \ref{Clocexp2} 
we only assume (anti-)Poisson condition for the $c_l^n$'s and the convergence of equation  (\ref{rhof}) and this hypothesis 
seems weaker than the one for Theorem \ref{p_l} because, there, the hypothesis of the polynomial bounds for the  $c_l^n$'s that is expressed by equation (\ref{c<d}) guarantees absolute convergence of (\ref{rhof}).  
It turns out, however, that the polynomial bounds assumed in Theorem \ref{p_l} are too vaguely stated to be able to distinguish between absolute and conditional convergence of (\ref{rhof}), so that, in practice, it seems hard to use Theorem \ref{Clocexp2} in order to sharpen the conditions on the $c_l^n$'s.

Now, in order to have a better feeling of the subtleties involved in relating the (anti-)Poisson condition to classical (anti-)localization, recall from Definition \ref{p_ap} that the (anti-)Poisson condition refers to the asymptotic limit of weakly oscillatory symbols ($j\to\infty$ keeping finite $l$'s). Thus, the (anti-)Poisson  condition in practice refers to the $j\to\infty$ asymptotics of smooth functions with an arbitrary $l$-cutoff. Such functions, when $J_3$-invariant, lie in $Poly_{\mathbb{C}}([-1,1])$ and Corollary \ref{pollocc} states that the (anti-)Poisson  condition is equivalent to polynomial (anti-)localization,  
{\it cf.} Definition \ref{polloc}. 
However, classical (anti-)localization of a symbol correspondence sequence requires all its $r$-convergent $\Pi$-distribution sequences converging to Dirac's distributions on 
$C^\infty_{\mathbb{C}}([-1,1])$, {\it cf.} Definition \ref{loc}, and generally this is related to the much subtler asymptotics of highly oscillatory symbols ($j,l\to\infty$), as well\footnote{We refer to \cite{prios} for discussions of various subtleties associated to highly oscillatory symbols, in particular  subtleties associated to the asymptotic limit of 
 twisted products of such symbols.}.

In summary, classical (anti-)localization of a general symbol correspondence sequence $\boldsymbol{W}_{\!\mathcal{C}}$ (equivalent to classical expectation for $\boldsymbol{W}_{\!\mathcal{C}}$) is a stronger property than $\boldsymbol{W}_{\!\mathcal{C}}$ being of (anti-)Poisson type. This is not the case  for sequences of  mapping-positive or isometric symbol correspondences, for which these two properties are equivalent, because (\ref{c<}) and (\ref{SWc}) impose sufficient bounds on the ``weights'' of highly oscillatory ($l\to\infty$) components of their  symbols, {\it cf.} (\ref{Wjc}) and (\ref{polb}). But as the Toeplitz examples suggest, although a positive-dual $\boldsymbol{W}_{\!\mathcal{C}}$ of (anti-)Poisson type may not have such bounding weights (\ref{c<d}), this class of 
symbol correspondence sequence may still satisfy weaker forms of asymptotic localization ($\mu$-analytical).   

Finally, we could probably say that this paper consists in a first precise study, albeit still very limited in scope, of some subtleties involved in the semiclassical asymptotics of highly oscillatory symbols, in the context of spin systems\footnote{In the context of affine mechanical systems, studies of subtleties in the semiclassical asymptotics of highly oscillatory symbols actually abound. Here we just mention the pioneering work of Berry on semiclassical Wigner functions of pure states \cite{berry} and some later studies on their dynamics \cite{RO} and their singularities \cite{CDR, DMR, DR}.}

\section{Appendix} 

\subsection{A proof of Lemma \ref{delta_c}}\label{Chebproof}
This is a basic well-known result for which we could not find a sufficiently pedestrian proof in the literature, so here we provide one, for completeness and reader's convenience. 

First, if $\rho_n\to \delta(z-\mu)$ on $C^{0}_{\mathbb C}([-1,1])$, then trivially $\mu_n\to\mu$ and $\sigma^2_n\to 0$. Now, let's suppose $\mu_n\to\mu$ and $\sigma^2_n\to 0$.

Given $f\in C^0_{\mathbb C}([-1,1])$, let $||f||_0$ be its sup-norm. For any $\eta > 0$, let  $\epsilon > 0$ be s.t.
\begin{equation}\label{epsiloneta}
|z-\mu|<\epsilon \implies |f(z)-f(\mu)|<\eta/3 \ .
\end{equation}

First, we have that 
\begin{equation}\label{eq11}
    \begin{aligned}
    \left|\int_{-1}^{1}f(z)\rho_n(z)dz-f(\mu)\right| 
    & \le \left|\int_{|z-\mu|\ge \epsilon}f(z)\rho_n(z)dz\right| + \left|\int_{|z-\mu|<\epsilon}f(z)\rho_n(z)dz-f(\mu)\right|\\
    & \le ||f||_0\int_{|z-\mu|\ge \epsilon}\rho_n(z)dz+\left|\int_{|z-\mu|<\epsilon}f(z)\rho_n(z)dz-f(\mu)\right| 
    \end{aligned}
\end{equation}
where the domains of integration on the r.h.s. are contained in $[-1,1]$.

By hypothesis $\mu_n\to\mu$, there exists $n_0\in\mathbb N$ such that, $\forall n> n_0$, $|\mu_n-\mu| < \epsilon/2$. But if $|z-\mu_n|< \epsilon/2$ for $n>n_0$, then $|z-\mu| \le |z-\mu_n|+|\mu_n-\mu| < \epsilon$.
Thus, $\forall n> n_0$, $|z-\mu|\ge \epsilon \implies |z-\mu_n|\ge \epsilon/2$ and, by Chebyshev's inequality,
\begin{equation*}
  n>n_0 \ \implies \ \int_{|z-\mu|\ge \epsilon}\rho_n(z)dz \ \le \ \int_{|z-\mu_n|\ge \epsilon/2}\rho_n(z)dz \ \le \ \dfrac{4\sigma^2_{n}}{\epsilon^2} \ .
\end{equation*}
Therefore, by hypothesis $\sigma^2_{n}\to 0$, there exists $n_1\geq n_0$ such that
\begin{eqnarray}
 n> n_1  &\implies& ||f||_0\int_{|z-\mu|\ge \epsilon}\rho_n(z)dz \ < \ \eta/3 \ , \label{eq22} \\
 &\implies& 1 \  \geq \  \int_{|z-\mu|<\epsilon}\rho_n(z)dz \  \geq \  1 - \frac{\eta}{3||f||_0} \ . \label{eq22b}
\end{eqnarray}

On the other hand, from (\ref{epsiloneta}) we get
\begin{equation*}
    \big(f(\mu)\!-\!\eta/3\big)\!\!\!\int_{|z-\mu|<\epsilon}\!\!\!\!\!\rho_n(z)dz <  \int_{|z-\mu|<\epsilon}\!\!\!\!\!\!f(z)\rho_n(z)dz\\
    < \big(f(\mu)\!+\!\eta/3\big)\!\!\!\int_{|z-\mu|<\epsilon}\!\!\!\!\!\rho_n(z)dz
\end{equation*}
which from (\ref{eq22b}) implies that, for $n>n_1$,  
\begin{equation*}
    \big(f(\mu)-\eta/3\big)\left(1 - \frac{\eta}{3||f||_0}\right)   < \  \int_{|z-\mu|<\epsilon}f(z)\rho_n(z)dz \  < \  f(\mu)+\eta/3  \ ,
\end{equation*}
and therefore 
\begin{equation}\label{eq33}
 n> n_1 \  \implies \  \  \left|\int_{|z-\mu|<\epsilon}f(z)\rho_n(z)dz-f(\mu)\right| \  < \  2\eta/3 \ . 
\end{equation}
Thus, from (\ref{eq11}), (\ref{eq22}) and (\ref{eq33}), we have (\ref{locC0}), that is,  
\begin{equation*} 
\forall\eta>0,\ \exists n_1\in\mathbb N \quad \mbox{s.t.} \quad  n>n_1\implies  \left|\int_{-1}^{1}f(z)\rho_n(z)dz-f(\mu)\right| <\eta \ .
\end{equation*}

\subsection{A proof of Lemma \ref{edm}}\label{Edproof}
We could not find a complete proof of Edmonds formula in the literature, so here we provide one.
At first, we follow Brussard and Tolhoek \cite{brus} and Flude \cite{flud} to prove that 
\begin{equation}\label{edm_1}
\lim_{j\to\infty}C_{\mu, m-\mu,\,\, m}^{\,l,\,\,\,j-\tau,\,\,\,\,j} = (-1)^{l-\tau}d^l_{\mu,\tau}(\theta),
\end{equation}
where $d^l_{\mu,\tau}$ is the Wigner (small) $d$-function and $\theta \in [0,\pi]$ is such that $\cos\theta = \lim_{j\to\infty}(m/j)$, holds for fixed $l$, $\mu$,  $\tau$ and  if either (i) $j-|m|\to \infty$ or (ii) $j-|m|\to 0$. 

For (i), we start with (\textit{cf.} \cite{bied, prios, varsh}):
\begin{equation}\label{ex_cg}
\begin{aligned}
C_{\mu, m-\mu,\,\, m}^{\,l,\,\,\,j-\tau,\,\,\,\,j} & = \sqrt{\dfrac{2j+1}{2j-\tau+l+1}}
\sqrt{(l-\tau)!(l+\tau)!(l+\mu)!(l-\mu)!}\\
&\,\,\,\,\,\,\,\times\sum_{z}\dfrac{(-1)^z\sqrt{\Omega_1\Omega_2\Omega_3}}{z!(l-\tau-z)!(l-\mu-z)!(\tau+\mu+z)!}
\end{aligned}
\end{equation}
where
\begin{equation*}
\Omega_1 \!\!=\!\! \dfrac{(2j-\tau-l)!}{(2j-\tau+l)!}, \, \Omega_2 \!\!=\!\! \dfrac{(j-\tau-m+\mu)!(j-m)!}{[(j-l-m+\mu+z)]^2}, \, \Omega_3 \!\!=\!\! \dfrac{(j-\tau+m-\mu)!(j+m)!}{[(j-\tau+m-\mu-z)!]^2}  .
\end{equation*}
The summation over $z$ is finite, so we can take the limit of each term. Thus, 
\begin{equation*}
\Omega_1\Omega_2\Omega_3 \sim  
\left(\dfrac{j-m}{2j}\right)^{2l-\tau-\mu-2z}\left(\dfrac{j+m}{2j}\right)^{\tau+\mu+2z} \ ,
\end{equation*}
where in the expressions for $\Omega_1$, $\Omega_2$, $\Omega_3$, we have used the Stirling approximation 
\begin{equation}\label{stir}
n! \sim \sqrt{2\pi}n^{n+1/2}e^{-n} \ .
\end{equation}

Supposing $m/j$ converges, we can take a $\theta\in [0,\pi]$ such that  $\cos\theta=\lim_{j\to\infty}(m/j)$, so that \ $\sin\frac{\theta}{2}=\sqrt{\frac{j-m}{2j}}$ and $\cos\frac{\theta}{2} = \sqrt{\frac{j+m}{2j}}$. Then, we get (\textit{cf.} \cite{varsh} for the exact expression of the Wigner $d$-function)
\begin{equation*}
\begin{aligned}
\lim_{j\to\infty}C_{\mu, m-\mu,\,\, m}^{\,l,\,\,\,j-\tau,\,\,\,\,j} & = \sqrt{(l-\tau)!(l+\tau)!(l+\mu)!(l-\mu)!}\\
& \,\,\,\,\,\,\,	\times\sum_{z}\dfrac{(-1)^z\left(\sin\frac{\theta}{2}\right)^{2l-\tau-\mu-2z}\left(\cos\frac{\theta}{2}\right)^{\tau+\mu+2z}}{z!(l-\tau-z)!(l-\mu-z)!(\tau+\mu+z)!}\\
& = (-1)^{l-\tau}d^l_{\mu,\tau}(\theta) \ .
\end{aligned}
\end{equation*}

Now, we must note that $j-|m|$ is a sequence of integer numbers, so the convergence condition (ii) implies that there is a $j_0$ such that for all $j>j_0$ we have $j-|m| = 0 \Rightarrow m = \pm j$. For $m = j$, we have $\mu\ge |\tau|$ and $m/j\to 1\Rightarrow \theta = 0\Rightarrow d^l_{\mu,\tau} = \delta_{\mu,\tau}$; for $m = -j$, we have $\mu\le|\tau|$ and $m/j\to -1\Rightarrow \theta = \pi\Rightarrow d^l_{\mu,\tau} = (-1)^{l-\tau}\delta_{-\mu,\tau}$. Writing, again, the exact formula 
\begin{equation*}
C_{j,-j\pm\mu, \pm\mu}^{j,\,\,\,\,j-\tau,\,\,\, l}\!\! =\!\! \sqrt{\dfrac{2l+1}{2j-\tau+l+1}}\!\sqrt{\dfrac{(l-\tau)!(l\pm\mu)!}{(l+\tau)!(-\tau\pm\mu)!(l\mp\mu)!}}\!\sqrt{\dfrac{(2j)!(2j-\tau\mp\mu)!}{(2j-\tau+l)!(2j-\tau-l)!}}
\end{equation*}
and then using (\ref{stir}) in the last factor, plus the symmetry properties of Clebsch-Gordan coefficients (\textit{cf.} \cite{bied, prios, varsh}), we conclude that 
\begin{equation*}
\lim_{j\to\infty}C_{\mu, j-\mu,\,\, j}^{\,l,\,j-\tau,\,\,\,\,j}  = (-1)^{l-\tau}\delta_{\mu,\tau} \quad , \quad
\lim_{j\to\infty}C_{\mu, -j-\mu,\,\, -j}^{\,l,\,\,\,j-\tau,\,\,\,\,j}  = \delta_{-\mu,\tau}\ \ .
\end{equation*}

Now, we expand condition (ii) to condition (iii) $j-|m|<a$ for some $a\in\mathbb{N}$ when $\tau=\mu=0$. We first assume $j-|m|\to a \in\mathbb{N}_0$. From \cite{varsh}, we have the following recursive relation:
\begin{equation}\label{rec}
\begin{aligned}
C_{0, \,\pm(j-a-1),\, \pm(j-a-1)}^{\,\,l,\,\,\,\,\,\,\,\,\,\,\,\,\,j,\,\,\,\,\,\,\,\,\,\,j} & = \sqrt{\dfrac{l(l+1)}{(2j-a)(a+1)}}C_{\pm1, \pm(j-a-1),\, \pm(j-a)}^{\,l,\,\,\,\,\,\,\,\,\,\,\,\,\,\,j,\,\,\,\,\,\,j}\\
&\,\,\,\,\,\,\,+C_{0, \pm(j-a),\, \pm(j-a)}^{\,\,l,\,\,\,\,\,\,\,j,\,\,\,\,\,j}
\end{aligned}
\end{equation}
Assuming the coefficients in the rhs of (\ref{rec}) satisfy the already proved (\ref{edm_1}), we obtain
\begin{equation*}
\lim_{j\to\infty}C_{0, \,\pm(j-a-1),\, \pm(j-a-1)}^{\,\,l,\,\,\,\,\,\,\,\,\,\,\,\,\,j,\,\,\,\,\,\,\,\,\,\,j} = (-1)^ld^l_{0,0}(\theta)
\end{equation*}
where $\theta = 0$ for plus sign and $\theta = \pi$ for minus sign. The hypothesis holds for $a = 0$, so, by induction, the formula holds for all $a\in\mathbb{N}_0$. Finally,  $j-|m|$ does not need to converge to $a\in\mathbb{N}_0$, it is sufficient for it to remain finite, because all the Clebsch-Gordan coefficients converge to the same value.

Putting together all the results, we have proved that, if $\lim_{j\to\infty}(m/j) = \cos\theta = 1-2r$, then
\begin{equation*}
\lim_{n\to\infty}(-1)^{k_n-1}C_{m, -m,\,\, 0}^{\,\,\,j,\,\,\,\,\,\,\,j,\,\,\,l}\sqrt{\dfrac{n+1}{2l+1}} = \lim_{n\to\infty}(-1)^lC_{0, m, m}^{l, \,\,\,\,j,\,\, j} = d^l_{0,0}(\theta) = P_l(1-2r) \ ,
\end{equation*}
which is Edmonds formula (\ref{edf}).

\bibliographystyle{plain}

\end{document}